\documentclass{aa_edt}

\usepackage[varg]{txfonts} 

\usepackage{natbib}
\bibpunct{(}{)}{;}{a}{}{,} 

\usepackage{graphicx}

\usepackage[english]{babel}

\usepackage{mathtools}		
\usepackage[amssymb,alsoload=hep]{SIunits}
\usepackage{subfig}

\usepackage{url}
\usepackage[mathcal]{euscript}

\usepackage{hyperref}

\renewcommand{\eqref}[1]{equation~(\ref{#1})}

\newcommand{\figref}[1]{Fig.~\ref{#1}} 

\newcommand{\figsref}[1]{Figs.~\ref{#1}} 

\newcommand{\phn}{\phantom{4}}
\newcommand{\phd}{\phantom{.}}
\newcommand{\phs}{\phantom{-}}

\begin{document} 

\title{AME - Asteroseismology Made Easy}
\subtitle{Estimating stellar properties by use of scaled models\thanks{\url{http://sac.au.dk/scientific-data/ame}}}

\author{Mia Lundkvist\inst{1}\fnmsep\thanks{\email{lundkvist@phys.au.dk}}
\and Hans Kjeldsen\inst{1}
\and Victor Silva Aguirre\inst{1}}

\institute{Stellar Astrophysics Centre (SAC), Department of Physics and Astronomy, Aarhus University, Ny Munkegade 120, DK-8000 Aarhus C, Denmark}

\date{Received Jan 13 2014 /
Accepted Apr 7 2014}

\abstract
{Stellar properties and in particular stellar radii of exoplanet host-stars are essential for measuring properties of exoplanets. Therefore it is becoming increasingly important to be able to supply reliable stellar radii fast. Grid-modelling is an obvious choice to do this, but that only offers a low degree of transparency to the non-experts.}
{Here we present a new method to obtain stellar properties for stars exhibiting solar-like oscillations in an easy, fast, and transparent way. The method, called Asteroseismology Made Easy (AME), can determine stellar masses, mean-densities, radii, and surface gravities, as well as estimate ages. In this writing we present AME as a visual and powerful tool which could be useful; in particular in the light of the large number of exoplanets being found.}
{AME consists of a set of figures from which the stellar parameters are deduced. These figures are made from a grid of stellar evolutionary models that cover masses ranging from $0.7 M_\sun$ to $1.6 M_\sun$ in steps of $0.1 M_\sun$ and metallicities in the interval $-0.3 \ \mathrm{dex} \leq [\mathrm{Fe}/\mathrm{H}] \leq +0.3 \ \mathrm{dex}$ in increments of $0.1 \ \mathrm{dex}$. The stellar evolutionary models are computed using the Modules for Experiments in Stellar Astrophysics (MESA) code with simple input physics.}
{We have compared the results from AME with results for three groups of stars; stars with radii determined from interferometry (and measured parallaxes), stars with radii determined from measurements of their parallaxes (and calculated angular diameters), and stars with results based on the modelling of their individual oscillation frequencies. We find that a comparison of the radii from interferometry to those from AME yield a weighted mean of the fractional differences of just $2\%$. This is also the level of deviation that we find when we compare the parallax-based radii to the radii determined from AME.}
{The comparison between independently determined stellar parameters and those found using AME show that our method can provide reliable stellar masses, radii, and ages, with median uncertainties in the order of $4\%$, $2\%$, and $25\%$ respectively.}


\titlerunning{AME - Asteroseismology Made Easy}
\authorrunning{M. Lundkvist et al.} 

\maketitle


\section{Introduction}
\label{sec:intro}

NASA's \textit{Kepler Mission} \citep{ref:kepler_koch} has provided photometric light curves of high precision which has enabled the detection of solar-like oscillations in more than 500 stars \citep{ref:kepler_chaplin}. Asteroseismic scaling relations are generally used to determine properties of such stars \citep[for a recent review on those issues see][]{ref:ast_sca_chaplin}. If one considers stars in hydrostatic equilibrium, it follows from homology that since the dynamical time scale is proportional to ${\bar{\rho}}^{-1/2}$ (where $\bar{\rho}$ is the mean-density) the oscillation frequencies will be proportional to ${\bar{\rho}}^{1/2}$. In the asymptotic approximation \citep[see e.g. page 218 of][]{ref:ast_book} the frequency spectrum for a star can be written as a function of radial order ($n$) and angular degree ($\ell$) as 
\begin{equation}
	\label{eq:asymp_rel}
	\nu_{n,\ell} = \Delta\nu \cdot \left( n + \ell /2 + \varepsilon \right) - \ell \left( \ell + 1 \right) \cdot D_0 \ .
\end{equation}
Since the frequencies scale with the square root of the mean-density, the main regularity in the asymptotic relation (expression~(\ref{eq:asymp_rel})), the large separation ($\Delta\nu$), will therefore be proportional to $\sqrt{\bar{\rho}}$. The parameters $D_0$ and $\varepsilon$ will depend on the detailed internal structure of a given star and can e.g. be used to estimate its age.

The large separation can be found from the sound-speed integral:
\begin{equation}
	\label{eq:sound_speed_integral}
	\Delta\nu = \left( 2 \int \limits_0^R \frac{\mathrm{d}r}{c(r)} \right) ^{-1} \ ,
\end{equation}
where $R$ is the stellar radius and $c(r)$ the sound-speed in the stellar interior. We will in this study use this most fundamental asteroseismic parameter $\Delta\nu$ to estimate the stellar mean-density. It has been common to also use the frequency at maximum oscillation power ($\nu_\mathrm{max}$) to measure the gravity of a given star since observationally it has been shown to be proportional to the acoustic cut-off frequency in the atmosphere (which scales with $g/\sqrt{T_\mathrm{eff}}$). While the observational link to the acoustic cut-off frequency is established \citep{ref:stello_2009} we still need to better understand the theoretical background for this relation \citep{ref:belkacem_2011,ref:belkacem_2013}. An issue in relation to the use of $\nu_\mathrm{max}$ is that there is a tight relation between $\nu_\mathrm{max}$ and $\Delta\nu$ \citep[see][]{ref:stello_2009}. This is the reason why we in the present study only consider $\Delta\nu$ and not $\nu_\mathrm{max}$. 

Obtaining stellar parameters such as mass, radius, and age through detailed modelling (modelling based on individual oscillation frequencies) for just one star showing solar-like pulsations can be a very time-consuming task. As an alternative, one or more of the analysis pipelines described in for instance \citet{ref:pip_seek} (SEEK), \citet{ref:pip_mathur}, \citet{ref:pip_radius} (RADIUS), \citet{ref:pip_yb}, and \citet{ref:pip_ybplus} can be utilised. However, this grid-based modelling does not offer a very transparent procedure and inconsistencies can occur between the mean-density derived using the large separation ($\bar{\rho} \propto \Delta \nu ^2$) and the mean-density found using the grid-determined mass and radius ($\bar{\rho} \propto M/R^3$). This is primarily an effect of mass and radius being determined individually \citep{ref:pip_seek} and the fact that grid-based modelling sometimes yield bimodal parameter distributions.

\citet{ref:pip_white} discuss the potential of constraining stellar masses and ages for main-sequence (MS) and sub-giant solar-like oscillators using diagrams based on their pulsation properties, such as the Christensen-Dalsgaard and the $\varepsilon$ diagrams. However, $\varepsilon$ can be hard to constrain \citep{ref:white_epsilon}, and the Christensen-Dalsgaard diagram requires the value of the small separation ($\delta \nu$) to be known and this is usually not available. The reason for this is that to find $\delta\nu$ some modes with $\ell=2$ need to be discernible in the power spectrum, which is often not the case.

Therefore we have developed a method to obtain self-consistent mean-densities along with other stellar parameters in an easy and fast way for MS and sub-giant stars showing solar-like oscillations. We consider this method to be transparent because it is easy to follow, from considering the figures, what effect a small change in one of the input parameters would have on the result. By scaling to the Sun we have made our method less sensitive to the stellar models and evolution code used. We call this method AME - Asteroseismology Made Easy. The aim of this paper is to present AME as a powerful and visual tool which does not need the frequency of maximum power to yield basic stellar properties.


\section{The AME models}
\label{sec:ame_models}

AME allows for the determination of stellar mean-density ($\bar{\rho}$), mass ($M$), and age ($\tau$), and by inference surface gravity ($g$) and radius ($R$) of MS and sub-giant stars showing solar-like oscillations. The method is based on a grid of models spanning a range in metallicity and mass from which we have made plots that can be used to determine the aforementioned parameters. We wanted to make our determination of the stellar parameters as insensitive as possible to the choice of evolution code used. Therefore, we have scaled all results to the Sun and we have tested the effects on the Hertzsprung-Russell diagram (H-R diagram) of using different evolution codes.


\subsection{The importance of the evolution code}
\label{subsec:evolcodes}

We have tested the sensitivity of our method to the choice of stellar evolution code and solar chemical composition. We did this to make sure that the results that come out of AME are not affected to any significant degree by either of the two. In order to facilitate this test, we computed a series of evolution tracks and compared them qualitatively.

We calculated a series of four models using each of the three evolution codes MESA \citep[Modules for Experiments in Stellar Astrophysics,][]{ref:mesa,ref:mesa_new}, ASTEC \citep[Aarhus STellar Evolution Code,][]{ref:astec}, and GARSTEC \citep[Garching Stellar Evolution Code,][]{ref:garstec}. These four models consisted of two using the solar abundances by \citet{ref:GS98} (hereafter GS98) and two using the newer abundances by \citet{ref:Asp09} (hereafter A09). For each set of abundances we computed an evolution track for a $1.0 M_\sun$ and for a $1.2 M_\sun$ star.

The tracks were calculated using as similar input physics as possible for the three evolution codes. The input microphysics that we used were the 2005 update of the OPAL EOS \citep{ref:opal_1996,ref:opal_2002}, the OPAL opacities \citep{ref:opal_opacity} supplemented by the \citet{ref:lowT_2005} opacities at low temperatures, the NACRE nuclear reaction rates \citep{ref:nacre_reac_rates} with the updated $^{14}\mathrm{N}(p,\gamma)^{15}\mathrm{O}$ reaction rate by \citet{ref:14N}, and no diffusion or settling.

For the macrophysics, we neglected convective overshoot and used an Eddington $T-\tau$ relation as the atmospheric boundary condition. We treated convection using the mixing-length theory of convection; in the case of GARSTEC in the formulation of \citet{ref:mlt_garstec}, and in the cases of ASTEC and MESA in the formulation of \citet{ref:mixinglength}.

\begin{figure}
   \centering
   \includegraphics[width=\hsize]{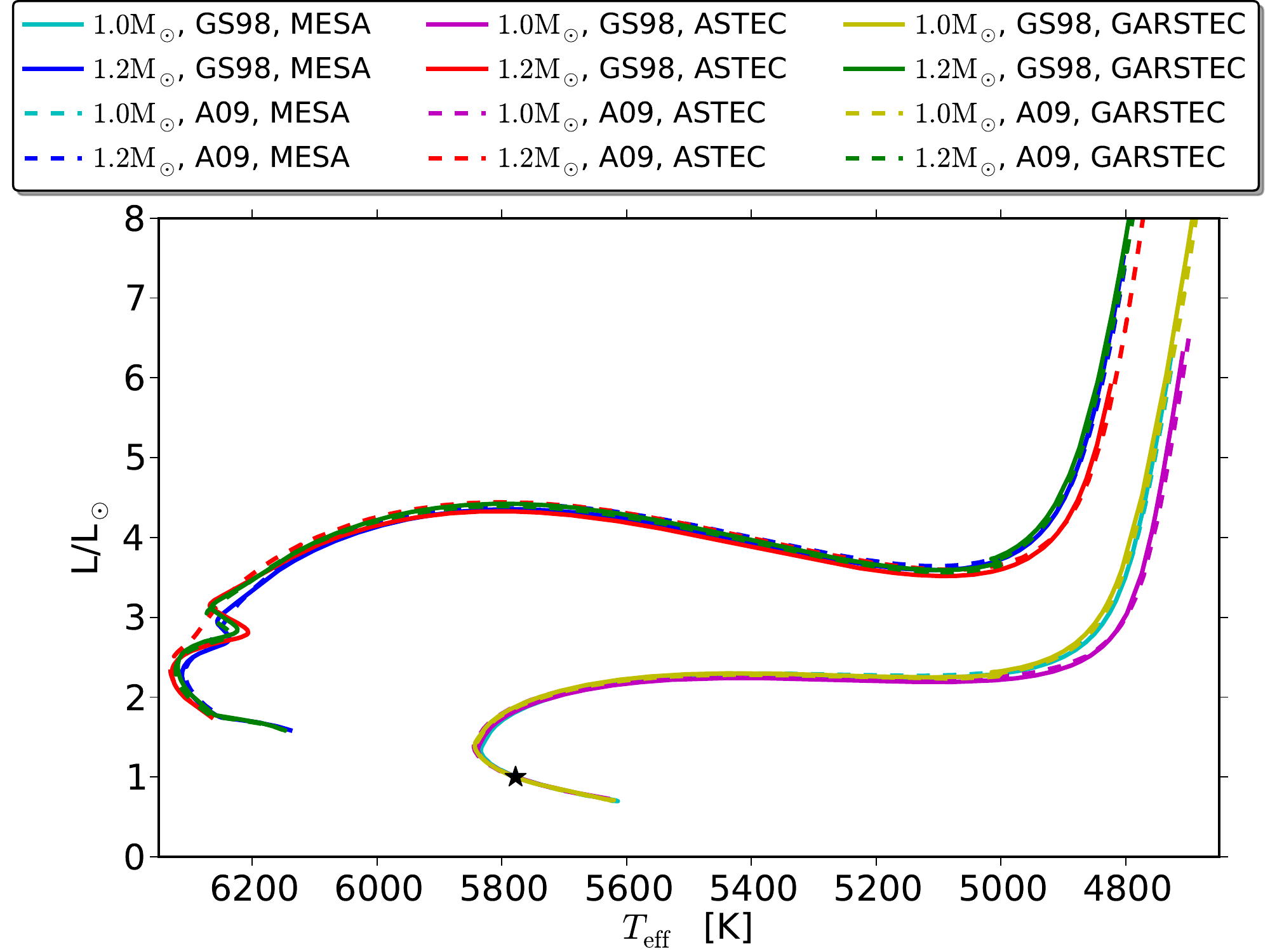}
   \caption{H-R diagram comparing evolution tracks made with MESA, ASTEC, and GARSTEC for stars of mass $1.0 M_\sun$ and $1.2 M_\sun$ with two different chemical compositions; GS98 and A09. The star is the Sun.}
   \label{fig:hr_codes}
\end{figure}

Before the tracks were made, each of the codes was calibrated to the Sun using each of the GS98 and A09 abundances. This calibration was performed by adjusting the mixing-length parameter ($\alpha_\mathrm{ML}$) and the initial helium ($Y_0$) and heavy-element ($Z_0$) mass-fractions until a model with the mass of the Sun had the solar radius and luminosity, and a surface value of $Z/X$ in accordance with the abundances chosen, at the age of the Sun. The target values of $Z/X$ were (for the present-day photosphere):  $\left(Z/X \right)_\mathrm{GS98} = 0.0231$ and $\left(Z/X \right)_\mathrm{A09} = 0.0181$ \citep{ref:Asp09}. The twelve resulting tracks can be found in \figref{fig:hr_codes}.

There are two places where the evolution tracks in \figref{fig:hr_codes} differ slightly. This is at the bottom of the Red Giant Branch (RGB) and in the $1.2 M_\sun$ case around the MS turn-off. The difference at the bottom of the RGB is not really relevant here since the stars in this region are not the prime focus of AME. The difference around the MS turn-off is to be expected since the change from a radiative core on the MS to a convective core happens at a mass just around $1.2 M_\sun$ and as a consequence there is a high sensitivity to any small differences between the codes at this particular mass. This is why this mass was included in the test in the first place, and the codes do not differ drastically. The GARSTEC tracks show clear signs of convective cores for both solar compositions (GS98 and A09) and the ASTEC and MESA tracks also show convective cores in both cases, but these are very small when using the A09 abundances. The difference in the size of the convective core when using the GS98 and A09 abundances can be understood since the GS98 composition has a higher abundance of the CNO-elements than the A09 composition. This higher CNO-element abundance will cause the CNO cycle to be more important which will lead to a slightly higher temperature dependence in the core since the CNO cycle depends more strongly on the temperature than the pp-chain. Thereby, the temperature gradient in the core will be slightly larger and therefore a star with the GS98 abundances will be more prone to have a convective core \citep[see also][]{ref:abundances_matter}.

Thus, the three codes give quite similar results. This is reassuring because it means that the decision to use MESA to make the stellar models for AME does not significantly affect the results. Also, we chose to use the A09 abundances to make the AME grid of models, and it is evident from \figref{fig:hr_codes} that this choice does not have a great impact on the results either. However, it is not clear from \figref{fig:hr_codes} how the ages of the different tracks compare. We have found that for the most evolved star that AME has been used on, the ages coming from the different tracks agree at the $5\%$ level. This means that also the ages from AME are fairly insensitive to the choice of evolution code and set of abundances, at least to the uncertainty level of the ages from AME (see Sect.~\ref{subsec:dis_mod_stars}).


\subsection{Grid of scaled models}
\label{subsec:scaledmodels}

We used stellar models obtained with MESA to generate a grid of models with masses spanning $0.7M_\odot - 1.6M_\odot$ in steps of $0.1M_\odot$ and metallicities in the range $-0.3  \ \text{dex} \leq [\mathrm{Fe}/\mathrm{H}] \leq +0.3  \ \text{dex}$ in increments of $0.1 \ \text{dex}$.

We chose to keep the stellar models simple which meant that we used MESA with the input physics described in Sect.~\ref{subsec:evolcodes} with the exception that we used the MESA photospheric tables \citep{ref:mesa} instead of an Eddington grey atmosphere. The value of the mixing length parameter was kept fixed at the value found by calibrating to the Sun; $\alpha_\mathrm{ML} = 1.71$.

The hydrogen- ($X$), helium- ($Y$), and heavy-element mass-fractions ($Z$) obtained from the solar calibration assuming the A09 abundances were used to compute the stellar mass-fractions assuming
\begin{equation}
	\label{eq:FeH}
	[\mathrm{Fe}/\mathrm{H}] = [\mathrm{M}/\mathrm{H}] = \log \left( \frac{Z}			{X} \right) _* - \log \left( \frac{Z}{X} \right) _\sun \ ,
\end{equation}
where the subscript $_*$ indicates the star, and the initial solar mixture is known from our solar calibration; $X_\sun = 0.71247$ and $Z_\sun = 0.01605$ (we use the modelled solar composition when the Sun arrived on the ZAMS in equation~(\ref{eq:FeH}) since our models contain no diffusion or settling). We used a helium evolution of $\frac{\Delta Y}{\Delta Z} = \frac{Y-Y_\mathrm{BBN}}{Z-Z_\mathrm{BBN}} = 1.41$ \citep{ref:balser_2006} with a Big Bang Nucleosynthesis helium mass-fraction of $Y_\mathrm{BBN} = 0.2488$ \citep{ref:steigman_2010} and heavy-element mass-fraction of $Z_\mathrm{BBN} = 0$. This, in combination with the well-known relation $X + Y + Z = 1$, allowed us to determine the stellar mass-fractions for a given metallicity.

All of the models in the grid were evolved until they reached a large separation of $20 \ \micro\hertz$ since this does not happen until far beyond the MS which is the prime focus of AME.


\subsection{Implications of the chosen input physics}
\label{subsec:inputphysics}

Since we have chosen to use simple input physics for our grid of models, we have made choices which may lead to significant systematic errors in our results \citep[see for instance][for a discussion of the mixing length parameter]{ref:importance_solar_alpha}. These choices include to neglect diffusion and overshoot and only consider one value of the mixing length parameter and the helium evolution.

In order to investigate the likely size and direction of any such systematics, we have tested the influence of changed input physics on our results. We did this by computing evolutionary tracks for five stellar masses ($M = 1.0, 1.1, 1.2, 1.3, 1.4 \ M_\sun$) where we changed one of our assumptions at a time (using just a single metallicity). We computed two sets of tracks that used a different mixing length parameter (one higher than our solar calibration and one lower; $\alpha_\mathrm{ML} = 1.51$ and $\alpha_\mathrm{ML} = 1.91$), one set of tracks that included overshoot (with a value $f_\mathrm{ov} = 0.015$), one set that incorporated element diffusion, and two sets that used a different helium evolution (again, one with a lower value and one with a higher value than the one we chose; $\tfrac{\Delta Y}{\Delta Z} = 1.00$ and $\tfrac{\Delta Y}{\Delta Z} = 1.82$). Subsequently, we compared the masses and radii derived for $20$ artificial stars using the tracks with our chosen input physics to the results obtained when using each of the sets of tracks with an altered input physics. The median difference (in the sense \textit{changed physics} - \textit{chosen physics}) can be seen in Table~\ref{tab:inputphysics}.

\begin{table*}
\caption{Median mass and radius differences caused by changes in the input physics.}
\label{tab:inputphysics}      
\centering          
\begin{tabular}{lll}     
\hline\hline       
Modification to input physics
& \multicolumn{1}{c}{$\mathrm{med(}\Delta M \mathrm{)} \ [\%]$}
& \multicolumn{1}{c}{$\mathrm{med(}\Delta R \mathrm{)} \ [\%]$} \\ 
\hline
$\Delta \alpha = \pm 0.20$		& $\mp 2.5$		& $\mp 0.5$ \\
$f_\mathrm{ov} = 0.015$			& $+0.2$		& $+0.05$ \\
Diffusion						& $+0.6$		& $+0.3$ \\
$\Delta \left( \tfrac{\Delta Y}{\Delta Z} \right) = \pm 0.41$ & $\mp 1.8$ & $\mp 0.6$ \\
\hline                  
\end{tabular}
\end{table*}

As can be seen from the table, the systematic effects that may be caused by neglecting convective overshoot or diffusion are rather small for the selected mass range while they are significant when choosing only a single value for $\alpha_\mathrm{ML}$ and $\tfrac{\Delta Y}{\Delta Z}$. Yet we have chosen to do this to keep the input physics and the method itself simple. As a consequence, these systematic effects should be borne in mind (and note that we solely quote internal uncertainties throughout this paper).

\citet{ref:valle} has obtained similar values for the effect on mass and radius from changing the value of the mixing length parameter or the helium evolution. However, they consider a lower mass range ($0.8 \leq M/M_\sun \leq 1.1$). This may explain why they find diffusion to be more significant than we do since diffusion is known to play an important role in the Sun, but seem to be counteracted by processes like radiative levitation for F-stars \citep{ref:turcotte}. This is also one of the reasons why we chose to neglect diffusion \citep[radiative levitation is not included in MESA, see][]{ref:mesa}.

The systematic uncertainties caused by the negligence of convective overshoot is not considered by \citet{ref:valle}, but we find that for the masses we investigated, it does not significantly influence the mass or the radius. However, it can affect the age. Including overshoot will cause typical age differences which are vanishingly small compared to the age uncertainties in AME (see Sect.~\ref{subsec:dis_mod_stars}), but during the MS turn-off, the age differences can be on the order of $15\%$ which is quite significant \citep[see for instance][who find a similar value]{ref:age_msturnof_diff}. However this phase of evolution is very fast and the chance of catching a star during this evolutionary step is fairly low. Therefore we have assumed that convective overshoot can be disregarded.

It is worth noting that while the systematics are not negligible they are still smaller than the median uncertainties on mass, radius, and age returned by AME which are $4\%$, $2\%$, and $25\%$ respectively (see Sect.~\ref{subsec:dis_mod_stars}). Also, we have found that our results are in excellent agreement with the stellar parameters found in the literature (see Sects.~\ref{sec:results} and~\ref{sec:discussion}).


\section{The AME figures}
\label{sec:ame_figs}

\begin{figure}
	\centering
	\includegraphics[width=\hsize]{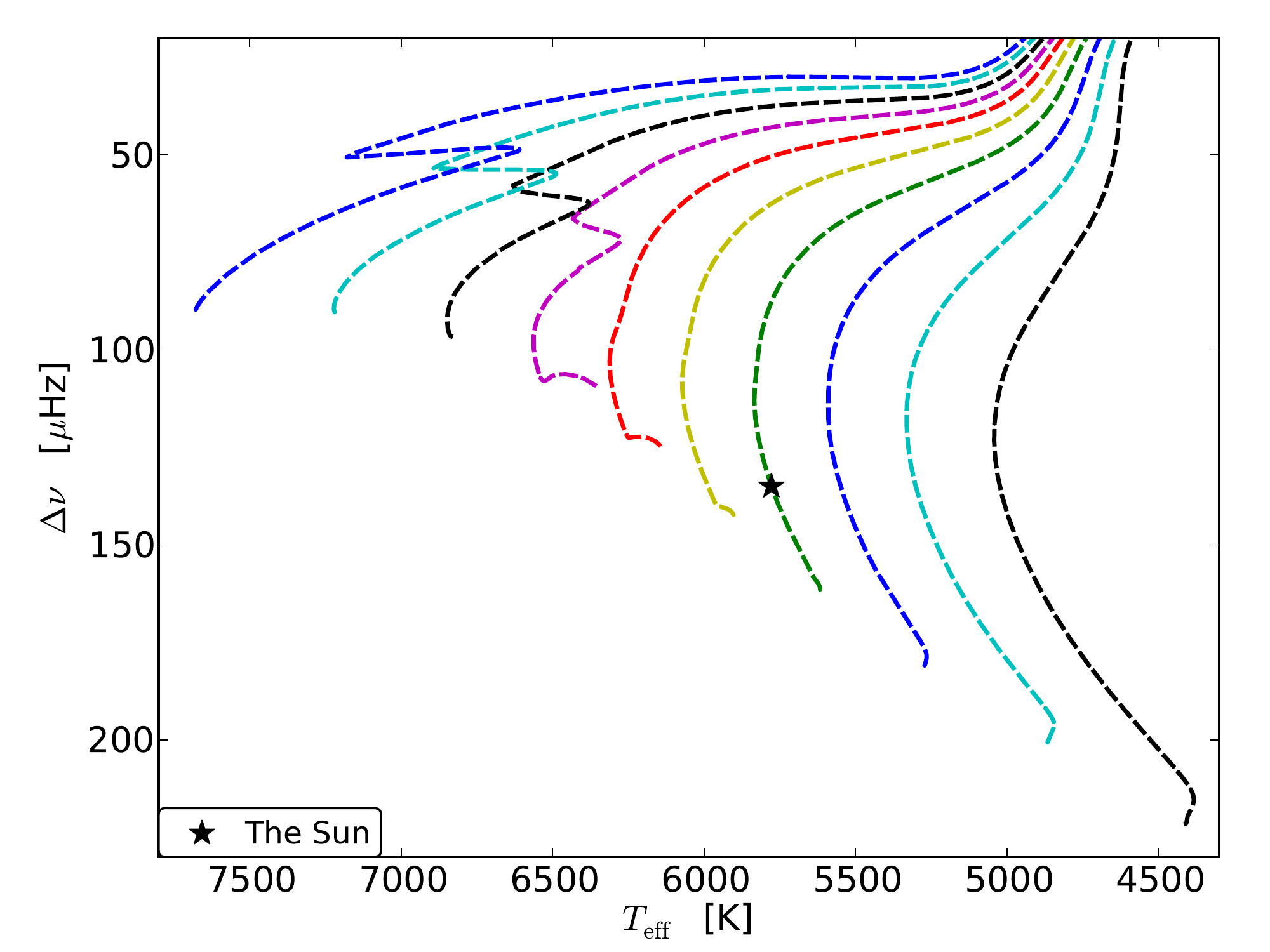}
	\caption{Evolutionary tracks for models with varying mass and solar metallicity plotted in a variation of a classic H-R diagram ($\Delta \nu$ plotted against $T_\mathrm{eff}$). The star shows the location of the Sun and the different line colours represent different masses starting with the black $0.7 M_\sun$ line on the far right going in increment of $0.1 M_\sun$ to the blue $1.6 M_\sun$ line on the far left.}
	\label{fig:JCD_request}
\end{figure}

The foundation of AME is the grid of models created using MESA, which we described in Sect.~\ref{subsec:scaledmodels}. We have used some of the output parameters from each model in the grid to create three types of plots. The idea of using plots to obtain stellar parameters came from the observation that evolutionary tracks in an H-R diagram look approximately similar for lines of neighbouring mass and metallicity. This is illustrated in \figref{fig:JCD_request} which shows  $\Delta\nu$ as a function of $T_\mathrm{eff}$ for models of solar metallicity. Based on this, we expected to be able to more or less make lines of different masses and metallicities overlap if we adjusted the axes - both in a plot like \figref{fig:JCD_request} and when plotting other stellar properties against each other.

The model outputs that we have used to make the AME plots are the large separation, the mass, the age, the temperature, the metallicity, and the mean-density. We scaled the found large separation by the ratio of the observed solar large separation to the modelled solar large separation (from the MESA solar calibration, calculated from the sound-speed integral in expression~(\ref{eq:sound_speed_integral})); $\frac{\Delta \nu_{\sun, \mathrm{obs}}}{\Delta \nu_{\sun, \mathrm{mod}}} = \frac{134.95 \ \micro\hertz}{141.8768 \ \micro\hertz} = 0.95118$, where the observed value of the solar large separation is from \citet{ref:largesep_sun}. This was done in order to correct for the difference between the $\Delta \nu$ found from (\ref{eq:sound_speed_integral}) which does not take near-surface effects into account and the observed $\Delta \nu$ found as $\Delta \nu = \nu_{n+1,0}-\nu_{n,0}$ which is affected by this. We have checked that the ratio found for the Sun is consistent with the ratios of $\Delta\nu$ found from model frequencies to $\Delta\nu$ found from the sound-speed integral (equation~(\ref{eq:sound_speed_integral})) for other stars to around $1\%$ and that this is independent of mass, chemical composition, and evolutionary state. A similar discrepancy has been found by \citet{ref:explain_par_diff} between the $\Delta\nu$ found from the oscillation frequencies and the one calculated from the scaling relation $\Delta\nu \propto \sqrt{\bar{\rho}}$ (see Sect.~\ref{sec:intro}).

In order to take the small difference into account and not to end up with an erroneously low error bar on the stellar parameters, we have set a lower limit on the uncertainty on $\Delta \nu$ of $1\%$. Thus if a star has a lower uncertainty on the large separation than $1\%$, we inflate this uncertainty to be $1\%$ before we use AME. This means that the plots (created using models) can (and should) be used with observed values of the large separation. By always scaling to the Sun it should also be possible to compensate for any systematic effect which is shared by all the models in the grid.

From the three types of plots, the mass, mean-density, and age can be determined based on only the large separation ($\Delta \nu$), the effective temperature ($T_\mathrm{eff}$), and the metallicity ($[\mathrm{Fe}/\mathrm{H}]$) of the star, using linear interpolation between the lines in the plots. From the mean-density and the mass, the radius and surface gravity can be derived using equations~(\ref{eq:sca_R})~and~(\ref{eq:sc_g}) respectively:
\begin{align}
	\label{eq:sca_R}
	\frac{R}{R_\odot} &= \left( \frac{M}{M_\odot} \right)^{1/3} \left(\frac{\bar{\rho}}{\bar{\rho}_\odot}\right) ^{-1/3} \ , \\
	\label{eq:sc_g}
		\frac{g}{g_\odot} &= \left( \frac{M}{M_\odot} \right)^{1/3} \left(\frac{\bar{\rho}}{\bar{\rho}_\odot}\right) ^{2/3} \ .
\end{align}

There are several versions of each type of plot which each corresponds to a region of the $\left( \Delta \nu, T_\mathrm{eff}, [\mathrm{Fe}/\mathrm{H}] \right)$ parameter space with a small overlap. An example of each type of plot can be seen in \figsref{fig:m_lm} (mass), \ref{fig:rho_lm}~(mean-density), and \ref{fig:tau_lm}~(age) in Sect.~\ref{sec:alphaCenA}. The full set of figures consisting of three mass plots, six mean-density plots and six age plots can be found in the appendix~\ref{om_sec:morefigs}.


\subsection{The figure axes}
\label{subsec:fig_axes}

The axes on the plots have been chosen to make the masses separate out in the type of plot used to determine stellar mass, and the mass and metallicity to have as little effect as possible when establishing the stellar mean-density and age. Thus the axes are the results of iterations based on physical expectations and the attempt to make the lines collapse. Below, we describe the axes of each plot-type.

\subsubsection{The axes of the mass plots}
\label{subsubsec:axes_mass}

For the type of plot used to determine the masses (an example can be seen in \figref{fig:m_lm}) we have used axes of the form $T_\mathrm{eff} \cdot 10^{a \cdot [\mathrm{Fe}/\mathrm{H}]}$ and $\Delta \nu \cdot 10^{-b \cdot [\mathrm{Fe}/\mathrm{H}]}$. The large separation decreases monotonically through the evolution of a solar-like oscillator whilst lines of constant density across evolution tracks in an H-R~diagram have different temperatures. Therefore a temperature versus large separation plot should separate out the different masses. The small metallicity dependence given by the variables $a$ and $b$ is used to collapse the lines of differing metallicity, but identical mass. The coefficients vary from plot to plot, but in each case the values which were found empirically to best collapse the lines were chosen.
	
\subsubsection{The axes of the mean-density plots}
\label{subsubsec:axes_density}

In the type of plot used to infer the mean-density, the main relation that we have used is the well-known asteroseismic scaling relation $\Delta \nu \propto \sqrt{\bar{\rho}}$ \citep{ref:sca_kjeldsen,ref:bk_dnu_rho} which was also mentioned in Sect.~\ref{sec:intro}. Following it, $\bar{\rho}/\Delta\nu^2$ is expected to be reasonably constant for all stars. Therefore, this is used on the x-axis since it results in a large precision on the mean-density. It is used in combination with a small metallicity dependence as in the mass plots, and a slight mass dependence. These serve the purpose of collapsing the lines to the largest extent possible.

On the other axis we have used a combination of the form $\Delta \nu \cdot M^c \cdot 10^{-d \cdot [\mathrm{Fe}/\mathrm{H}]}$. Thereby, except for the small metallicity dependence which serves the same purpose as previously, the axis is a product of the large separation and the mass. The exponent of the mass ($c$) varies a little from plot to plot (again, depending on what is found to best collapse the lines), but it is in the order of one. Based on the approximate proportionality on the MS of radius and mass, the product of the mass to a power in the vicinity of unity and the large separation will be roughly similar for all stars on the MS, which explains why this is on the y-axis.
	
\subsubsection{The axes of the age plots}
\label{subsubsec:axes_age}

The x-axis of the age plots has the form $\Delta \nu \cdot M^f \cdot 10^{g \cdot [\mathrm{Fe}/\mathrm{H}]}$ where $g$ can be both negative and positive, but is around zero and $f$ is positive and around one. This is therefore much the same axis as was used on the y-axis of the mean-density plots.

The y-axis features the age of the star and the mass to an approximate exponent of four in addition to a now well-known term in metallicity to help collapse the lines. The use of the product $\tau \cdot M^{4}$ was based on the nuclear time scale which is given by $t_\mathrm{nuc} \propto M L^{-1}$ \citep[equation (1.90) of][]{ref:hkt_book} and the luminosity-mass relation for low-mass stars $L \propto M^{5.5}$ \citep[page 29 of][]{ref:hkt_book}. Combining these two and taking the nuclear time scale as an estimate of the total stellar age, it is found that the product of stellar age and mass to the power of $\sim 4.5$ is expected to be more or less similar for all stars. This is why we used it on the y-axis of the age plots.


\section{The 3 steps in using AME; the case of $\alpha$ Cen A}
\label{sec:alphaCenA}

\begin{table*}
\caption{Literature values for stars with results from detailed modelling.}
\label{tab:litt_values}      
\centering          
\begin{tabular}{llllllll}     
\hline\hline       
Star 	& \multicolumn{1}{c}{$\Delta \nu \ [\micro\hertz]$}
& \multicolumn{1}{c}{$T_\mathrm{eff} \ [\kelvin]$}
& \multicolumn{1}{c}{$[\mathrm{Fe}/\mathrm{H}] [\text{dex}]$}
& \multicolumn{1}{c}{$\mathrm{M}/\mathrm{M}_\odot$}
& \multicolumn{1}{c}{$\mathrm{R}/\mathrm{R}_\odot$}
& \multicolumn{1}{c}{$\tau \ [\giga\mathrm{yr}]$}
& References \\ 
\hline
Kepler-37				& $\phn 178.7 \pm 1.4$\tablefootmark{*}			& $5417 \pm 75$ 
						& $-0.32 \pm 0.07$ 				& $0.802 \pm 0.068$ 
						& $0.770 \pm 0.026$ 			& $\sim 6$
						& 1 \\
$\alpha$ Cen B			& $\phn 161.1 \pm 0.1$\tablefootmark{*}			& $5260 \pm 50$
						& $\phs 0.24 \pm 0.05$			& $0.934 \pm 0.0061$
						& $0.863 \pm 0.005$ 			& $\phn 6.52 \pm 0.30$
						& 2, 3, 4, 5, 6 \\
Kepler-10				& $\phn 118.2 \pm 0.2$\tablefootmark{*}			& $5643 \pm 75$
						& $-0.15 \pm 0.07$				& $0.913 \pm 0.022$
						& $1.065 \pm 0.009$ 			& $10.41 \pm 1.36$
						& 7, 8, 9 \\
$\alpha$ Cen A			& $\phn 105.5 \pm 0.1$\tablefootmark{*}			& $5810 \pm 50$
						& $\phs  0.22 \pm 0.05$ 		& $1.105 \pm 0.0070$
						& $1.224 \pm 0.003$ 			& $\phn 6.52 \pm 0.30$
						& 2, 3, 4, 5, 6 \\
Perky					& $\phn \phd \phn 104 \pm 0.5$\tablefootmark{*}	& $6000 \pm 200$ 
						& $-0.09 \pm 0.1$ 				& $\phn 1.11 \pm 0.05$ 
						& $\phn 1.23 \pm 0.02$			& $\phn 4.87 \pm 0.50$
						& 10 \\
Kepler-68				& $101.51 \pm 0.09$\tablefootmark{*}				& $5793 \pm 44$ 
						& $\phs  0.12 \pm 0.04$ 		& $1.079 \pm 0.051$
						& $1.243 \pm 0.019$ 			& $\phn \phn 6.3 \pm 1.7$
						& 11 \\
Kepler-65				& $\phn \phn 90.0 \pm 0.5$\tablefootmark{*}		& $6211 \pm 66$ 
						& $\phs  0.17 \pm 0.06$ 		& $\phn 1.25 \pm 0.06$
						& $\phn 1.41 \pm 0.03$ 			& $\phn \phn 2.9 \pm 0.7$
						& 12 \\
$\mu$ Arae				& $\phn \phn \phn \phd 90 \pm 1.1$ & $5813 \pm 40$
						& $\phs  0.32 \pm 0.05$ 		& $\phn 1.10 \pm 0.02$
						& $\phn  1.36 \pm 0.06$ 		& $\phn 6.34 \pm 0.80$
						& 13, 14 \\
Dushera					& $\phn \phn \phd \phn 88 \pm 0.6$\tablefootmark{*}	& $6200 \pm 200$ 
						& $\phs \phn 0.0 \pm 0.15$ 		& $\phn 1.15 \pm 0.04$ 
						& $\phn 1.39 \pm 0.01$			& $\phn 3.80 \pm 0.37$
						& 10 \\
Kepler-50				& $\phn \phn 76.0 \pm 0.9$		& $6225 \pm 66$
						& $\phs  0.03 \pm 0.06$			& $\phn 1.24 \pm 0.05$
						& $\phn  1.58 \pm 0.02$ 		& $\phn \phn 3.8 \pm 0.80$ 
						& 12 \\
Kepler-36				& $\phn \phn 67.9 \pm 1.2$		& $5911 \pm 66$		
						& $-0.20 \pm 0.06 $				& $1.071 \pm 0.043 \phn$									& $1.626 \pm 0.019$ 			& $\phn \phn 6.8 \pm 1.0$
						& 15 \\
HAT-P-7					& $\phn 59.22 \pm 0.59$\tablefootmark{*}			& $6350 \pm 80$
						& $\phs  0.26 \pm 0.08$ 		& $\phn 1.53 \pm 0.005$	
						& $1.997 \pm 0.003$				& $\phn 1.81 \pm 0.08$
						& 16, 17, 18 \\
$\beta$ Hyi				& $\phn 57.24 \pm 0.16 $\tablefootmark{*}		& $5840 \pm 59$
						& $-0.04 \pm 0.10$				& $1.085 \pm 0.028$
						& $1.809 \pm 0.015$ 			& $\phn 6.40 \pm 0.56$
						& 19, 20, 21, 22 \\
Kepler-7				& $\phn \phn 56.4 \pm 1.7$		& $5933 \pm 44$
						& $\phs  0.11 \pm 0.03$			& $1.359 \pm 0.031$
						& $1.966 \pm 0.013$ 			& $\phn \phn 3.3 \pm 0.4$
						& 18, 23, 24 \\
Procyon					& $\phn \phn 56.0$\tablefootmark{*} & $6494 \pm 48$
						& $\phs  0.02 \pm 0.10$			& $1.461 \pm 0.025$
						& $2.059 \pm 0.015$ 			& $\phn 1.87 \pm 0.13$
						& 19, 25, 26 \\
$\eta$ Boo				& $\phn \phn 39.9 \pm 0.1$\tablefootmark{*}	& $6030 \pm 80$
						& $\phs  0.24 \pm 0.07$ 		& $\phn 1.57 \pm 0.07$
						& $2.701 \pm 0.040$ 			& $\phn 2.67 \pm 0.10$
						& 19, 27 \\
\hline                  
\end{tabular}

\tablebib{	(1)~\citet{{ref:kepler37}};
			(2)~\citet{ref:cenAB}; (3)~\citet{ref:cenAB_masses}; (4)~\citet{ref:cenAB_diameter}; (5)~\citet{ref:cenAB_deltanuA}; (6)~\citet{ref:cenAB_deltanuB};
			(7)~\citet{ref:Kepler10}; (8)~\citet{ref:torres_2012}; (9)~\citet{ref:kepler10_again};
			(10)~\citet{ref:vsa_2013_ages};
			(11)~\citet{ref:kepler68};
			(12)~\citet{ref:kepler50and65};
			(13)~\citet{ref:muArae}; (14)~\citet{ref:muArae_spec};
			(15)~\citet{ref:Kepler36};
			(16)~\citet{ref:hatp7_spec}; (17)~Lund et al. (in prep.); (18)~\citet{ref:huber_2013};
			(19)~\citet{ref:bruntt}; (20)~\citet{ref:betaHyi_dnu}; (21)~\citet{ref:betaHyi_MR}; (22)~\citet{ref:betaHyi_age};
			(23)~\citet{ref:kepler7_spec}; (24)~\citet{ref:kepler7};
			(25)~\citet{ref:procyon}; (26)~\citet{ref:procyon_age};
			(27)~\citet{ref:etaBoo}.
       }

\tablefoot{
			\tablefoottext{*}{Stars where the $\Delta \nu$ uncertainty used in AME has been inflated to $1\%$ by the authors.}
			}

\end{table*}

AME has been used on a number of stars which either have radii determined from interferometry (and measured parallaxes), radii based on parallax measurements (and calculations of the stellar angular diameters), or parameters found from detailed modelling (in some cases in combination with other methods). One of these stars is the solar-like star $\alpha$~Cen~A whose values of its spectroscopic parameters, large separation, and mass, radius, and age can be found in Table~\ref{tab:litt_values} along with the properties of the other stars with results from detailed modelling (listed in order of decreasing large separation). In order to illustrate the use of AME, we describe in Sects.~\ref{subsec:det_mass}, \ref{subsec:det_density}, and \ref{subsec:det_age} below how we used AME to obtain the stellar parameters for $\alpha$~Cen~A. Furthermore, in Sect.~\ref{subsec:uncertainties} we explain how we determined the internal uncertainties on the stellar parameters.


\subsection{Step 1: Determining the stellar mass}
\label{subsec:det_mass}

\begin{figure}
	\centering
	\includegraphics[width=\hsize]{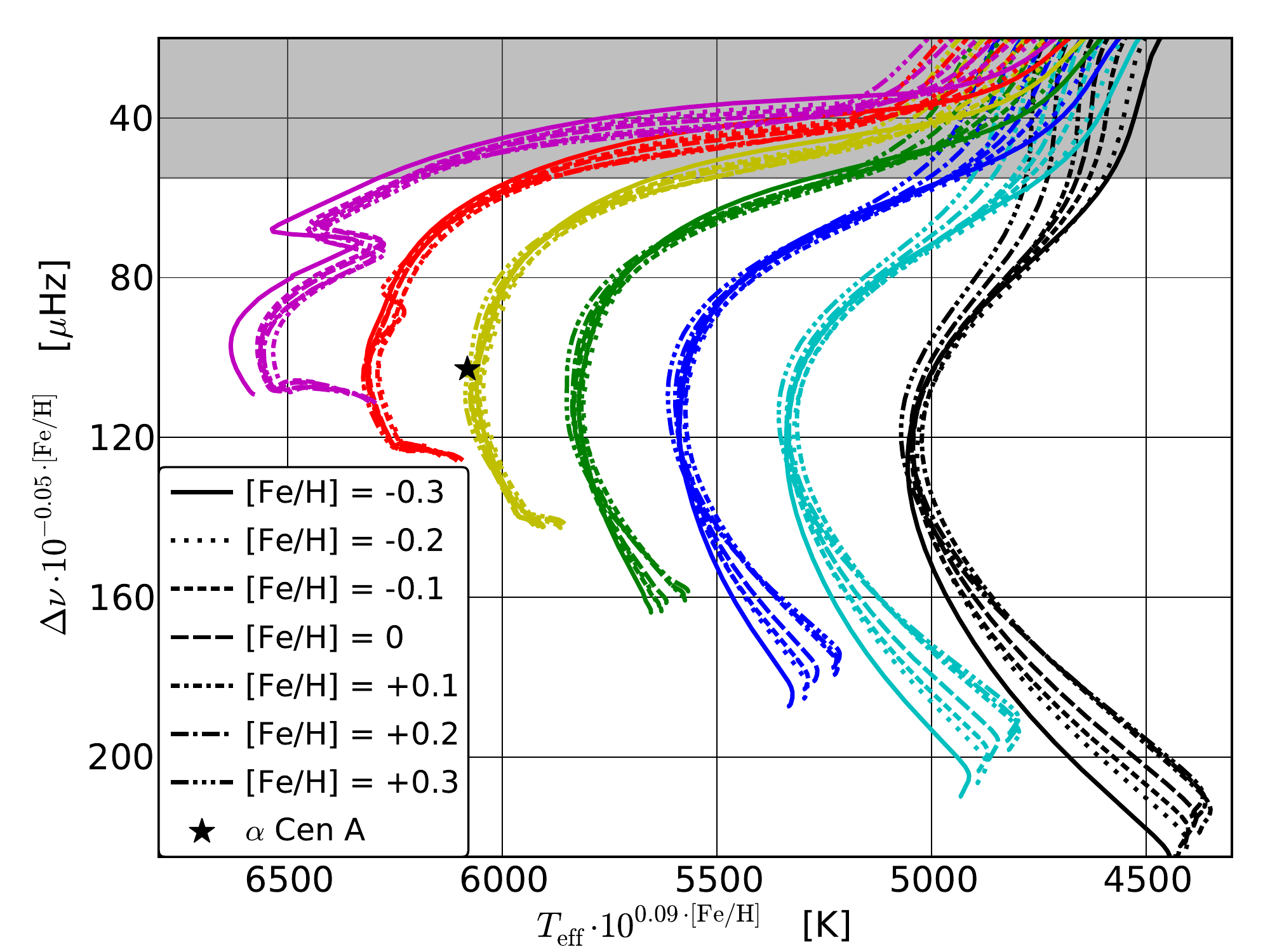}
	\caption{Plot used to determine the mass of $\alpha$~Cen~A. The colours give the masses of the lines ranging from $0.7 M_\sun$ (black) in the right of the figure in steps of $0.1 M_\sun$ to $1.3 M_\sun$ (magenta/purple) in the left of the figure. The different line types give the metallicity. The star shows the location of $\alpha$~Cen~A in the figure. The shaded area in the top indicates the region where another mass plot would be more suitable.}
	\label{fig:m_lm}
\end{figure}

The first step that we performed using AME on $\alpha$~Cen~A was to find the mass. Depending primarily on the known values of the effective temperature and large separation, the appropriate version of the mass plot had to be used, and we chose the relevant one for $\alpha$~Cen~A. The plot can be seen in \figref{fig:m_lm} with the location of $\alpha$~Cen~A marked by a star. This plot was chosen firstly because it encompassed the parameters of $\alpha$~Cen~A, and secondly because the star did not fall in the grey-shaded region of the plot.

It can easily been seen from this figure that $\alpha$~Cen~A lies almost on top of the $1.1 M_\sun$ (the yellow) lines. By considering the metallicity of the star, we found the mass to be $M = 1.10 M_\sun$ using linear interpolation between the lines of different masses and metallicities. It should be noted that this value is in excellent agreement with the value listed in Table~\ref{tab:litt_values} which is found using the binarity of the $\alpha$~Cen system.


\subsection{Step 2: Finding the mean stellar density}
\label{subsec:det_density}

\begin{figure}
	\centering
	\includegraphics[width=\hsize]{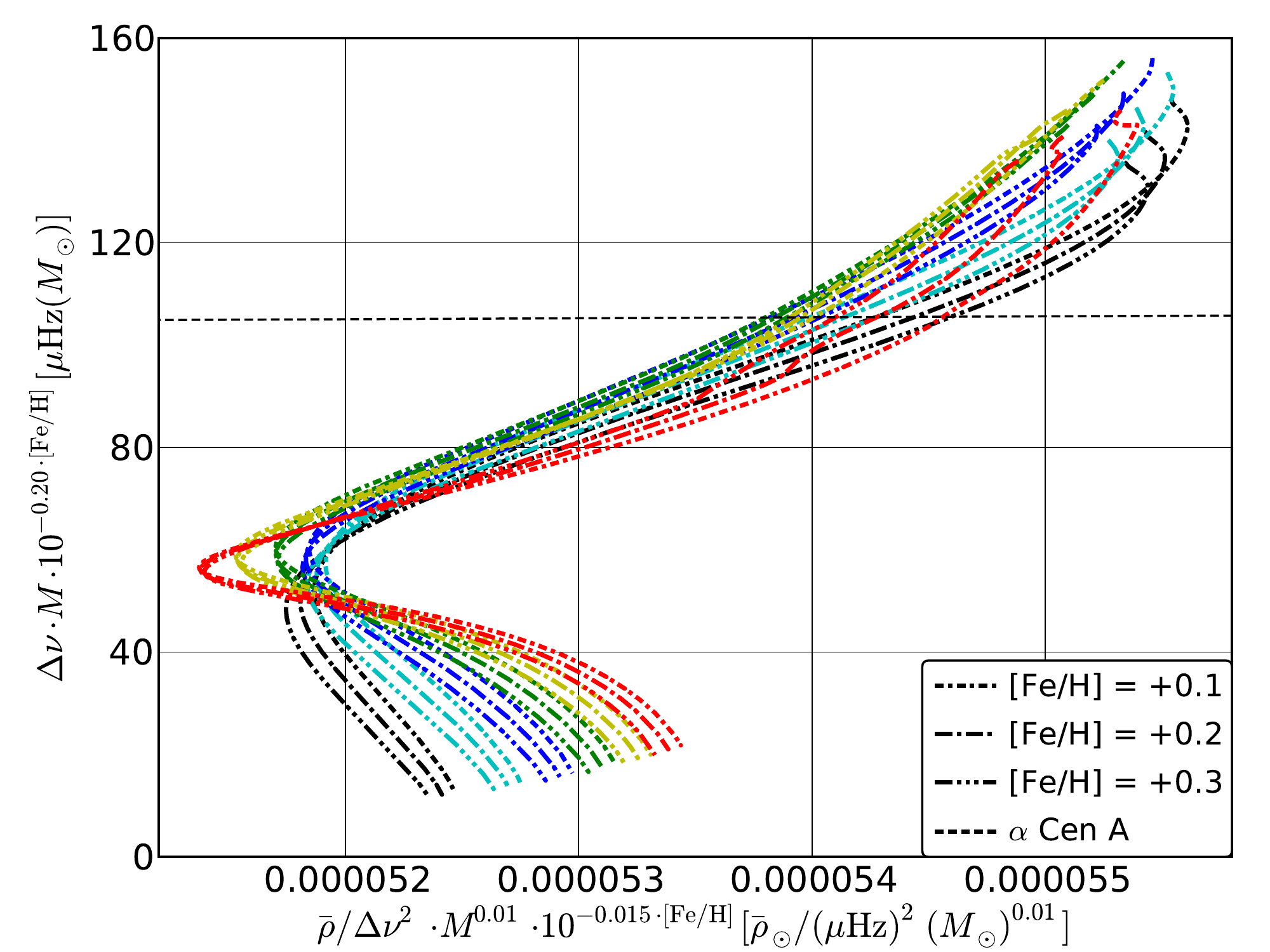}
	\caption{Plot used to determine the mean-density of $\alpha$~Cen~A. The colours represent different masses; $0.7 M_\sun$ (black), $0.8 M_\sun$ (light blue), $0.9 M_\sun$ (blue), $1.0 M_\sun$ (green), $1.1 M_\sun$ (yellow), and $1.2 M_\sun$ (red), whilst the line styles give the metallicity of a given line. The location of $\alpha$~Cen~A in the plot is marked by the horizontal, black, dashed line.}
	\label{fig:rho_lm}
\end{figure}

The second step was to find the mean-density of $\alpha$~Cen~A. This could be done by using the suitable mean-density plot after we had determined the mass. The mean-density plot that we used can be seen in \figref{fig:rho_lm}. This particular plot was chosen because it covered the mass and metallicity of $\alpha$~Cen~A.

From the expression on the y-axis, we calculated the y-axis location of $\alpha$~Cen~A in the plot. This is marked by a horizontal, black, dashed line in \figref{fig:rho_lm}. The mean-density of $\alpha$~Cen~A was now determined by first interpolating the crossing of the dashed line with the lines of surrounding masses ($1.1 M_\sun$ and $1.2 M_\sun$) and metallicities ($+0.2 \ \text{dex}$ and $+0.3 \ \text{dex}$) to find the position of the crossing between the horizontal, dashed line and a line with the metallicity and mass of $\alpha$~Cen~A. This gave us the position on the x-axis of $\alpha$~Cen~A and thereby the value of $\bar{\rho} / \Delta \nu^2 \cdot M^{0.01} \cdot 10^{-0.015 \cdot [\mathrm{Fe}/\mathrm{H}]}$. Using the large separation and the metallicity from Table~\ref{tab:litt_values} along with the mass found from the previous plot, we then calculated the mean-density of $\alpha$~Cen~A to be $\bar{\rho} = 0.6042 \bar{\rho}_\sun$.

From the mass and mean-density, the radius and surface gravity of $\alpha$~Cen~A were computed using expressions~(\ref{eq:sca_R}) and~(\ref{eq:sc_g}) respectively. We found the radius to be $R = 1.22 R_\sun$ and a surface gravity of $g = 0.738 g_\sun$. Using a solar surface gravity of for instance $\log g_\sun = 4.43789 \ \mathrm{dex} \ [\mathrm{cgs}]$\footnote{Calculated from the MESA values of the gravitational constant, the solar mass, and the solar radius.}, this value can be converted into classical units; $\log g = 4.31 \ \mathrm{dex} \ [\mathrm{cgs}]$. It is noteworthy that the found radius agrees very well with the one from the literature (see Table~\ref{tab:litt_values}).


\subsection{Step 3: Obtaining the stellar age}
\label{subsec:det_age}

\begin{figure}
	\centering
	\includegraphics[width=\hsize]{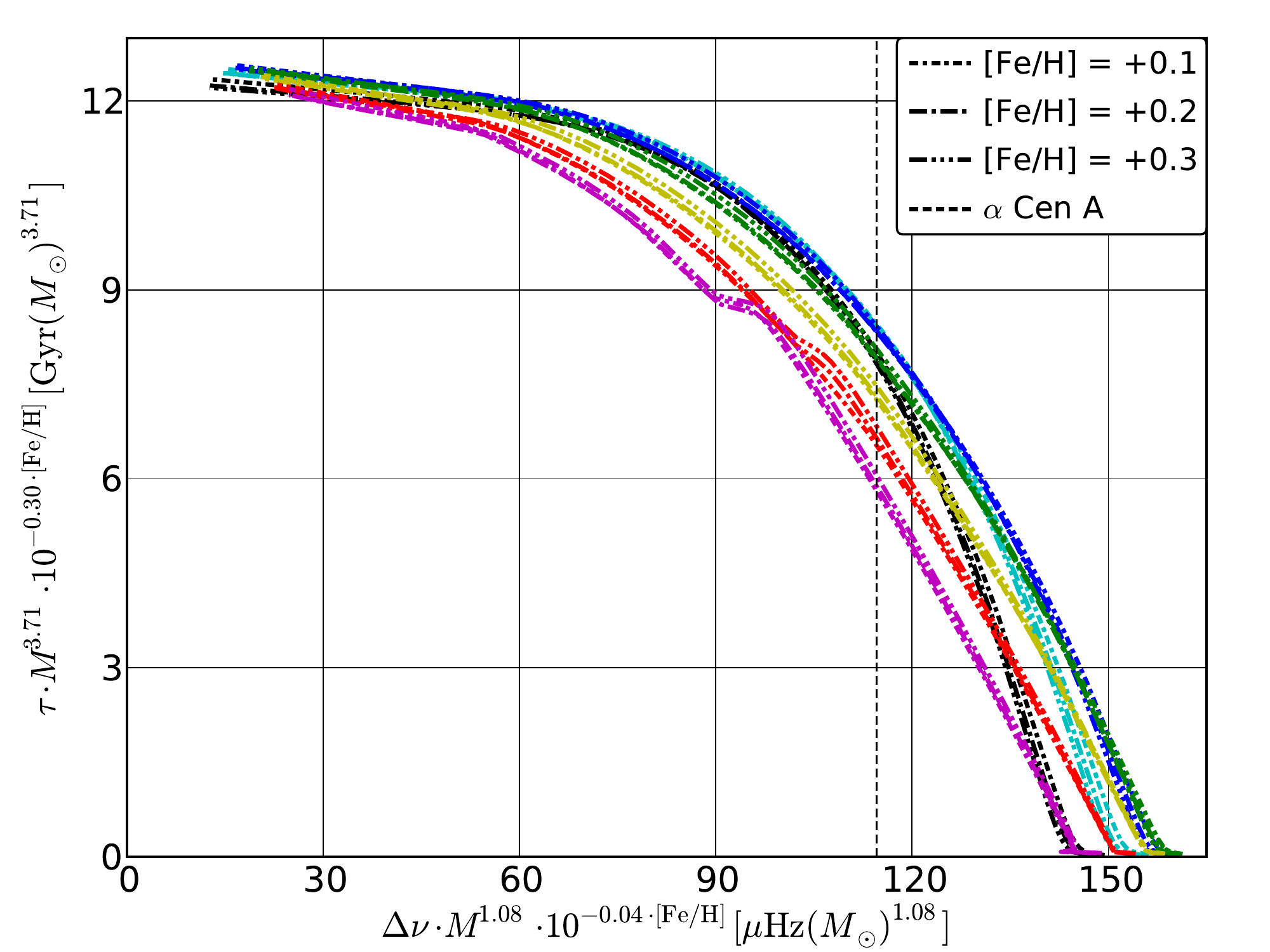}
	\caption{Plot used to determine the age of $\alpha$~Cen~A. The colours represent different masses; $0.7 M_\sun$ (black), $0.8 M_\sun$ (light blue), $0.9 M_\sun$ (blue), $1.0 M_\sun$ (green), $1.1 M_\sun$ (yellow) $1.2 M_\sun$ (red), and $1.3 M_\sun$ (magenta/purple), whilst the line types indicate the metallicity of a given line. The location of $\alpha$~Cen~A in the plot is marked by the vertical, black, dashed line.}
	\label{fig:tau_lm}
\end{figure}

The third and last step that we performed when using AME on $\alpha$~Cen~A was to determine its age. This largely followed the same procedure as outlined above to find the mean-density. First, we established the x-axis position of $\alpha$~Cen~A in the appropriate age plot. This age plot was, as before, chosen because it covered the mass and metallicity of $\alpha$~Cen~A. The plot can be seen in \figref{fig:tau_lm} with the x-axis position of $\alpha$~Cen~A shown as the vertical, black, dashed line.

Second, we found the crossing between the vertical, dashed line and the position of a hypothetical line with the mass and metallicity of $\alpha$~Cen~A by interpolation. This gave us the location of $\alpha$~Cen~A on the secondary axis and thereby the value of $\tau \cdot M^{3.71} \cdot 10^{-0.30 \cdot [\mathrm{Fe}/\mathrm{H}]}$. Using, as before, the information from Table~\ref{tab:litt_values} and the previously determined mass, we were able to determine the age to be $\tau = 6.0 \ \giga\mathrm{yr}$. This is consistent with the value found in the literature (see Table~\ref{tab:litt_values}).


\subsection{The uncertainties}
\label{subsec:uncertainties}

The $1\sigma$ internal uncertainties on the stellar parameters found with AME have been estimated by considering the centre and the corners of the relevant error box. For the mass, we considered the $\left( \Delta \nu, T_\mathrm{eff}, [\mathrm{Fe}/\mathrm{H}] \right)$ error box whilst the error box applicable to the mean-density and age was $\left( \Delta \nu, M, [\mathrm{Fe}/\mathrm{H}] \right)$ in both cases. The $1 \sigma$ uncertainties for the derived quantities (radius and surface gravity) were found in a similar manner, but taking into account the correlation between mass and mean-density (recall that the mass was used to determine the mean-density). We describe below in some detail the determination of the mass uncertainty for $\alpha$~Cen~A and the other uncertainties follow by extension.

The values of the large separation, effective temperature, and metallicity and their associated error bars for $\alpha$~Cen~A can be found in Table~\ref{tab:litt_values}. Note here, that $\alpha$~Cen~A is one of many star in this paper where we have inflated the $\Delta\nu$ uncertainty to $1\%$ in order to account for deviations from the $\frac{\Delta \nu_{\sun, \mathrm{obs}}}{\Delta \nu_{\sun, \mathrm{mod}}}$ scaling used for the large separations in the models.

We first fixed $[\mathrm{Fe}/\mathrm{H}]$ at its centre value, $[\mathrm{Fe}/\mathrm{H}] = +0.22 \ \mathrm{dex}$ and calculated the values of $T_\mathrm{eff} \cdot 10^{0.09 \cdot [\mathrm{Fe}/\mathrm{H}]}$ and $\Delta\nu \cdot 10^{-0.05 \cdot [\mathrm{Fe}/\mathrm{H}]}$ (the axes in \figref{fig:m_lm}) using the values of $T_\mathrm{eff}$ and $\Delta \nu$ from Table~\ref{tab:litt_values}. First without the $1\sigma$ uncertainties (the central value) and subsequently using $T_\mathrm{eff} - \sigma T_\mathrm{eff}$ and $T_\mathrm{eff} + \sigma T_\mathrm{eff}$, as well as $\Delta\nu - \sigma\Delta\nu$ and $\Delta\nu + \sigma\Delta\nu$. This gave us $6081 \pm 52 \ \kelvin$ for the first expression and $102.9 \pm 1.0 \ \micro\hertz$ for the second (see the first row of Table~\ref{tab:aCenA_uncertainties}). From the central value ($6081 \ \kelvin$, $102.9 \ \micro\hertz$) we estimate the central mass ($1.10 M_\sun$ in this case) and from the possible combinations of the extreme values we do the same ($6081 - 52 \ \kelvin$ and $102.9-1.0 \ \micro\hertz$, $6081 - 52 \ \kelvin$ and $102.9+1.0 \ \micro\hertz$, $6081 + 52 \ \kelvin$ and $102.9 - 1.0 \ \micro\hertz$, and $6081 + 52 \ \kelvin$ and $102.9+1.0 \ \micro\hertz$). To get the uncertainty on the central mass we then take half of the full difference between the maximum and the minimum mass estimates obtained. This gives us a mass of $1.10 \pm 0.02 M_\sun$ which can be seen in the right column in the first row of Table~\ref{tab:aCenA_uncertainties}.

\begin{table*}
\caption{Estimate of the uncertainty on the mass for $\alpha$~Cen~A.}             
\label{tab:aCenA_uncertainties}      
\centering          
\begin{tabular}{llll}     
\hline\hline       
$[\mathrm{Fe}/\mathrm{H}] [\text{dex}]$
& \multicolumn{1}{c}{$T_\mathrm{eff} \cdot 10^{0.09 \cdot [\mathrm{Fe}/\mathrm{H}]} \ [\kelvin]$}
& \multicolumn{1}{c}{$\Delta\nu \cdot 10^{-0.05 \cdot [\mathrm{Fe}/\mathrm{H}]} \ [\micro\hertz]$}
& \multicolumn{1}{c}{$M/M_\sun$} \\ 
\hline
Centre value: $0.22$	& $6081 \pm 52$	& $102.9 \pm 1.0$	& $1.10 \pm 0.02$ \\
Upper value: $0.27$		& $6144 \pm 53$	& $102.3 \pm 1.0$	& $1.12 \pm 0.02$ \\
Lower value: $0.17$		& $6018 \pm 52$	& $103.5 \pm 1.0$	& $1.08 \pm 0.02$ \\
\hline                  
\end{tabular}
\end{table*}

Hereafter, we recalculated the values using instead the $+1\sigma$ and $-1\sigma$ values of $[\mathrm{Fe}/\mathrm{H}]$ in turn. The results can be seen in the last two rows of Table~\ref{tab:aCenA_uncertainties}. The mass estimates and the associated errors from the uncertainties in $T_\mathrm{eff}$ and $\Delta \nu$ are given in the right column.

In order to calculate the final $1\sigma$ uncertainty, we took half of the full difference between the \textit{upper value} mass estimate and the \textit{lower value} mass estimate given in Table~\ref{tab:aCenA_uncertainties}. This we added in quadrature (we assume the measurements to be independent) to the maximal uncertainty on an individual mass estimate (to err on the side of caution); in the case of $\alpha$~Cen~A they are the same; $0.2 M_\sun$. Thereby we obtained a final mass and uncertainty estimate of
\begin{equation}
	\label{eq:now_i_have_a_number}
	\frac{M}{M_\sun} = 1.10 \pm \sqrt{ \left(\frac{1.12 - 1.08}{2}\right)^2 + 0.02^2} = 1.10 \pm 0.03 \ .
\end{equation}

The result of carrying out this operation for the other parameters of $\alpha$~Cen~A can be found in Sect.~\ref{sec:results}.

It should be noted that we have confirmed that the uncertainties scale linearly as could be expected. This means that using the $3\sigma$ error bars on the input parameters ($\Delta\nu$, $T_\mathrm{eff}$, and $[\mathrm{Fe}/\mathrm{H}]$) leads to an increase of a factor of $\sim 3$ on the uncertainties of the stellar mass, mean-density, radius, surface gravity, and age.


\section{Results}
\label{sec:results}

AME has been used on seven stars with known interferometric radii, $20$ stars with radii based on parallaxes determined with the Hipparcos satellite, and $16$ solar-like stars with results from detailed modelling (in some cases in combination with results from other approaches). The results from AME for these three groups of stars will be presented in separate sections below.


\subsection{Stars with interferometry}
\label{subsec:res_inter_stars}

\begin{table*}
\caption{Literature and AME values for the seven stars with interferometric radii. All the stars have had the $\Delta \nu$ uncertainty used in AME increased to $1\%$.}
\label{tab:values_interf_stars}      
\centering          
\begin{tabular}{lllllll}     
\hline\hline       
Star
& \multicolumn{1}{c}{$\Delta \nu \ [\micro\hertz]$}
& \multicolumn{1}{c}{$T_\mathrm{eff} \ [\kelvin]$}
& \multicolumn{1}{c}{$[\mathrm{Fe}/\mathrm{H}] [\text{dex}]$}
& \multicolumn{1}{c}{$\left( \mathrm{R}/\mathrm{R}_\odot \right) _\mathrm{inter}$}
& \multicolumn{1}{c}{$\left( \mathrm{R}/\mathrm{R}_\odot \right) _\mathrm{AME}$}
& \multicolumn{1}{c}{References}
\\ 
\hline
KIC-$8006161$	& $149.3 \pm 0.4$				& $5295 \pm 51$ 
				& $\phs \phn 0.34 \pm 0.06$		& $0.952 \pm 0.021$ 
				& $0.93 \pm 0.02$ 				& 1, 2 \\
$16$~Cyg~B		& $117.0 \pm 0.1$				& $5809 \pm 39$
				& $\phs 0.052 \pm 0.021$		& $\phn 1.12 \pm 0.02$
				& $1.11 \pm 0.01$ 				& 3, 4 \\
KIC-$6225718$	& $105.8 \pm 0.3$				& $6153 \pm 89$ 
				& \phn$-0.17 \pm 0.06$ 			& $1.306 \pm 0.047$ 
				& $1.20 \pm 0.02$ 				& 1, 2 \\               
KIC-$6106415$	& $104.3 \pm 0.3$				& $5908 \pm 72$ 
				& \phn$-0.09 \pm 0.06$ 			& $1.289 \pm 0.037$ 
				& $1.19 \pm 0.02$ 				& 1, 2 \\
$16$~Cyg~A		& $103.5 \pm 0.1$				& $5839 \pm 42$
				& $\phs 0.096 \pm 0.026$		& $\phn 1.22 \pm 0.02$
				& $1.22 \pm 0.01$ 				& 3, 4 \\
$\theta$~Cyg	& $\phn 84.0 \pm 0.2$			& $6749 \pm 44$
				& \phn$-0.03 \pm 0.15$\tablefootmark{a} & $\phn 1.48 \pm 0.02$
				& $1.49 \pm 0.02$ 				& 3, 5, 6 \\
KIC-$8751420$	& $\phn 34.6 \pm 0.1$			& $5236 \pm 37$ 
				& \phn$-0.15 \pm 0.06$ 			& $2.703 \pm 0.071$ 
				& $2.76 \pm 0.08$ 				& 1, 2 \\
				
\hline                  
\end{tabular}

\tablebib{	(1)~\citet{ref:bruntt_2012};
			(2)~\citet{ref:huber_interferometry};
			(3)~\citet{ref:white_interferometry};
			(4)~\citet{ref:metal_16CygAB};
			(5)~\citet{ref:largesep_thetaCyg_too};
			(6)~\citet{ref:metal_thetaCyg}.
       }
       
\tablefoot{
			\tablefoottext{a}{The uncertainty is estimated by the authors based on information in \citet{ref:metal_thetaCyg}.}
			}

\end{table*}

\begin{figure}
	\centering
	\includegraphics[width=\hsize]{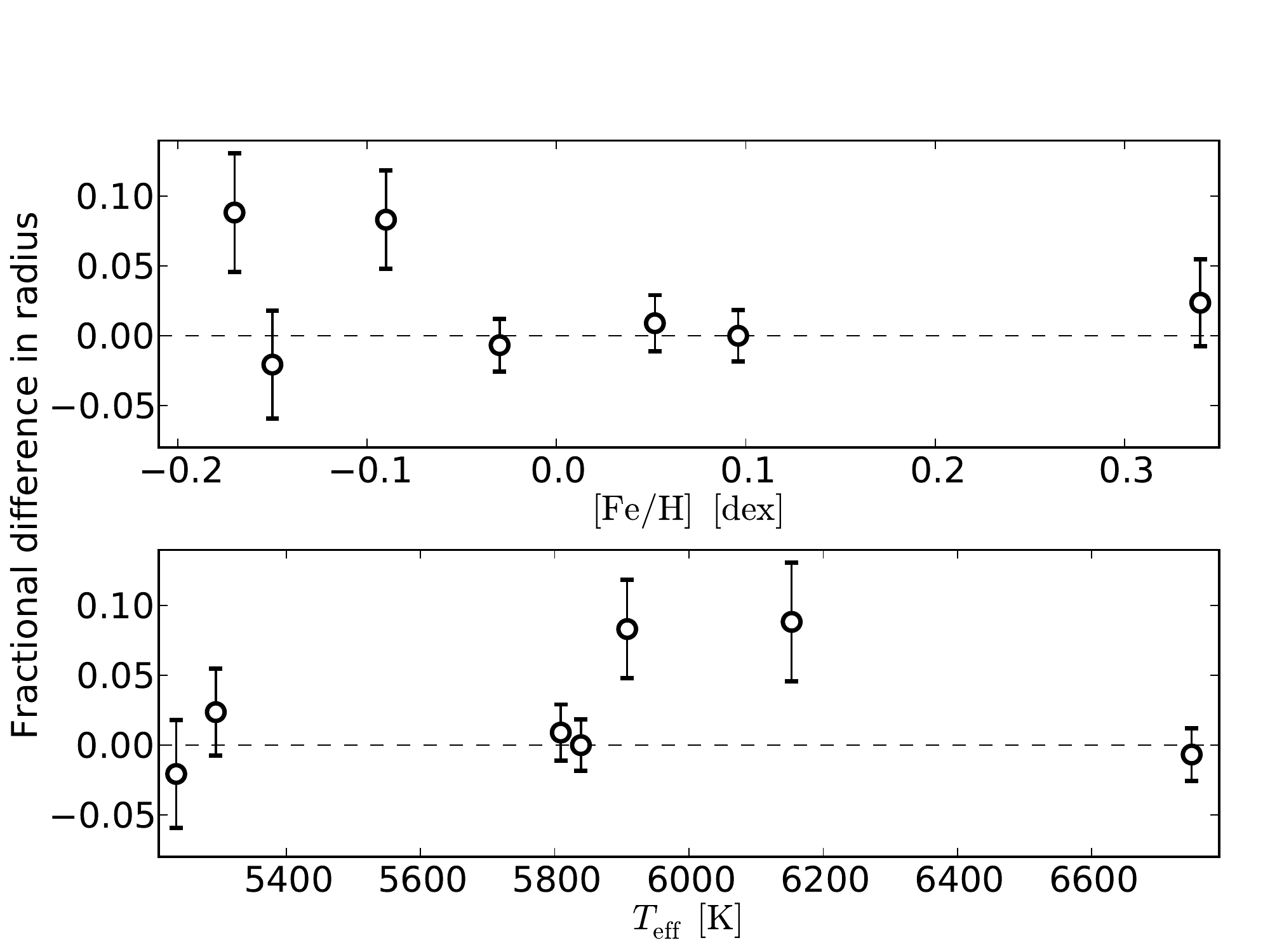}
	\caption{Fractional difference in radius as a function of metallicity and effective temperature for the seven stars with radii determined from interferometry. The open symbols give the fractional difference between the stellar radius determined from interferometry and the one determined from AME (the ratio of the difference in radius and the radius from AME). The errors have been added in quadrature. The dashed line marks a fractional difference of zero.}
	\label{fig:comp_rad_interf}
\end{figure}

We have used AME on seven stars with radii determined from interferometry and parallaxes. These seven stars are four of the stars from \citet{ref:huber_interferometry} and the three stars $\theta$~Cyg, $16$~Cyg~A, and $16$~Cyg~B which have interferometric radii found by \citet{ref:white_interferometry}. A comparison of the AME results with those from interferometry allows us to assess the reliability of the radii from AME, since interferometric radii are among the most model-independent radii that can be obtained. The input parameters needed for AME for the seven stars are given in Table~\ref{tab:values_interf_stars} along with the known interferometric radii and the radii obtained with AME. Figure~\ref{fig:comp_rad_interf} shows the fractional difference between the interferometric radius and the radius found from AME as a function of metallicity and effective temperature. By fractional difference we mean the difference between the interferometric radius and the AME radius divided by the AME radius.

It is clear from \figref{fig:comp_rad_interf} that the agreement between the AME and the interferometric radii is good. No trend is seen in the difference neither as a function of metallicity nor as a function of effective temperature and the weighted mean of the fractional differences is just $1.7\%$ (using as weights: $w = 1/\sigma^2$ where $\sigma$ denotes the size of the relevant error bar).


\subsection{Stars with parallaxes}
\label{subsec:res_par_stars}

\begin{table*}
\caption{Literature and AME values for the $20$ stars with radii based on parallaxes and angular diameters. All parameters except for the AME radii are from \citet{ref:vsa_2012_radii}.}             
\label{tab:values_parall_stars}      
\centering          
\begin{tabular}{llllll}     
\hline\hline       
Star 	& \multicolumn{1}{c}{$\Delta \nu \ [\micro\hertz]$}
& \multicolumn{1}{c}{$T_\mathrm{eff} \ [\kelvin]$}
& \multicolumn{1}{c}{$[\mathrm{Fe}/\mathrm{H}] [\text{dex}]$}
& \multicolumn{1}{c}{$\left( \mathrm{R}/\mathrm{R}_\odot \right)_\mathrm{paral}$}
& \multicolumn{1}{c}{$\left( \mathrm{R}/\mathrm{R}_\odot \right) _\mathrm{AME}$} \\ 
\hline
KIC-$3632418$	& $\phn 60.8 \pm 0.2$\tablefootmark{*}			& $6286 \pm 70$ 
				& $-0.01 \pm 0.1$ 				& $ 1.911 \phd \substack{+0.025 \\ -0.026}$ 
				& $1.85 \pm 0.02$ \\
KIC-$3733735$	& $\phn 92.3 \pm 0.3$\tablefootmark{*}			& $6824 \pm 131$ 
				& $-0.10 \pm 0.1$ 				& $1.427 \phd \substack{+0.019 \\ -0.020}$ 
				& $1.38 \pm 0.02$ \\
KIC-$4914923$	& $\phn 88.7 \pm 0.3$\tablefootmark{*}			& $5828 \pm 56$ 
				& $\phs 0.17 \pm 0.1$ 			& $1.408 \phd \substack{+0.022 \\ -0.023}$ 
				& $1.38 \pm 0.02$ \\
KIC-$5371516$	& $\phn 55.4 \pm 0.2$\tablefootmark{*}			& $6360 \pm 115$ 
				& $\phs 0.13 \pm 0.1$ 			& $2.066 \phd \substack{+0.022 \\ -0.021}$ 
				& $2.03 \pm 0.03$ \\
KIC-$5774694$	& $140.2 \pm 4.0$				& $6380 \pm 70$ 
				& $\phs 0.01 \pm 0.1$ 			& $1.000 \phd \substack{+0.015 \\ -0.016}$ 
				& $1.00 \pm 0.03$ \\
KIC-$6106415$	& $104.3 \pm 0.3$\tablefootmark{*}			& $6061 \pm 89$ 
				& $-0.06 \pm 0.1$ 				& $1.240 \pm 0.018$
				& $1.22 \pm 0.02$ \\
KIC-$6225718$	& $105.8 \pm 0.3$\tablefootmark{*}			& $6338 \pm 88$ 
				& $-0.15 \pm 0.1$ 				& $1.256 \pm 0.014$
				& $1.22 \pm 0.02$ \\
KIC-$7747078$	& $\phn 54.0 \pm 0.2$\tablefootmark{*}			& $5856 \pm 112$ 
				& $-0.26 \pm 0.1$ 				& $1.952 \pm 0.039$
				& $1.93 \pm 0.03$ \\
KIC-$7940546$	& $\phn 58.9 \pm 0.2$\tablefootmark{*}			& $6287 \pm 74$ 
				& $-0.04 \pm 0.1$ 				& $1.944 \phd \substack{+0.024 \\ -0.031}$ 
				& $1.89 \pm 0.02$ \\
KIC-$8006161$	& $149.3 \pm 0.4$\tablefootmark{*}			& $5355 \pm 107$ 
				& $\phs 0.34 \pm 0.1$ 			& $0.927 \pm 0.014$ 
				& $0.94 \pm 0.03$ \\
KIC-$8228742$	& $\phn 62.1 \pm 0.2$\tablefootmark{*}			& $6130 \pm 107$ 
				& $-0.14 \pm 0.1$ 				& $1.855 \pm 0.027$
				& $1.78 \pm 0.03$ \\
KIC-$8751420$	& $\phn 34.6 \pm 0.1$\tablefootmark{*}			& $5243 \pm 162$ 
				& $-0.20 \pm 0.1$ 				& $2.722 \phd \substack{+0.048 \\ -0.057}$
				& $2.73 \pm 0.14$ \\
KIC-$9139151$	& $117.3 \pm 0.3$\tablefootmark{*}				& $6141 \pm 114$ 
				& $\phs 0.15 \pm 0.1$ 			& $1.178 \pm 0.018$
				& $1.17 \pm 0.02$ \\
KIC-$9139163$	& $\phn 81.1 \pm 0.2$\tablefootmark{*}			& $6525 \pm 111$ 
				& $\phs 0.15 \pm 0.1$ 			& $1.571 \pm 0.010$
				& $1.55 \pm 0.02$ \\
KIC-$9206432$	& $\phn 84.7 \pm 0.3$\tablefootmark{*}			& $6614 \pm 135$ 
				& $\phs 0.23 \pm 0.1$ 			& $1.544 \pm 0.015$
				& $1.52 \pm 0.02$ \\
KIC-$10068307$	& $\phn 54.0 \pm 0.2$\tablefootmark{*}			& $6197 \pm 97$ 
				& $-0.13 \pm 0.1$ 				& $2.060 \phd \substack{+0.028 \\ -0.033}$ 
				& $2.00 \pm 0.03$ \\
KIC-$10162436$	& $\phn 55.8 \pm 0.2$\tablefootmark{*}			& $6245 \pm 110$ 
				& $-0.08 \pm 0.1$ 				& $2.015 \phd \substack{+0.025 \\ -0.027}$ 
				& $1.96 \pm 0.03$ \\
KIC-$10454113$	& $105.1 \pm 0.3$\tablefootmark{*}				& $6134 \pm 113$ 
				& $-0.06 \pm 0.1$ 				& $1.251 \pm 0.017$
				& $1.22 \pm 0.02$ \\
KIC-$11253226$	& $\phn 77.0 \pm 0.2$\tablefootmark{*}			& $6715 \pm 97$ 
				& $-0.03 \pm 0.1$ 				& $1.628 \phd \substack{+0.017 \\ -0.018}$ 
				& $1.58 \pm 0.02$ \\
KIC-$12258514$	& $\phn 75.0 \pm 0.2$\tablefootmark{*}			& $6064 \pm 121$ 
				& $\phs 0.13 \pm 0.1$ 			& $1.630 \phd \substack{+0.029 \\ -0.031}$ 
				& $1.60 \pm 0.05$ \\
				
\hline                  
\end{tabular}

\tablefoot{
			\tablefoottext{*}{Stars where the $\Delta \nu$ uncertainty used in AME has been inflated to $1\%$ by the authors.}
			}

\end{table*}

\begin{figure*}
\centering
\subfloat[][]{\includegraphics[width=9cm]{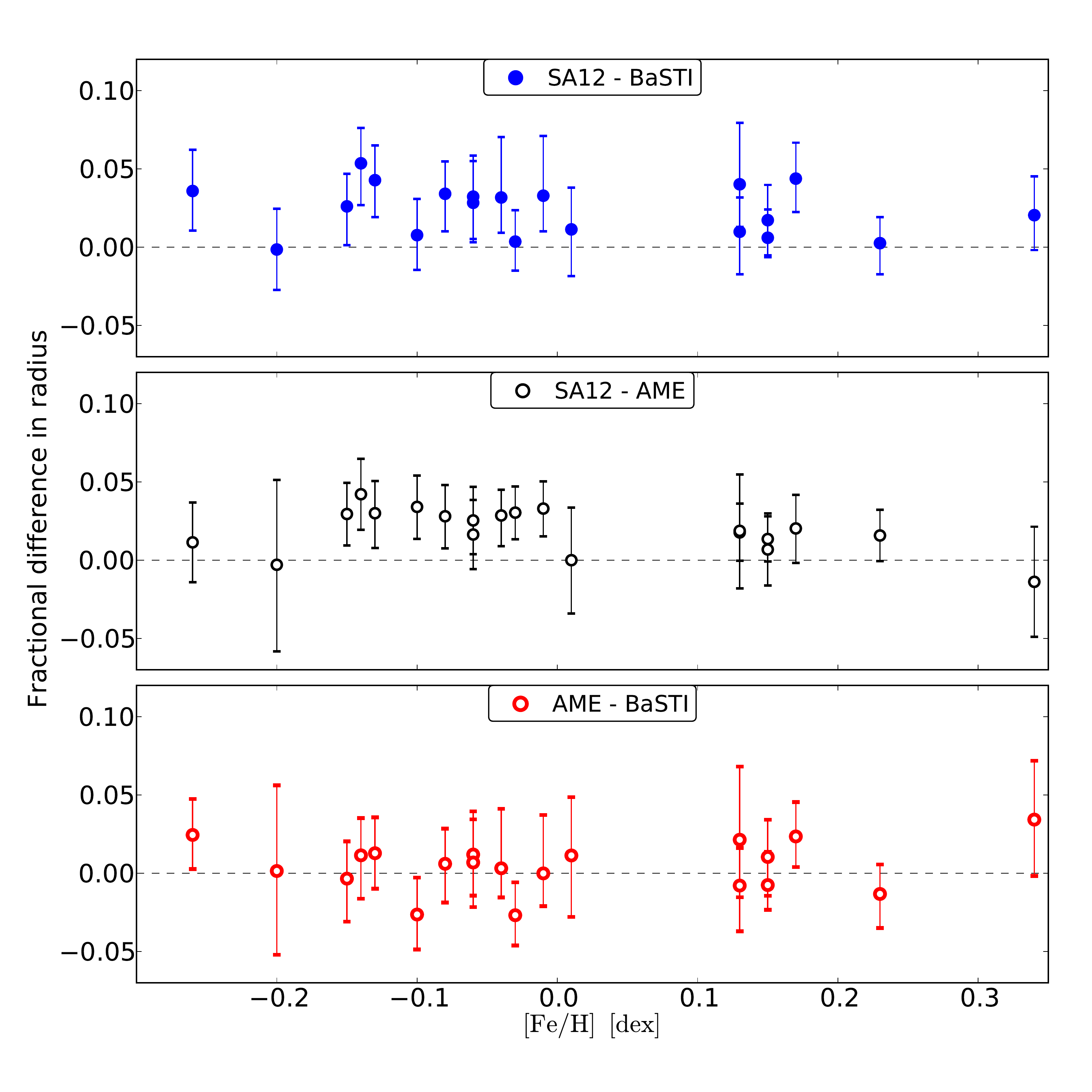} \label{fig:comp_feh_r_parall}}
\hfill
\subfloat[][]{\includegraphics[width=9cm]{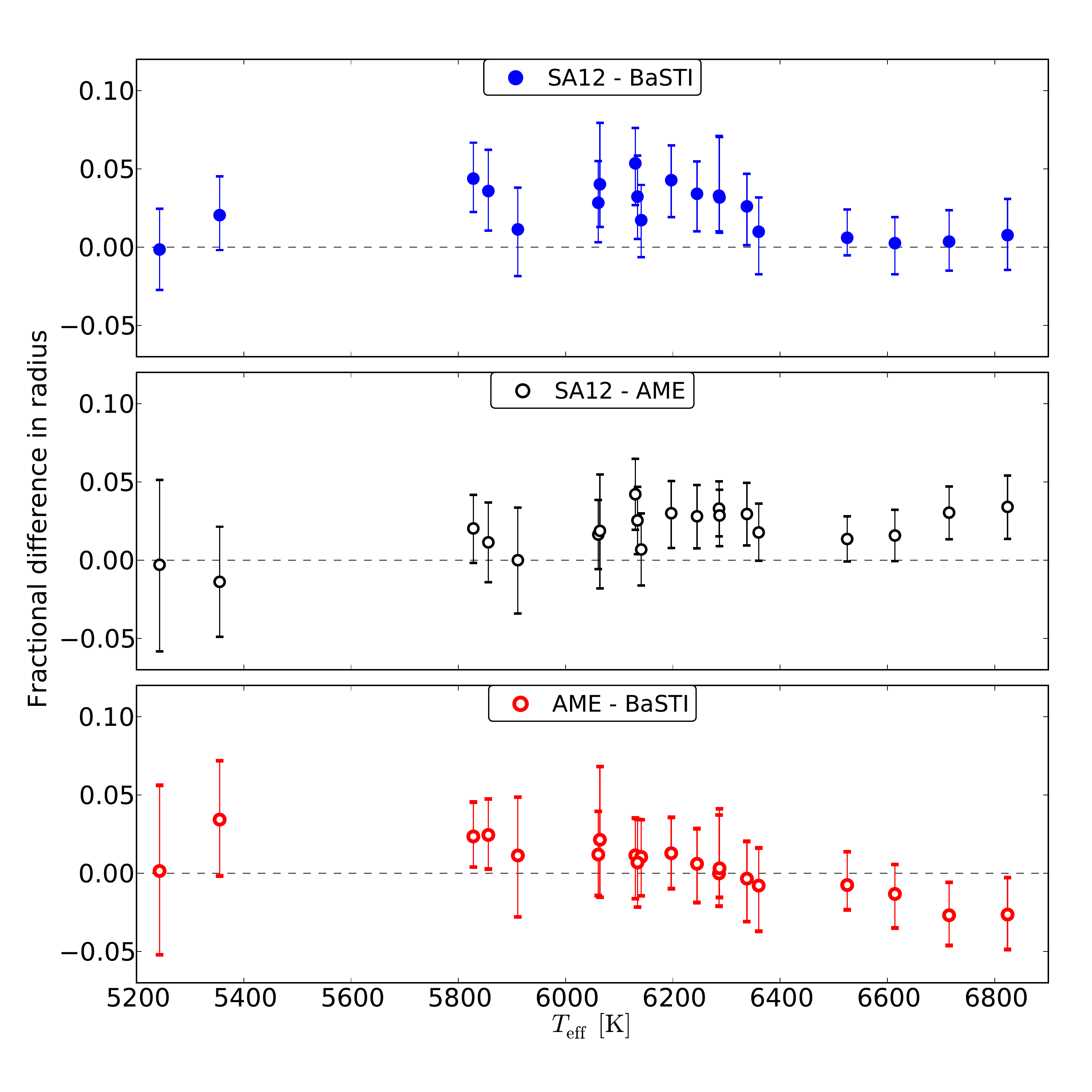} \label{fig:comp_teff_r_parall}}
\caption{Fractional difference in radius as a function of metallicity \protect\subref{fig:comp_feh_r_parall} and effective temperature \protect\subref{fig:comp_teff_r_parall} for the $20$ stars with radii based on parallaxes. The symbols give the fractional difference between the stellar radius determined from the parallax (SA12), the stellar radius determined from AME, and the one determined using the BaSTI grid in turn (we made the difference fractional by dividing it with the radius from AME). The individual errors on the radii have been added in quadrature. The dashed line marks a difference of zero.}
\label{fig:comp_r_parall}
\end{figure*}

AME has been used on a sample of $20$ stars from \citet{ref:vsa_2012_radii}. These stars have radii determined from measurements of their parallaxes and calculations of their angular diameters \citep[for details see][]{ref:vsa_2012_radii}. We wanted to use AME on these stars in order to facilitate a comparison between the reasonably model-independent radii from parallaxes and the ones from AME. Table~\ref{tab:values_parall_stars} gives the relevant parameters from \citet{ref:vsa_2012_radii} for the $20$ stars as well as the radii returned by AME. Note that four stars are in common between this sample and the interferometric sample (compare Table~\ref{tab:values_parall_stars} to Table~\ref{tab:values_interf_stars}). In the middle panel of \figref{fig:comp_feh_r_parall} and \figref{fig:comp_teff_r_parall} we show the fractional difference between the parallax-based radius and the AME radius (again the ratio of the difference and the AME radius) as a function of metallicity and effective temperature.

The agreement between the AME and the parallax-based radii is better than $2 \sigma$ for all targets and no clear slope is seen in the figures neither as a function of effective temperature no as a function of metallicity. However, it is evident that the AME radii are generally smaller than the parallax ones, which can be seen as a positive offset in the plots. In order to look into this, we determined the radii for all the stars using the BaSTI grid of stellar models which uses the same input parameters as AME and thus not the $\nu_\mathrm{max}$ scaling mentioned in Sect.~\ref{sec:intro}.

Briefly, the BaSTI grid was originally computed for the \citet{ref:victor_2} revision of the Geneva-Copenhagen Survey, and has now been extended in metallicity range for asteroseismic purposes. The input physics is that described in \citet{ref:victor_1}, while the final stellar parameters are determined performing Monte Carlo simulations over the input parameters ($T_\mathrm{eff}$, $[\mathrm{Fe}/\mathrm{H}]$, and $\Delta\nu$). The interested reader can find further details in \citet{ref:vsa_2013_ages}.

The results of comparing these radii to the parallax-based and AME radii can be found in the top and bottom panels of \figref{fig:comp_r_parall}. It can be seen that the fractional differences between the radii derived from the parallax and the BaSTI radii show distributions similar to the fractional differences between the parallax-based and AME radii. When comparing the AME and BaSTI results (bottom panels), it can be seen that these agree better than the others; the AME-BaSTI plot only has four stars with a fractional difference in excess of $1\sigma$ while this number is more than twice as high in the parallax-BaSTI and the parallax-AME plots. We will return to the offset in Sect.~\ref{subsec:dis_par_stars}.


\subsection{Stars with detailed modelling}
\label{subsec:res_mod_stars}

\begin{table*}
\caption{Stellar parameters from AME for the $16$ stars with properties determined from detailed modelling.}             
\label{tab:AME_values}      
\centering          
\begin{tabular}{llllll}     
\hline\hline       
Star
& \multicolumn{1}{c}{$M/M_\sun$}
& \multicolumn{1}{c}{$\bar{\rho}/\bar{\rho}_\sun$}
& \multicolumn{1}{c}{$\tau \ [\giga\mathrm{yr}]$}
& \multicolumn{1}{c}{$R/R_\sun$}
& \multicolumn{1}{c}{$g/g_\sun$} \\ 
\hline
Kepler-37		& $0.80 \pm 0.05$		& $1.7721 \pm 0.045$
				& $\phn 6.4 \pm 6.5$	& $0.77 \pm 0.02$
				& $1.359 \pm 0.036$ \\			
$\alpha$ Cen B	& $0.97 \pm 0.04$		& $1.4414 \pm 0.032$
				& $\phn 1.5 \pm 2.6$	& $0.88 \pm 0.01$
				& $1.263 \pm 0.025$ \\
Kepler-10		& $0.86 \pm 0.04$		& $0.7504 \pm 0.020$
				& $14.4 \pm 3.7$		& $1.05 \pm 0.02$
				& $0.785 \pm 0.018$ \\
$\alpha$ Cen A	& $1.10 \pm 0.03$		& $0.6042 \pm 0.013$
				& $\phn 6.0 \pm 1.2$	& $1.22 \pm 0.01$
				& $0.738 \pm 0.013$ \\
Perky			& $1.03 \pm 0.10$		& $0.5865 \pm 0.017$
				& $\phn 6.5 \pm 4.1$	& $1.21 \pm 0.03$
				& $0.708 \pm 0.033$ \\
Kepler-68		& $1.06 \pm 0.03$		& $0.5561 \pm 0.013$
				& $\phn 7.4 \pm 1.3$	& $1.24 \pm 0.01$
				& $0.689 \pm 0.013$ \\
Kepler-65		& $1.27 \pm 0.04$		& $0.4462 \pm 0.013$
				& $\phn 2.8 \pm 0.8$	& $1.42 \pm 0.01$
				& $0.632 \pm 0.014$ \\
$\mu$ Arae		& $1.19 \pm 0.04$		& $0.4378 \pm 0.012$
				& $\phn 5.3 \pm 1.0$	& $1.40 \pm 0.02$
				& $0.611 \pm 0.013$\\
Dushera			& $1.17 \pm 0.12$		& $0.4237 \pm 0.015$
				& $\phn 4.3 \pm 3.1$	& $1.40 \pm 0.04$
				& $0.594 \pm 0.035$ \\
Kepler-50		& $1.25 \pm 0.11$		& $0.3156 \pm 0.017$
				& $\phn 3.7 \pm 1.9$	& $1.58 \pm 0.03$
				& $0.499 \pm 0.027$ \\
Kepler-36		& $1.03 \pm 0.04$		& $0.2405 \pm 0.010$
				& $\phn 8.2 \pm 1.5$	& $1.62 \pm 0.03$
				& $0.391 \pm 0.012$ \\
HAT-P-7			& $1.55 \pm 0.05$		& $0.1959 \pm 0.009$
				& $\phn 1.9 \pm 0.5$	& $1.99 \pm 0.02$
				& $0.390 \pm 0.013$ \\
$\beta$ Hyi		& $1.14 \pm 0.05$		& $0.1701 \pm 0.005$
				& $\phn 6.4 \pm 1.4$	& $1.89 \pm 0.03$
				& $0.321 \pm 0.008$ \\
Kepler-7		& $1.26 \pm 0.03$		& $0.1671 \pm 0.012$
				& $\phn 4.9 \pm 0.6$	& $ 1.96 \pm 0.04$
				& $0.328 \pm 0.015$ \\
Procyon			& $1.39 \pm 0.06$		& $0.1761 \pm 0.009$
				& $\phn 2.8 \pm 0.7$	& $1.99 \pm 0.02$
				& $0.351 \pm 0.015$ \\
$\eta$ Boo		& $1.53 \pm 0.04$		& $0.0841 \pm 0.003$
				& $\phn 2.6 \pm 0.3$	& $2.63 \pm 0.02$
				& $0.221 \pm 0.006$ \\
\hline
\end{tabular}
\end{table*}

AME has been used on $16$ solar-like stars for which results from detailed modelling were available. This, as in the other cases, allows for a comparison of the properties obtained from AME with those available in the literature albeit in this case with parameters which are more model-dependent than the radii we compared to in the previous sections. However, it should be noted that $\alpha$~Cen~A and~B have their masses determined from their binarity and their radii from interferometry. Also, for the $16$ stars with detailed modelling results, we not only compare the radii, but also the masses and ages. Table~\ref{tab:AME_values} lists the stellar parameters that were obtained with AME for the detailed modelling targets (the literature values can be found in Table~\ref{tab:litt_values}).

In order to compare the output from AME with more than the parameters from detailed modelling which are based on more information and have assumptions that vary from star to star, we also ran all stars through the BaSTI grid of stellar models. We did this to get a set of consistent results for all the stars based on inputs similar to AME (for Perky and Dushera the BaSTI results were taken from \citet{ref:vsa_2013_ages}).

\begin{figure*}
	\includegraphics[width=17cm]{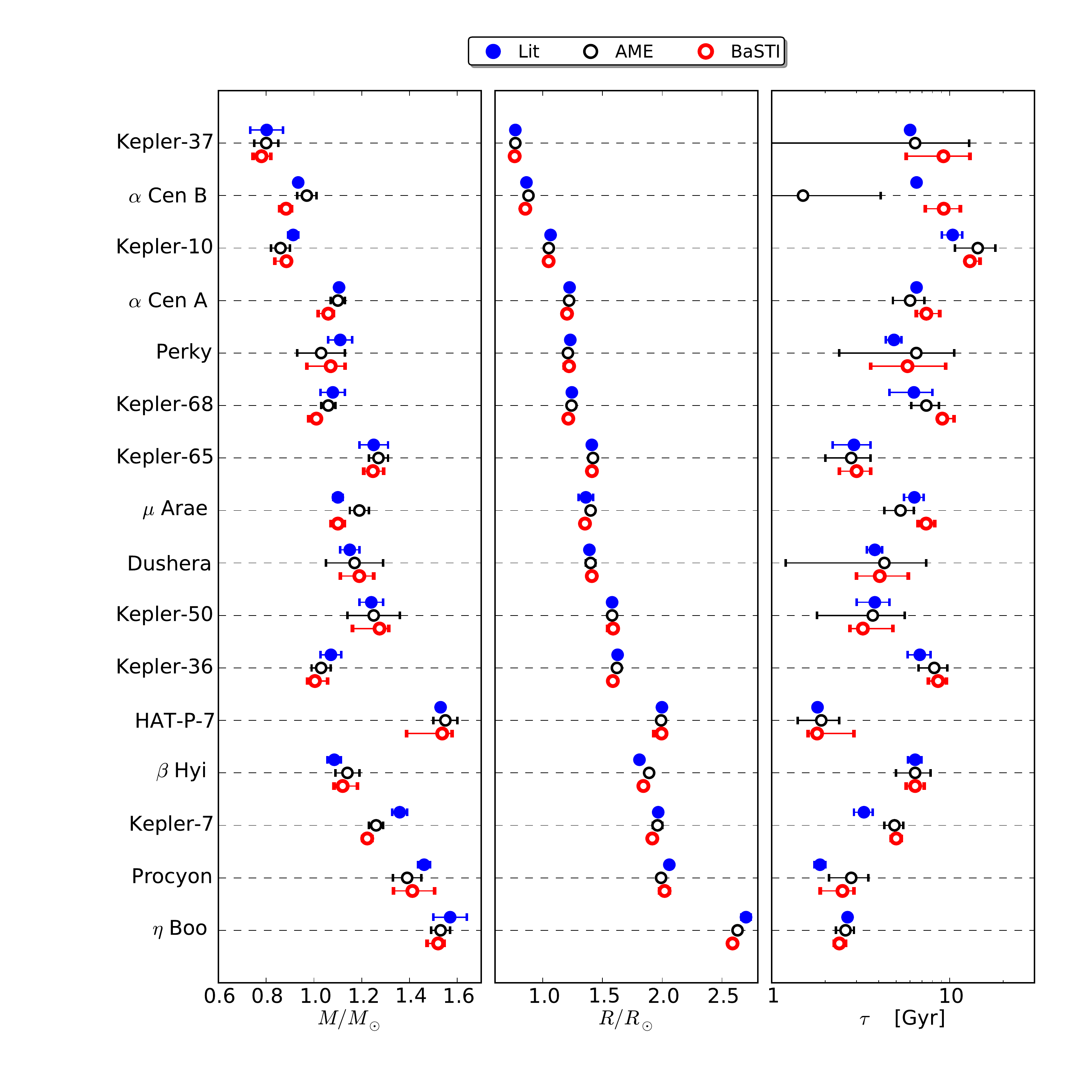}
    \caption{Comparison of the mass, radius, and age values found using AME, in the literature (Lit), and using the BaSTI grid for the stars with parameters known from detailed modelling. The values and their error bars have been plotted. Where the error bars cannot be seen, they are smaller than the symbol used to plot the point except in the case of Kepler-37. Here, no uncertainty is quoted in the literature for its age which is why this point has no error bar in the plot.}
	\label{fig:comparison}
\end{figure*}

Figure~\ref{fig:comparison} shows a comparison between the mass, radius, and age results obtained with AME and those found in the literature (Table~\ref{tab:litt_values}) and using the BaSTI grid. Note that the age of Kepler-37 is given as $\sim 6 \ \giga\mathrm{yr}$ by \citet{ref:kepler37} and consequently no error bars on the age have been plotted in \figref{fig:comparison} for this star.

\begin{figure*}
\centering
\subfloat[][]{\includegraphics[width=9cm]{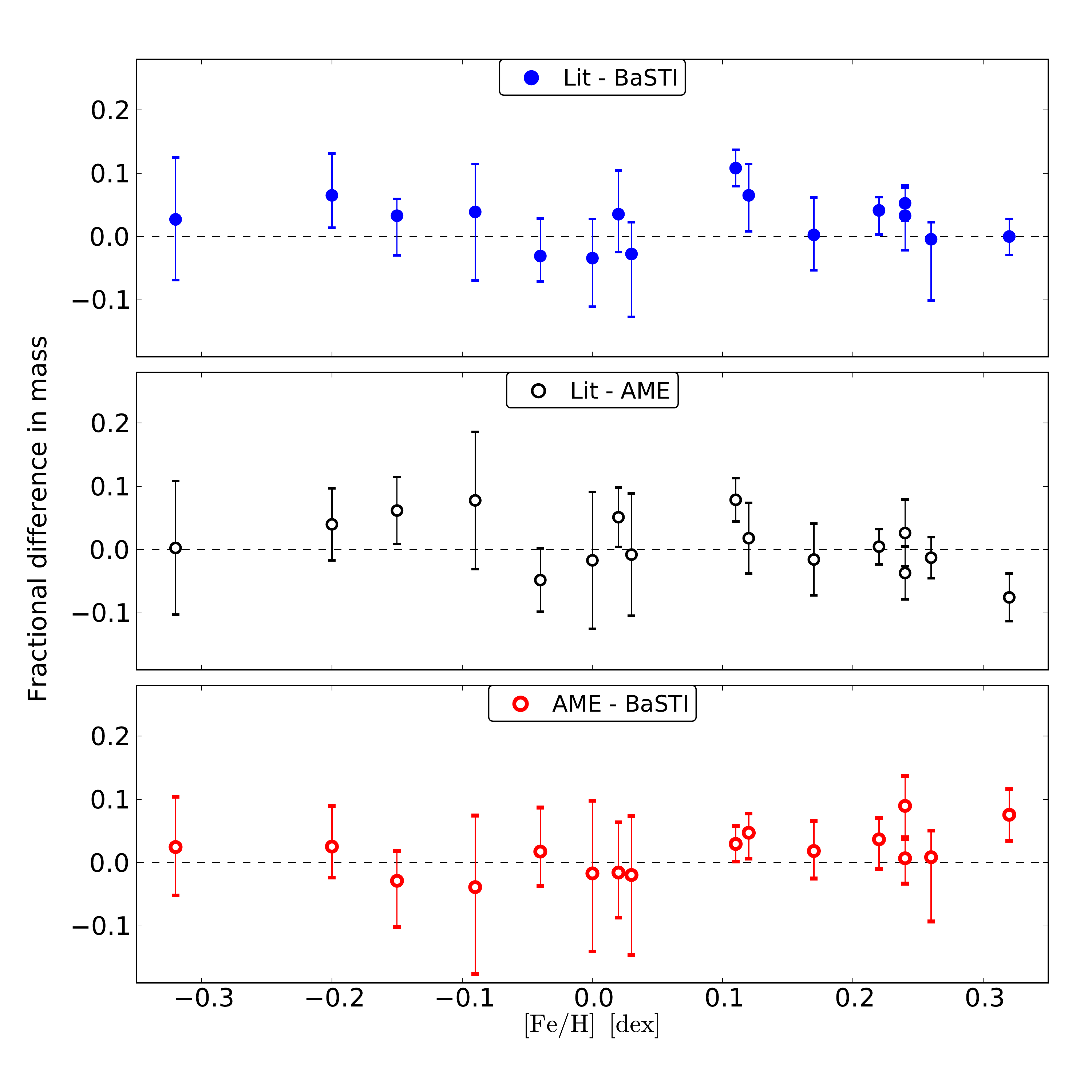} \label{fig:comp_feh_m}}
\hfill
\subfloat[][]{\includegraphics[width=9cm]{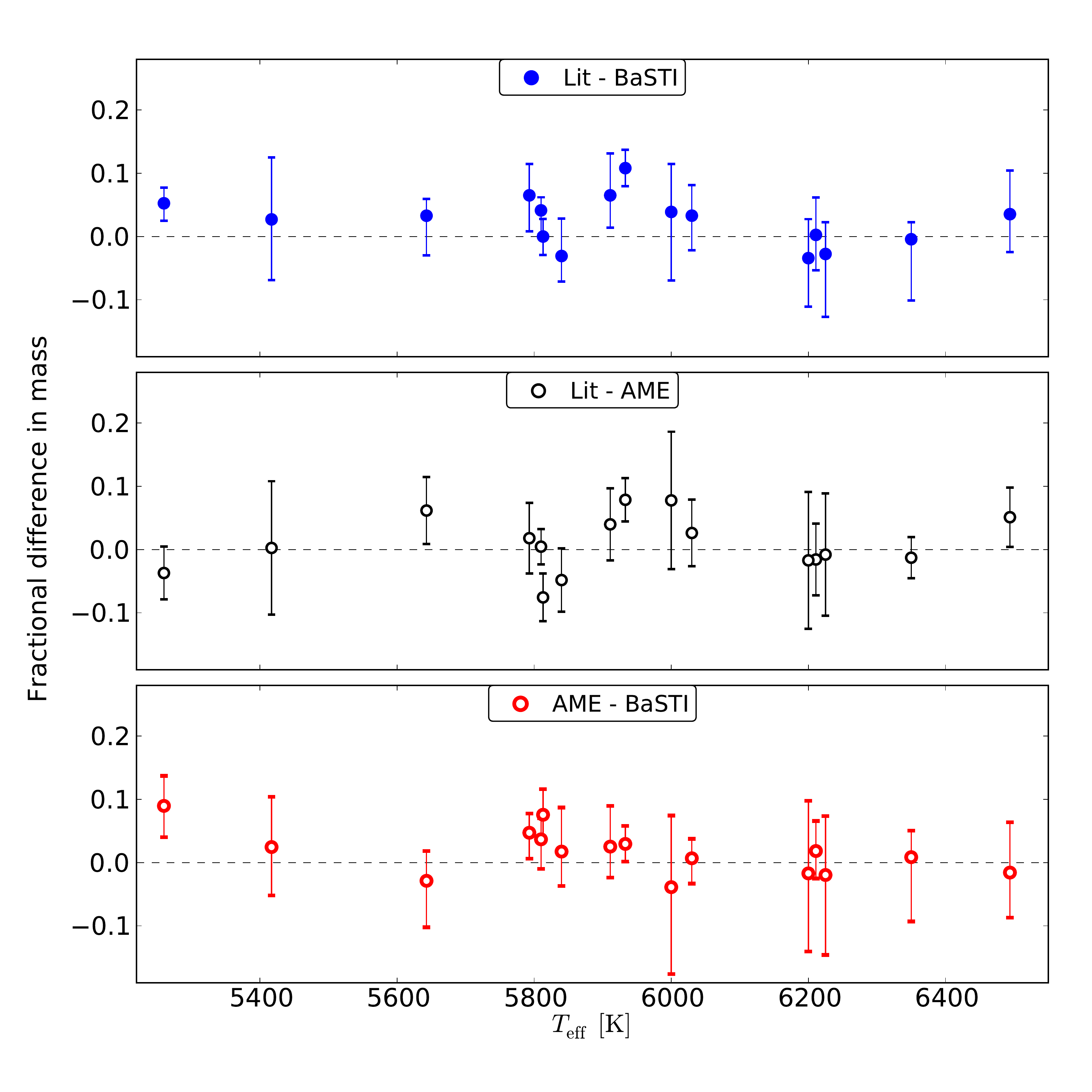} \label{fig:comp_teff_m}}
\caption{Comparison of the difference in mass as a function of metallicity (panel \protect\subref{fig:comp_feh_m}) and effective temperature (panel \protect\subref{fig:comp_teff_m}) for the stars with results from detailed modelling. The different coloured points show the Lit-AME, Lit-BaSTI, and AME-BaSTI differences respectively. The differences have been made fractional by dividing the difference with the mass obtained with AME. The errors have been added in quadrature.}
\label{fig:comp_mass}
\end{figure*}

\begin{figure*}
\centering
\subfloat[][]{\includegraphics[width=9cm]{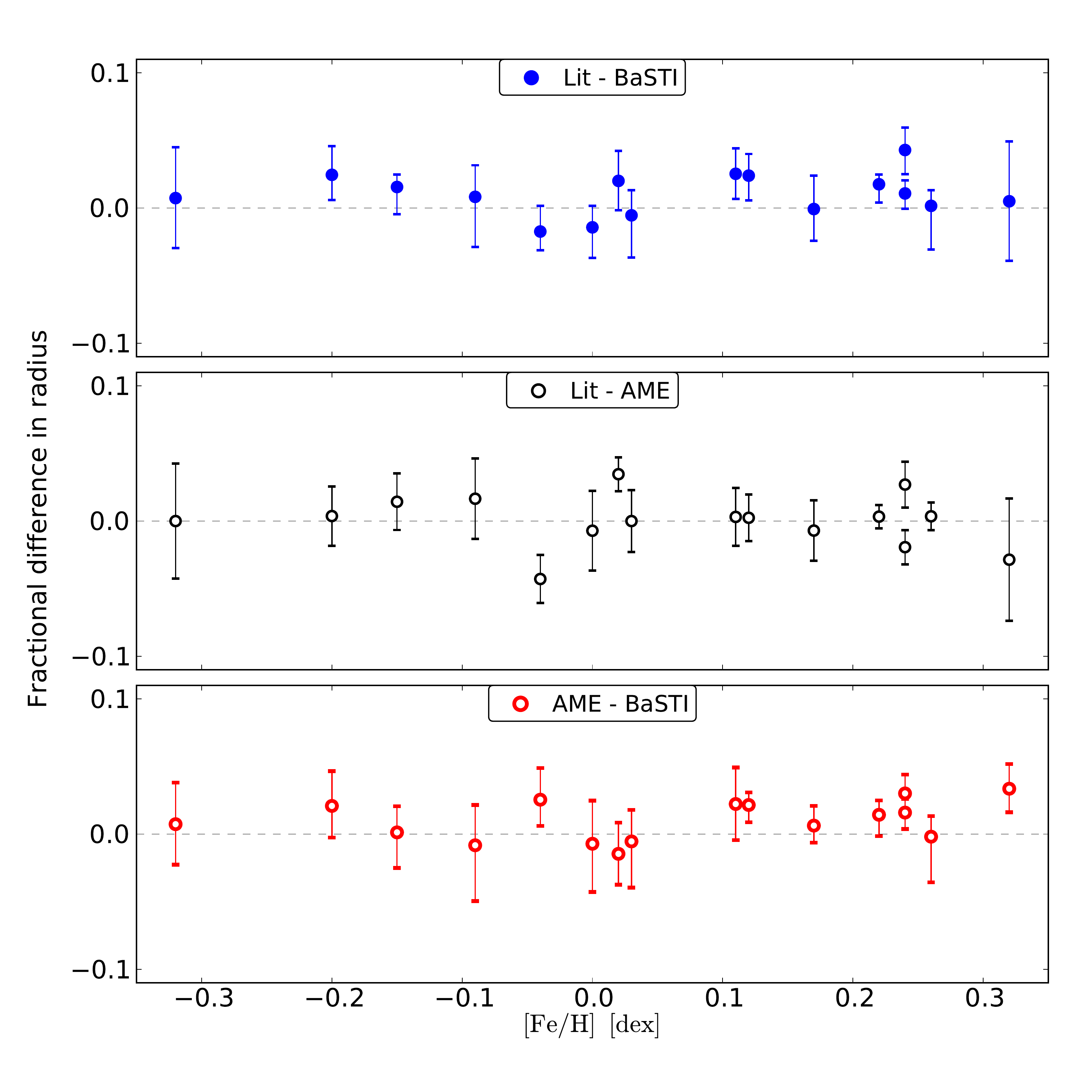} \label{fig:comp_feh_r}}
\hfill
\subfloat[][]{\includegraphics[width=9cm]{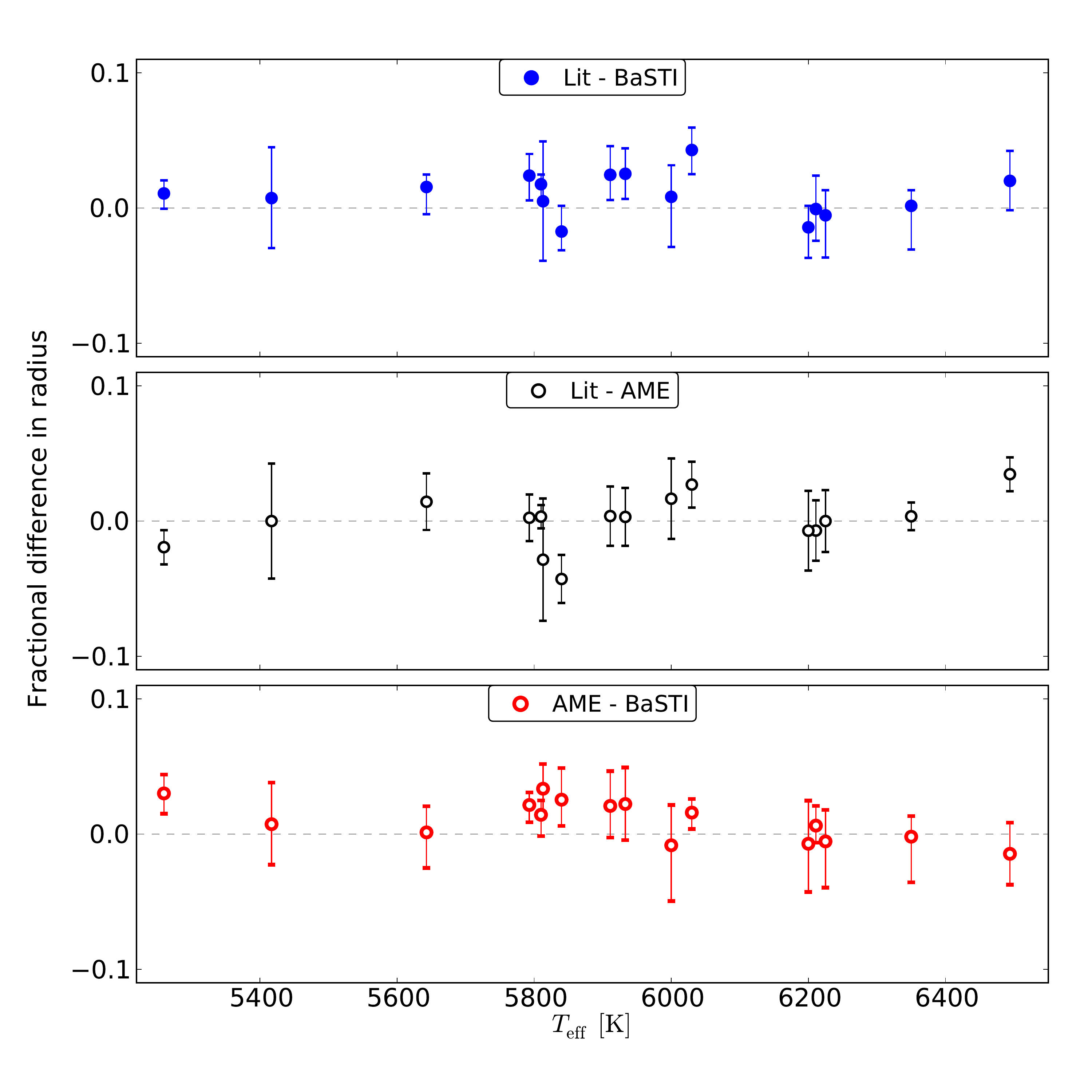} \label{fig:comp_teff_r}}
\caption{Comparison of the difference in radius as a function of metallicity (panel \protect\subref{fig:comp_feh_r}) and effective temperature (panel \protect\subref{fig:comp_teff_r}) for Lit-AME, Lit-BaSTI, and AME-BaSTI. All the differences have been scaled with the AME radius thus making the differences fractional. The errors have been added in quadrature.}
\label{fig:comp_radius}
\end{figure*}

It can be seen from \figref{fig:comparison} that the stellar parameters found using AME are in general consistent with the parameters found in the literature and using the BaSTI grid. This is very satisfactory because it indicates that AME can be reliably used to determine stellar properties. Figures~\ref{fig:comp_mass} and~\ref{fig:comp_radius} show the fractional differences in mass and radius between two sets of results at a time (Lit-BaSTI, Lit-AME, and AME-BaSTI) as a function of metallicity and effective temperature. In order to make the differences fractional, they have all been scaled with the AME results.

Most of the stars fall at or below the $1\sigma$ level which means that the results from the different methods generally agree quite well within their individual uncertainties. No trend or offset is apparent in either of the plots and this is encouraging because it shows that AME can reproduce the BaSTI and detailed modelling results without any bias. Also, it is noteworthy that the groups of points Lit-BaSTI and Lit-AME show similar distributions. This implies that the AME and BaSTI results agree equally well with the literature. Furthermore, the AME and BaSTI set of results also agree well with each other. It is reassuring that AME can produce results which are similar to the BaSTI results since AME and BaSTI use similar amounts of information as input for their results. In summary, we find that the masses and radii from AME and BaSTI agree very well, signifying that our method is at the same level as a grid-based approach.

\begin{figure}
	\centering
	\includegraphics[width=\hsize]{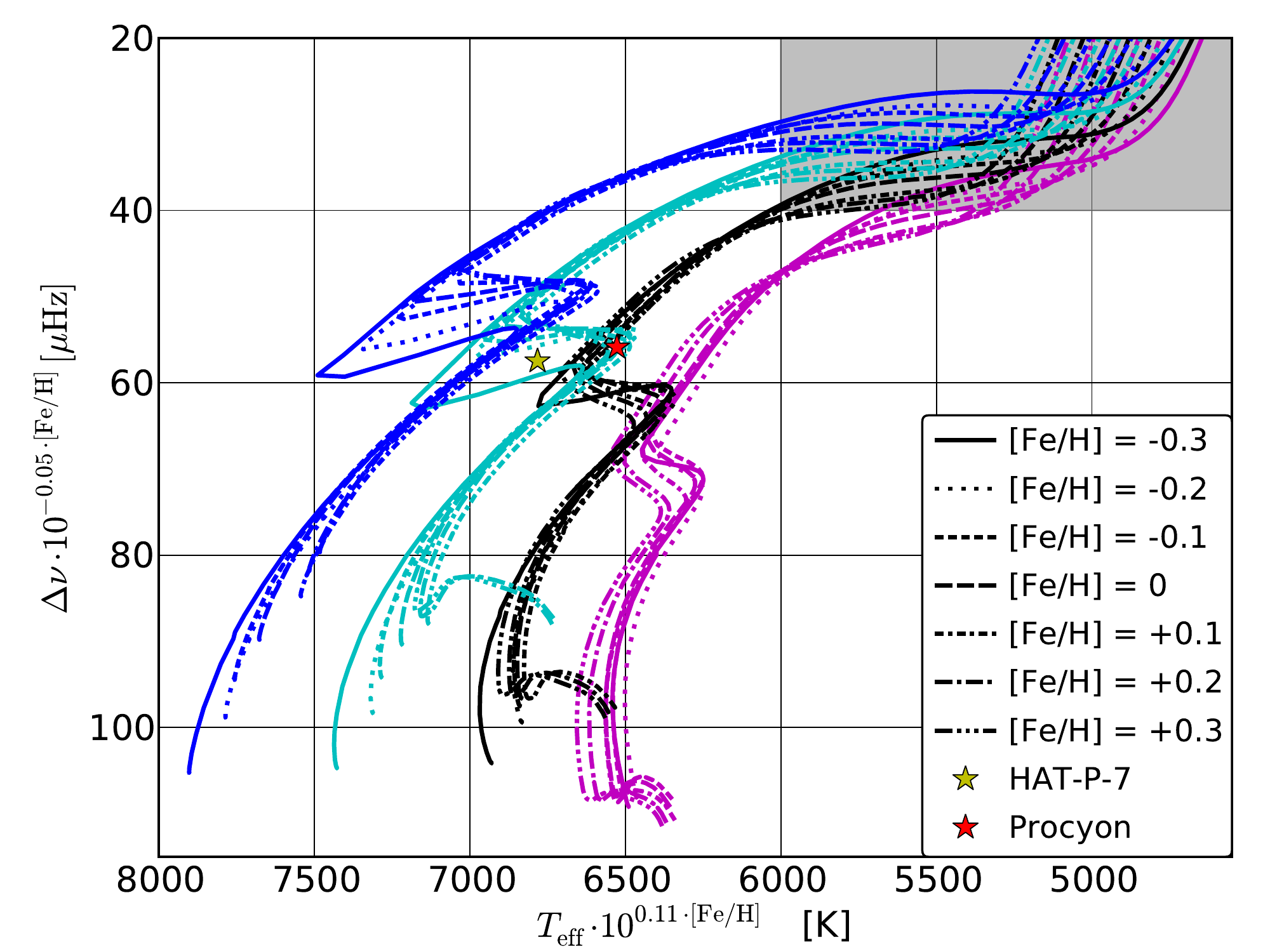}
	\caption{Plot used to determine the mass for high mass stars. The different line colours signify the masses of the lines ranging from $1.3 M_\sun$ (magenta/purple) to $1.6 M_\sun$ (blue) in steps of $0.1 M_\sun$. The various line types indicate the metallicity. The shaded area shows where the lines start to become hard to separate. The yellow star indicates the position of HAT-P-7 whilst the red star gives the location of Procyon in the plot.}
	\label{fig:m_handp}
\end{figure}

A few of results given in Table~\ref{tab:AME_values} deserve further comments. In the case of Procyon, this star was found to be placed in an area of the mass plot where lines of different masses cross (see \figref{fig:m_handp}). The reason for this crossing is the MS turn-off. Since a star passes through this phase relatively fast compared to its MS and sub-giant-branch lifetimes (for our models, the evolution of a $1.4M_\sun$ star in the turn-off phase is more than twice as fast as in the MS phase and approximately $20\%$ faster than in the sub-giant phase), we assumed that Procyon was not in the turn-off, and consequently we disregarded those lines (the light blues lines) when we determined its mass. As a consequence of this, for stars in some parts of the plots, systematic differences could occur due to the non-negligible possibility of being in a different evolutionary phase. A similar assumption as in the case Procyon was made for HAT-P-7. Here, we chose to use tracks that were in the same evolutionary state for the determination of the mass (see \figref{fig:m_handp}). This meant that we used the pre-turn-off tracks in light blue and blue to establish the mass and thus ignored the black lines in this case.

We also had to ignore some of the lines when we determined the masses of two of the parallax stars from Table~\ref{tab:values_parall_stars}. These two stars are KIC-5371516 and KIC-9139151. In the case of the former, it was situated right on top of the light-blue MS turn-off tracks seen in \figref{fig:m_handp} and we as a consequence chose to determine its mass via linear extrapolation based on the purple and black tracks which were in the same evolutionary state at the location of the star. In case of the latter star, the situation was a little different, since this star was not located near the MS turn-off, but instead in the bottom part of the mass plot for low-mass stars seen in \figref{fig:m_lm} before the beginning of the $1.3 M_\sun$ tracks. Probably as an effect of the density of the grid, the mass of this star needed to be determined by extrapolation based on the $1.1 M_\sun$ (yellow) and $1.2 M_\sun$ (red) tracks. Had the grid been denser such that a track had been present for a $1.25 M_\sun$ star, this extrapolation had probably not been necessary (the mass of the star was found to be $1.23 M_\sun$).

As can be seen when inspecting Tables~\ref{tab:litt_values} and~\ref{tab:values_interf_stars}, a few of the metallicity values lie slightly beyond the AME grid. In these cases, we used linear extrapolation similarly to what was described above. Note though that great care was taken to make sure that the results were meaningful; they have been manually inspected and we did not attempt metallicities that were far from the grid.

It can happen that a star falls close to the limit of a plot. For instance the star Perky has a metallicity of $-0.09 \pm 0.1 \ \mathrm{dex}$ which means that when the uncertainties were considered, two different plots had to be used to estimate the mean-density and its associated error bar (the same was true for the age). In this case, we simply used the relevant plot according to the corner of the error box under consideration.

It may also occur that a star falls exactly on the limit between two plots, which is the case for the star KIC-$3733735$ (see Table~\ref{tab:values_parall_stars}) since it has a metallicity of $-0.10 \pm 0.1 \ \mathrm{dex}$. In this case, we used both relevant plots to determine the mean-density and used the plot with the appropriate metallicity depending on the corner of the error box to find the uncertainty. The plots used to establish the mean-density returned identical results (again, the same was true for the age). If this had not been the case, we would have used the mean value as the result and added the difference between the plots in quadrature to the uncertainty estimates.


\section{Discussion}
\label{sec:discussion}

We will comment on the results from the study of the seven stars with radii from interferometry in Sect.~\ref{subsec:dis_int_stars}, the results for the $20$ stars with radii based on measurements of their parallaxes in Sect.~\ref{subsec:dis_par_stars} and finally the results for the $16$ targets with results from detailed modelling in Sect.~\ref{subsec:dis_mod_stars}.


\subsection{Stars with interferometry}
\label{subsec:dis_int_stars}

The agreement between the interferometric and AME radii is excellent as can be seen from \figref{fig:comp_rad_interf}. Five of the seven stars have results that are within $1 \sigma$ while two stars have radii that differ by more than $1\sigma$ ($2.0\sigma$ and $2.2\sigma$ respectively). These two stars are  KIC-$6106415$ and KIC-$6225718$ and they have the smallest angular diameters of the stars measured by \citet{ref:huber_interferometry}. Thus, their interferometric uncertainties are more difficult to determine than those for the stars with larger angular diameters. The weighted mean of the fractional differences between the interferometric and the AME radii is $1.7\%$ as was mentioned earlier. Hence, the radii from AME are consistent with the interferometric radii over the range of metallicities, effective temperatures, and large separations represented by the interferometric targets. The consistency  between the AME and the largely model-independent interferometric radii implies that the radii from AME are likely trustworthy and accurate.


\subsection{Stars with parallaxes}
\label{subsec:dis_par_stars}

The offset seen in \figref{fig:comp_r_parall} points towards a systematic difference between the radii given in the literature for the parallax stars (SA12 radii) and the AME radii. In spite of the offset, the radii from AME are still within $5\%$ of the SA12 radii which is the accuracy claimed by \citet{ref:vsa_2012_radii}. In fact the weighted mean of the fractional differences is $2.3\%$. The weighted mean of the fractional differences when comparing the SA12 results to the BaSTI results is $2.1\%$ which is similar to the SA12-AME value, while the AME-BaSTI weighted mean of the fractional differences is $1.4\%$. Thus SA12-AME radii and SA12-BaSTI radii show similar discrepancies while the AME-BaSTI radii agree slightly better.

A similar offset to that seen between SA12 and AME, and SA12 and BaSTI in \figref{fig:comp_teff_r_parall}, both in size and shape, is found by \citet{ref:explain_par_diff} (see the top left panel of their Fig.~4). They find that the "boomerang" shaped discrepancy arises from the difference between determining $\Delta\nu$ from individual oscillation frequencies and from the $\Delta\nu \propto \sqrt{\bar{\rho}}$ scaling relation. We use a similar approach to find the large separation for the models in AME as the scaling relation approach (we use the sound-speed integral given in equation~\ref{eq:sound_speed_integral}) and the approach for the stars in \citet{ref:vsa_2012_radii} is to use calculated oscillations frequencies to obtain the large separation. As a consequence, a similar discrepancy can be expected - and found - for our comparison of the SA12 and the AME results (note that \citet{ref:explain_par_diff} plot the fractional difference in the "scaling relation-individual" sense where we have done the opposite and that their temperature axis is reversed with respect to ours). BaSTI finds the large separation of its models using the scaling relation, and as a consequence the same boomerang is expected - and seen - also for the SA12 - BaSTI comparison (see the top panel of \figref{fig:comp_teff_r_parall}). However, the shape and size of the offset in the case of SA12-AME is not completely the same as the SA12-BaSTI one. This may be due to the fact that we, as mentioned before, do not use the exact same method to find the $\Delta\nu$'s. From \figref{fig:comp_teff_r_parall} it appears that using the sound-speed integral to find the large separation (and scaling it as we have done) introduces a smaller offset than using the scaling relation to obtain it (as BaSTI does). For the warmest stars, we see the deviation in shape between the SA12-AME and the SA12-BaSTI results. There is an indication that the AME radii are underestimated for these warmest stars. However, first of all these stars are F-stars and therefore notoriously challenging. And second, AME uses scaling to the Sun and these stars are the least sun-like in the sample so it makes sense why the AME results show a small discrepancy from the SA12 results.


\subsection{Stars with detailed modelling}
\label{subsec:dis_mod_stars}

From \figref{fig:comparison} and Table~\ref{tab:AME_values} it is seen that the AME ages generally have fairly large error bars. The age values that have been found for the stars with detailed modelling have a median uncertainty of $25\%$ (with percentages ranging from $12\%$ to $173\%$) while the median uncertainty on the mass of these stars is $4\%$ and the median uncertainty on the radius for the full sample of stars is $2\%$. However, to be able to determine the age was not an initial goal of AME, and the reason for the sometimes quite large error bars on the AME age determinations is largely the strong dependence of the age on the mass (see the y-axis of \figref{fig:tau_lm}). Note that the median uncertainties quoted are internal uncertainties and as a consequence systematic errors arising from i.e. the input physics chosen are not taken into account.

Furthermore, the ages are the plotted AME parameter in \figref{fig:comparison} that differ most from the literature and BaSTI properties. The weighted mean of the fractional differences between the literature and AME ages is $15\%$ whilst it is $10\%$ between the AME and the BaSTI ages. This is of course a large difference, but still the AME ages can be used as estimates, especially when one considers the uncertainties, since the ages from AME are most often within $1\sigma$ of the BaSTI and the literature values. The star $\alpha$~Cen~B is a noticeable example of a star whose age differ by more than $1 \sigma$ from the value given in the literature. The reason for this deviation and the large uncertainty is likely a combination of its early evolutionary state (which places it where the curve on the graph in \figref{fig:tau_lm} is steep) and the fact that the mass deviates by $1\sigma$ from the one given in the literature (the mass has a high exponent on the y-axis of the age plot). The mass that AME finds is larger than the literature one and thus the age will be underestimated.

In \figref{fig:comparison} the masses and radii from AME agree well with the values from the other sources. Since these parameters are more precisely determined than the age, both in AME and in the literature and the BaSTI results, we have taken a closer look at these parameters by plotting them as functions of metallicity and effective temperature in Figs.~\ref{fig:comp_mass} and~\ref{fig:comp_radius}.

There are a few outliers in the Lit-AME group which can be seen if we consider the middle panels of these figures. In the middle panels of \figref{fig:comp_mass}, four stars have masses that differ by more than $1\sigma$. These are Kepler-10, Procyon, Kepler-7, and $\mu$~Arae. However, the weighted mean of the fractional differences of all the masses is just $4\%$ and only Kepler-7 has a mass value that differ by more than $2\sigma$.

Figures~\ref{fig:comp_feh_r} and~\ref{fig:comp_teff_r} show the radius differences as a function of metallicity and effective temperature respectively. The weighted mean of the fractional differences between the literature and the AME results (the middle panels) is only $1.3\%$, but there are still four stars with a fractional difference in excess of $1 \sigma$. These are $\alpha$~Cen~B and the three sub-giants $\beta$~Hyi, Procyon, and $\eta$~Boo. We already discussed $\alpha$~Cen~B, so we focus on the sub-giants now. A possible explanation for the trend that the results for the sub-giants deviate by more than $1\sigma$ is that the temperature sensitivity of the AME results for the sub-giants is a lot higher than for stars on the MS (see \figref{fig:JCD_request} where it is clear that the evolution tracks of models with different masses lie a lot closer at low $\Delta \nu$ compared to at high values of $\Delta \nu$). It is noteworthy that the BaSTI and AME results for Procyon agree very well (better than $1\sigma$) and that the agreement between the BaSTI and AME results for $\beta$~Hyi and $\eta$~Boo is also good (the fractional difference is just above $1\sigma$ in both cases).

The bottom panels of Figs.~\ref{fig:comp_mass} and~\ref{fig:comp_radius} show the fractional differences between the AME and BaSTI results. The agreement between the results is good. The weighted mean of fractional differences is $4\%$ in mass and $1.6\%$ in radius. Only a couple of stars have fractional differences in excess of $1\sigma$ and all of these deviate by less than $2\sigma$. For the mass differences, the stars that deviate by more than $1\sigma$ are Kepler-10, Kepler-68, $\alpha$~Cen~B, and $\mu$~Arae while the ones that differ by more than $1\sigma$ in radius are Kepler-68, $\alpha$~Cen~B, and $\mu$~Arae as in the mass case and then the two sub-giants $\beta$~Hyi and $\eta$~Boo. Considering the low weighted mean of the fractional differences, the results are very promising since AME and the BaSTI grid obtain their stellar properties based on a similar amount of information. Thereby, all in all AME displays good agreement with both results from the BaSTI grid and values found in the literature. This further shows that AME can be reliably used to obtain basic stellar parameters for stars exhibiting solar-like oscillations.


\section{Conclusion}
\label{sec:conclusion}

We have shown that AME as a tool to determine basic stellar parameters works very well. It offers a transparent and easy way to find stellar properties for stars exhibiting solar-like oscillations at a level where the large separation can be detected.

The full set of AME figures which are the backbone of the method can be found in the appendix~\ref{om_sec:morefigs}. They cover a range in masses from $0.7 M_\sun$ in steps of $0.1 M_\sun$ to $1.6 M_\sun$, and in metallicity the span is $-0.3 \ \mathrm{dex} \leq [\mathrm{Fe}/\mathrm{H}] \leq +0.3 \ \mathrm{dex}$ in increments of $0.1 \ \mathrm{dex}$.

We have used AME on a total of $43$ stars with stellar parameters determined in different ways. We have compared the results from AME to those found in the literature and in some cases to results obtained with the BaSTI grid. The overall agreement is good and the differences can be explained. AME was found to perform as well as the BaSTI grid in terms of agreement with the values quoted in the literature for the $16$ stars with results from detailed modelling. Furthermore, the radii from AME was found to be consistent with both radii from interferometry (and measured parallaxes) and radii determined from parallaxes (and computed  angular diameters).

In the future, the AME grid will be extended to cover a wider range of metallicities and masses and potentially made denser. We are currently working on a web interface to enable people to use AME without having to plot their stars in the diagrams themselves.


\begin{acknowledgements}
Funding for the Stellar Astrophysics Centre is provided by The Danish National Research Foundation (Grant agreement no.: DNRF106). The research is supported by the ASTERISK project (ASTERoseismic Investigations with SONG and Kepler) funded by the European Research Council (Grant agreement no.: 267864).
\end{acknowledgements}


\bibliographystyle{aa} 

\begin{thebibliography}{75}
\expandafter\ifx\csname natexlab\endcsname\relax\def\natexlab#1{#1}\fi

\bibitem[{{Aerts} {et~al.}(2010){Aerts}, {Christensen-Dalsgaard}, \&
  {Kurtz}}]{ref:ast_book}
{Aerts}, C., {Christensen-Dalsgaard}, J., \& {Kurtz}, D.~W. 2010,
  {Asteroseismology}, 1st edn. (Springer)

\bibitem[{{Angulo} {et~al.}(1999){Angulo}, {Arnould}, {Rayet}, {Descouvemont},
  {Baye}, {Leclercq-Willain}, {Coc}, {Barhoumi}, {Aguer}, {Rolfs}, {Kunz},
  {Hammer}, {Mayer}, {Paradellis}, {Kossionides}, {Chronidou}, {Spyrou},
  {degl'Innocenti}, {Fiorentini}, {Ricci}, {Zavatarelli}, {Providencia},
  {Wolters}, {Soares}, {Grama}, {Rahighi}, {Shotter}, \& {Lamehi
  Rachti}}]{ref:nacre_reac_rates}
{Angulo}, C., {Arnould}, M., {Rayet}, M., {et~al.} 1999, Nuclear Physics A,
  656, 3

\bibitem[{{Asplund} {et~al.}(2009){Asplund}, {Grevesse}, {Sauval}, \&
  {Scott}}]{ref:Asp09}
{Asplund}, M., {Grevesse}, N., {Sauval}, A.~J., \& {Scott}, P. 2009, Annual
  Review of Astronomy \& Astrophysics, 47, 481

\bibitem[{{Balser}(2006)}]{ref:balser_2006}
{Balser}, D.~S. 2006, \aj, 132, 2326

\bibitem[{{Barclay} {et~al.}(2013){Barclay}, {Rowe}, {Lissauer}, {Huber},
  {Fressin}, {Howell}, {Bryson}, {Chaplin}, {D{\'e}sert}, {Lopez}, {Marcy},
  {Mullally}, {Ragozzine}, {Torres}, {Adams}, {Agol}, {Barrado}, {Basu},
  {Bedding}, {Buchhave}, {Charbonneau}, {Christiansen},
  {Christensen-Dalsgaard}, {Ciardi}, {Cochran}, {Dupree}, {Elsworth},
  {Everett}, {Fischer}, {Ford}, {Fortney}, {Geary}, {Haas}, {Handberg},
  {Hekker}, {Henze}, {Horch}, {Howard}, {Hunter}, {Isaacson}, {Jenkins},
  {Karoff}, {Kawaler}, {Kjeldsen}, {Klaus}, {Latham}, {Li}, {Lillo-Box},
  {Lund}, {Lundkvist}, {Metcalfe}, {Miglio}, {Morris}, {Quintana}, {Stello},
  {Smith}, {Still}, \& {Thompson}}]{ref:kepler37}
{Barclay}, T., {Rowe}, J.~F., {Lissauer}, J.~J., {et~al.} 2013, \nat, 494, 452

\bibitem[{{Basu} {et~al.}(2010){Basu}, {Chaplin}, \& {Elsworth}}]{ref:pip_yb}
{Basu}, S., {Chaplin}, W.~J., \& {Elsworth}, Y. 2010, \apj, 710, 1596

\bibitem[{{Batalha} {et~al.}(2011){Batalha}, {Borucki}, {Bryson}, {Buchhave},
  {Caldwell}, {Christensen-Dalsgaard}, {Ciardi}, {Dunham}, {Fressin},
  {Gautier}, {Gilliland}, {Haas}, {Howell}, {Jenkins}, {Kjeldsen}, {Koch},
  {Latham}, {Lissauer}, {Marcy}, {Rowe}, {Sasselov}, {Seager}, {Steffen},
  {Torres}, {Basri}, {Brown}, {Charbonneau}, {Christiansen}, {Clarke},
  {Cochran}, {Dupree}, {Fabrycky}, {Fischer}, {Ford}, {Fortney}, {Girouard},
  {Holman}, {Johnson}, {Isaacson}, {Klaus}, {Machalek}, {Moorehead},
  {Morehead}, {Ragozzine}, {Tenenbaum}, {Twicken}, {Quinn}, {VanCleve},
  {Walkowicz}, {Welsh}, {Devore}, \& {Gould}}]{ref:Kepler10}
{Batalha}, N.~M., {Borucki}, W.~J., {Bryson}, S.~T., {et~al.} 2011, \apj, 729,
  27

\bibitem[{{Bedding} \& {Kjeldsen}(2010)}]{ref:bk_dnu_rho}
{Bedding}, T.~R. \& {Kjeldsen}, H. 2010, Communications in Asteroseismology,
  161, 3

\bibitem[{{Bedding} {et~al.}(2007){Bedding}, {Kjeldsen}, {Arentoft}, {Bouchy},
  {Brandbyge}, {Brewer}, {Butler}, {Christensen-Dalsgaard}, {Dall}, {Frandsen},
  {Karoff}, {Kiss}, {Monteiro}, {Pijpers}, {Teixeira}, {Tinney}, {Baldry},
  {Carrier}, \& {O'Toole}}]{ref:betaHyi_dnu}
{Bedding}, T.~R., {Kjeldsen}, H., {Arentoft}, T., {et~al.} 2007, \apj, 663,
  1315

\bibitem[{{Bedding} {et~al.}(2010){Bedding}, {Kjeldsen}, {Campante},
  {Appourchaux}, {Bonanno}, {Chaplin}, {Garcia}, {Marti{\'c}}, {Mosser},
  {Butler}, {Bruntt}, {Kiss}, {O'Toole}, {Kambe}, {Ando}, {Izumiura}, {Sato},
  {Hartmann}, {Hatzes}, {Barban}, {Berthomieu}, {Michel}, {Provost},
  {Turck-Chi{\`e}ze}, {Lebrun}, {Schmitt}, {Bertaux}, {Benatti}, {Claudi},
  {Cosentino}, {Leccia}, {Frandsen}, {Brogaard}, {Glowienka}, {Grundahl},
  {Stempels}, {Arentoft}, {Bazot}, {Christensen-Dalsgaard}, {Dall}, {Karoff},
  {Lundgreen-Nielsen}, {Carrier}, {Eggenberger}, {Sosnowska}, {Wittenmyer},
  {Endl}, {Metcalfe}, {Hekker}, \& {Reffert}}]{ref:procyon}
{Bedding}, T.~R., {Kjeldsen}, H., {Campante}, T.~L., {et~al.} 2010, \apj, 713,
  935

\bibitem[{{Belkacem} {et~al.}(2011){Belkacem}, {Goupil}, {Dupret}, {Samadi},
  {Baudin}, {Noels}, \& {Mosser}}]{ref:belkacem_2011}
{Belkacem}, K., {Goupil}, M.~J., {Dupret}, M.~A., {et~al.} 2011, \aap, 530,
  A142

\bibitem[{{Belkacem} {et~al.}(2013){Belkacem}, {Samadi}, {Mosser}, {Goupil}, \&
  {Ludwig}}]{ref:belkacem_2013}
{Belkacem}, K., {Samadi}, R., {Mosser}, B., {Goupil}, M.~J., \& {Ludwig}, H.-G.
  2013, ArXiv e-prints

\bibitem[{{B{\"o}hm-Vitense}(1958)}]{ref:mixinglength}
{B{\"o}hm-Vitense}, E. 1958, \zap, 46, 108

\bibitem[{{Bonaca} {et~al.}(2012){Bonaca}, {Tanner}, {Basu}, {Chaplin},
  {Metcalfe}, {Monteiro}, {Ballot}, {Bedding}, {Bonanno}, {Broomhall},
  {Bruntt}, {Campante}, {Christensen-Dalsgaard}, {Corsaro}, {Elsworth},
  {Garc{\'{\i}}a}, {Hekker}, {Karoff}, {Kjeldsen}, {Mathur}, {R{\'e}gulo},
  {Roxburgh}, {Stello}, {Trampedach}, {Barclay}, {Burke}, \&
  {Caldwell}}]{ref:importance_solar_alpha}
{Bonaca}, A., {Tanner}, J.~D., {Basu}, S., {et~al.} 2012, \apjl, 755, L12

\bibitem[{{Bouchy} \& {Carrier}(2002)}]{ref:cenAB_deltanuA}
{Bouchy}, F. \& {Carrier}, F. 2002, \aap, 390, 205

\bibitem[{{Brand{\~a}o} {et~al.}(2011){Brand{\~a}o}, {Do{\u g}an},
  {Christensen-Dalsgaard}, {Cunha}, {Bedding}, {Metcalfe}, {Kjeldsen},
  {Bruntt}, \& {Arentoft}}]{ref:betaHyi_age}
{Brand{\~a}o}, I.~M., {Do{\u g}an}, G., {Christensen-Dalsgaard}, J., {et~al.}
  2011, \aap, 527, A37

\bibitem[{{Bruntt} {et~al.}(2012){Bruntt}, {Basu}, {Smalley}, {Chaplin},
  {Verner}, {Bedding}, {Catala}, {Gazzano}, {Molenda-{\.Z}akowicz}, {Thygesen},
  {Uytterhoeven}, {Hekker}, {Huber}, {Karoff}, {Mathur}, {Mosser},
  {Appourchaux}, {Campante}, {Elsworth}, {Garc{\'{\i}}a}, {Handberg},
  {Metcalfe}, {Quirion}, {R{\'e}gulo}, {Roxburgh}, {Stello},
  {Christensen-Dalsgaard}, {Kawaler}, {Kjeldsen}, {Morris}, {Quintana}, \&
  {Sanderfer}}]{ref:bruntt_2012}
{Bruntt}, H., {Basu}, S., {Smalley}, B., {et~al.} 2012, \mnras, 423, 122

\bibitem[{{Bruntt} {et~al.}(2010){Bruntt}, {Bedding}, {Quirion}, {Lo Curto},
  {Carrier}, {Smalley}, {Dall}, {Arentoft}, {Bazot}, \& {Butler}}]{ref:bruntt}
{Bruntt}, H., {Bedding}, T.~R., {Quirion}, P.-O., {et~al.} 2010, \mnras, 405,
  1907

\bibitem[{{Carrier} \& {Bourban}(2003)}]{ref:cenAB_deltanuB}
{Carrier}, F. \& {Bourban}, G. 2003, \aap, 406, L23

\bibitem[{{Carrier} {et~al.}(2005){Carrier}, {Eggenberger}, \&
  {Bouchy}}]{ref:etaBoo}
{Carrier}, F., {Eggenberger}, P., \& {Bouchy}, F. 2005, \aap, 434, 1085

\bibitem[{{Carter} {et~al.}(2012){Carter}, {Agol}, {Chaplin}, {Basu},
  {Bedding}, {Buchhave}, {Christensen-Dalsgaard}, {Deck}, {Elsworth},
  {Fabrycky}, {Ford}, {Fortney}, {Hale}, {Handberg}, {Hekker}, {Holman},
  {Huber}, {Karoff}, {Kawaler}, {Kjeldsen}, {Lissauer}, {Lopez}, {Lund},
  {Lundkvist}, {Metcalfe}, {Miglio}, {Rogers}, {Stello}, {Borucki}, {Bryson},
  {Christiansen}, {Cochran}, {Geary}, {Gilliland}, {Haas}, {Hall}, {Howard},
  {Jenkins}, {Klaus}, {Koch}, {Latham}, {MacQueen}, {Sasselov}, {Steffen},
  {Twicken}, \& {Winn}}]{ref:Kepler36}
{Carter}, J.~A., {Agol}, E., {Chaplin}, W.~J., {et~al.} 2012, Science, 337, 556

\bibitem[{{Casagrande} {et~al.}(2011){Casagrande}, {Sch{\"o}nrich}, {Asplund},
  {Cassisi}, {Ram{\'{\i}}rez}, {Mel{\'e}ndez}, {Bensby}, \&
  {Feltzing}}]{ref:victor_2}
{Casagrande}, L., {Sch{\"o}nrich}, R., {Asplund}, M., {et~al.} 2011, \aap, 530,
  A138

\bibitem[{{Chaplin} {et~al.}(2013{\natexlab{a}}){Chaplin}, {Basu}, {Huber},
  {Serenelli}, {Casagrande}, {Silva Aguirre}, {Ball}, {Creevey}, {Gizon},
  {Handberg}, {Karoff}, {Lutz}, {Marques}, {Miglio}, {Stello}, {Suran},
  {Pricopi}, {Metcalfe}, {Monteiro}, {Molenda-Zakowicz}, {Appourchaux},
  {Christensen-Dalsgaard}, {Elsworth}, {Garcia}, {Houdek}, {Kjeldsen},
  {Bonanno}, {Campante}, {Corsaro}, {Gaulme}, {Hekker}, {Mathur}, {Mosser},
  {Regulo}, \& {Salabert}}]{ref:explain_par_diff}
{Chaplin}, W.~J., {Basu}, S., {Huber}, D., {et~al.} 2013{\natexlab{a}}, ArXiv
  e-prints

\bibitem[{{Chaplin} {et~al.}(2011){Chaplin}, {Kjeldsen},
  {Christensen-Dalsgaard}, {Basu}, {Miglio}, {Appourchaux}, {Bedding},
  {Elsworth}, {Garc{\'{\i}}a}, {Gilliland}, {Girardi}, {Houdek}, {Karoff},
  {Kawaler}, {Metcalfe}, {Molenda-{\.Z}akowicz}, {Monteiro}, {Thompson},
  {Verner}, {Ballot}, {Bonanno}, {Brand{\~a}o}, {Broomhall}, {Bruntt},
  {Campante}, {Corsaro}, {Creevey}, {Do{\u g}an}, {Esch}, {Gai}, {Gaulme},
  {Hale}, {Handberg}, {Hekker}, {Huber}, {Jim{\'e}nez}, {Mathur}, {Mazumdar},
  {Mosser}, {New}, {Pinsonneault}, {Pricopi}, {Quirion}, {R{\'e}gulo},
  {Salabert}, {Serenelli}, {Silva Aguirre}, {Sousa}, {Stello}, {Stevens},
  {Suran}, {Uytterhoeven}, {White}, {Borucki}, {Brown}, {Jenkins}, {Kinemuchi},
  {Van Cleve}, \& {Klaus}}]{ref:kepler_chaplin}
{Chaplin}, W.~J., {Kjeldsen}, H., {Christensen-Dalsgaard}, J., {et~al.} 2011,
  Science, 332, 213

\bibitem[{{Chaplin} \& {Miglio}(2013)}]{ref:ast_sca_chaplin}
{Chaplin}, W.~J. \& {Miglio}, A. 2013, \araa, 51, 353

\bibitem[{{Chaplin} {et~al.}(2013{\natexlab{b}}){Chaplin}, {Sanchis-Ojeda},
  {Campante}, {Handberg}, {Stello}, {Winn}, {Basu}, {Christensen-Dalsgaard},
  {Davies}, {Metcalfe}, {Buchhave}, {Fischer}, {Bedding}, {Cochran},
  {Elsworth}, {Gilliland}, {Hekker}, {Huber}, {Isaacson}, {Karoff}, {Kawaler},
  {Kjeldsen}, {Latham}, {Lund}, {Lundkvist}, {Marcy}, {Miglio}, {Barclay}, \&
  {Lissauer}}]{ref:kepler50and65}
{Chaplin}, W.~J., {Sanchis-Ojeda}, R., {Campante}, T.~L., {et~al.}
  2013{\natexlab{b}}, ArXiv e-prints

\bibitem[{{Christensen-Dalsgaard}(2008)}]{ref:astec}
{Christensen-Dalsgaard}, J. 2008, \apss, 316, 13

\bibitem[{{Christensen-Dalsgaard} \& {Houdek}(2010)}]{ref:abundances_matter}
{Christensen-Dalsgaard}, J. \& {Houdek}, G. 2010, \apss, 328, 51

\bibitem[{{Demory} {et~al.}(2011){Demory}, {Seager}, {Madhusudhan}, {Kjeldsen},
  {Christensen-Dalsgaard}, {Gillon}, {Rowe}, {Welsh}, {Adams}, {Dupree},
  {McCarthy}, {Kulesa}, {Borucki}, \& {Koch}}]{ref:kepler7}
{Demory}, B.-O., {Seager}, S., {Madhusudhan}, N., {et~al.} 2011, \apjl, 735,
  L12

\bibitem[{{Eggenberger} {et~al.}(2004){Eggenberger}, {Charbonnel}, {Talon},
  {Meynet}, {Maeder}, {Carrier}, \& {Bourban}}]{ref:cenAB}
{Eggenberger}, P., {Charbonnel}, C., {Talon}, S., {et~al.} 2004, \aap, 417, 235

\bibitem[{{Erspamer} \& {North}(2003)}]{ref:metal_thetaCyg}
{Erspamer}, D. \& {North}, P. 2003, \aap, 398, 1121

\bibitem[{{Ferguson} {et~al.}(2005){Ferguson}, {Alexander}, {Allard}, {Barman},
  {Bodnarik}, {Hauschildt}, {Heffner-Wong}, \& {Tamanai}}]{ref:lowT_2005}
{Ferguson}, J.~W., {Alexander}, D.~R., {Allard}, F., {et~al.} 2005, \apj, 623,
  585

\bibitem[{{Fogtmann-Schulz} {et~al.}(2013){Fogtmann-Schulz}, {Hinrup}, {Van
  Eylen}, {Christensen-Dalsgaard}, {Kjeldsen}, {Silva Aguirre}, \&
  {Tingley}}]{ref:kepler10_again}
{Fogtmann-Schulz}, A., {Hinrup}, B., {Van Eylen}, V., {et~al.} 2013, ArXiv
  e-prints

\bibitem[{{Formicola} {et~al.}(2004){Formicola}, {Imbriani}, {Costantini},
  {Angulo}, {Bemmerer}, {Bonetti}, {Broggini}, {Corvisiero}, {Cruz},
  {Descouvemont}, {F{\"u}l{\"o}p}, {Gervino}, {Guglielmetti}, {Gustavino},
  {Gy{\"u}rky}, {Jesus}, {Junker}, {Lemut}, {Menegazzo}, {Prati}, {Roca},
  {Rolfs}, {Romano}, {Rossi Alvarez}, {Sch{\"u}mann}, {Somorjai}, {Straniero},
  {Strieder}, {Terrasi}, {Trautvetter}, {Vomiero}, \& {Zavatarelli}}]{ref:14N}
{Formicola}, A., {Imbriani}, G., {Costantini}, H., {et~al.} 2004, Physics
  Letters B, 591, 61

\bibitem[{{Gai} {et~al.}(2011){Gai}, {Basu}, {Chaplin}, \&
  {Elsworth}}]{ref:pip_ybplus}
{Gai}, N., {Basu}, S., {Chaplin}, W.~J., \& {Elsworth}, Y. 2011, \apj, 730, 63

\bibitem[{{Gilliland} {et~al.}(2013){Gilliland}, {Marcy}, {Rowe}, {Rogers},
  {Torres}, {Fressin}, {Lopez}, {Buchhave}, {Christensen-Dalsgaard},
  {D{\'e}sert}, {Henze}, {Isaacson}, {Jenkins}, {Lissauer}, {Chaplin}, {Basu},
  {Metcalfe}, {Elsworth}, {Handberg}, {Hekker}, {Huber}, {Karoff}, {Kjeldsen},
  {Lund}, {Lundkvist}, {Miglio}, {Charbonneau}, {Ford}, {Fortney}, {Haas},
  {Howard}, {Howell}, {Ragozzine}, \& {Thompson}}]{ref:kepler68}
{Gilliland}, R.~L., {Marcy}, G.~W., {Rowe}, J.~F., {et~al.} 2013, \apj, 766, 40

\bibitem[{{Grevesse} \& {Sauval}(1998)}]{ref:GS98}
{Grevesse}, N. \& {Sauval}, A.~J. 1998, Space Science Reviews, 85, 161

\bibitem[{{Guzik} {et~al.}(2011){Guzik}, {Houdek}, {Chaplin}, {Kurtz},
  {Gilliland}, {Mullally}, {Rowe}, {Haas}, {Bryson}, {Still}, \&
  {Boyajian}}]{ref:largesep_thetaCyg_too}
{Guzik}, J.~A., {Houdek}, G., {Chaplin}, W.~J., {et~al.} 2011, ArXiv e-prints

\bibitem[{{Hansen} {et~al.}(2004){Hansen}, {Kawaler}, \&
  {Trimble}}]{ref:hkt_book}
{Hansen}, C.~J., {Kawaler}, S.~D., \& {Trimble}, V. 2004, {Stellar interiors :
  physical principles, structure, and evolution}

\bibitem[{{Huber} {et~al.}(2013){Huber}, {Chaplin}, {Christensen-Dalsgaard},
  {Gilliland}, {Kjeldsen}, {Buchhave}, {Fischer}, {Lissauer}, {Rowe},
  {Sanchis-Ojeda}, {Basu}, {Handberg}, {Hekker}, {Howard}, {Isaacson},
  {Karoff}, {Latham}, {Lund}, {Lundkvist}, {Marcy}, {Miglio}, {Silva Aguirre},
  {Stello}, {Arentoft}, {Barclay}, {Bedding}, {Burke}, {Christiansen},
  {Elsworth}, {Haas}, {Kawaler}, {Metcalfe}, {Mullally}, \&
  {Thompson}}]{ref:huber_2013}
{Huber}, D., {Chaplin}, W.~J., {Christensen-Dalsgaard}, J., {et~al.} 2013,
  \apj, 767, 127

\bibitem[{{Huber} {et~al.}(2012){Huber}, {Ireland}, {Bedding}, {Brand{\~a}o},
  {Piau}, {Maestro}, {White}, {Bruntt}, {Casagrande}, {Molenda-{\.Z}akowicz},
  {Silva Aguirre}, {Sousa}, {Barclay}, {Burke}, {Chaplin},
  {Christensen-Dalsgaard}, {Cunha}, {De Ridder}, {Farrington}, {Frasca},
  {Garc{\'{\i}}a}, {Gilliland}, {Goldfinger}, {Hekker}, {Kawaler}, {Kjeldsen},
  {McAlister}, {Metcalfe}, {Miglio}, {Monteiro}, {Pinsonneault}, {Schaefer},
  {Stello}, {Stumpe}, {Sturmann}, {Sturmann}, {ten Brummelaar}, {Thompson},
  {Turner}, \& {Uytterhoeven}}]{ref:huber_interferometry}
{Huber}, D., {Ireland}, M.~J., {Bedding}, T.~R., {et~al.} 2012, \apj, 760, 32

\bibitem[{{Iglesias} \& {Rogers}(1996)}]{ref:opal_opacity}
{Iglesias}, C.~A. \& {Rogers}, F.~J. 1996, \apj, 464, 943

\bibitem[{{Kervella} {et~al.}(2003){Kervella}, {Th{\'e}venin}, {S{\'e}gransan},
  {Berthomieu}, {Lopez}, {Morel}, \& {Provost}}]{ref:cenAB_diameter}
{Kervella}, P., {Th{\'e}venin}, F., {S{\'e}gransan}, D., {et~al.} 2003, \aap,
  404, 1087

\bibitem[{{Kippenhahn} {et~al.}(2013){Kippenhahn}, {Weigert}, \&
  {Weiss}}]{ref:mlt_garstec}
{Kippenhahn}, R., {Weigert}, A., \& {Weiss}, A. 2013, {Stellar Structure and
  Evolution}

\bibitem[{{Kjeldsen} \& {Bedding}(1995)}]{ref:sca_kjeldsen}
{Kjeldsen}, H. \& {Bedding}, T.~R. 1995, \aap, 293, 87

\bibitem[{{Kjeldsen} {et~al.}(2008){Kjeldsen}, {Bedding}, \&
  {Christensen-Dalsgaard}}]{ref:betaHyi_MR}
{Kjeldsen}, H., {Bedding}, T.~R., \& {Christensen-Dalsgaard}, J. 2008, \apjl,
  683, L175

\bibitem[{{Koch} {et~al.}(2010){Koch}, {Borucki}, {Basri}, {Batalha}, {Brown},
  {Caldwell}, {Christensen-Dalsgaard}, {Cochran}, {DeVore}, {Dunham},
  {Gautier}, {Geary}, {Gilliland}, {Gould}, {Jenkins}, {Kondo}, {Latham},
  {Lissauer}, {Marcy}, {Monet}, {Sasselov}, {Boss}, {Brownlee}, {Caldwell},
  {Dupree}, {Howell}, {Kjeldsen}, {Meibom}, {Morrison}, {Owen}, {Reitsema},
  {Tarter}, {Bryson}, {Dotson}, {Gazis}, {Haas}, {Kolodziejczak}, {Rowe}, {Van
  Cleve}, {Allen}, {Chandrasekaran}, {Clarke}, {Li}, {Quintana}, {Tenenbaum},
  {Twicken}, \& {Wu}}]{ref:kepler_koch}
{Koch}, D.~G., {Borucki}, W.~J., {Basri}, G., {et~al.} 2010, \apjl, 713, L79

\bibitem[{{Latham} {et~al.}(2010){Latham}, {Borucki}, {Koch}, {Brown},
  {Buchhave}, {Basri}, {Batalha}, {Caldwell}, {Cochran}, {Dunham}, {F{\H
  u}r{\'e}sz}, {Gautier}, {Geary}, {Gilliland}, {Howell}, {Jenkins},
  {Lissauer}, {Marcy}, {Monet}, {Rowe}, \& {Sasselov}}]{ref:kepler7_spec}
{Latham}, D.~W., {Borucki}, W.~J., {Koch}, D.~G., {et~al.} 2010, \apjl, 713,
  L140

\bibitem[{{Liebert} {et~al.}(2013){Liebert}, {Fontaine}, {Young}, {Williams},
  \& {Arnett}}]{ref:procyon_age}
{Liebert}, J., {Fontaine}, G., {Young}, P.~A., {Williams}, K.~A., \& {Arnett},
  D. 2013, \apj, 769, 7

\bibitem[{{Mathur} {et~al.}(2010){Mathur}, {Garc{\'{\i}}a}, {R{\'e}gulo},
  {Creevey}, {Ballot}, {Salabert}, {Arentoft}, {Quirion}, {Chaplin}, \&
  {Kjeldsen}}]{ref:pip_mathur}
{Mathur}, S., {Garc{\'{\i}}a}, R.~A., {R{\'e}gulo}, C., {et~al.} 2010, \aap,
  511, A46

\bibitem[{{P{\'a}l} {et~al.}(2008){P{\'a}l}, {Bakos}, {Torres}, {Noyes},
  {Latham}, {Kov{\'a}cs}, {Marcy}, {Fischer}, {Butler}, {Sasselov}, {Sip{\H
  o}cz}, {Esquerdo}, {Kov{\'a}cs}, {Stefanik}, {L{\'a}z{\'a}r}, {Papp}, \&
  {S{\'a}ri}}]{ref:hatp7_spec}
{P{\'a}l}, A., {Bakos}, G.~{\'A}., {Torres}, G., {et~al.} 2008, \apj, 680, 1450

\bibitem[{{Paxton} {et~al.}(2011){Paxton}, {Bildsten}, {Dotter}, {Herwig},
  {Lesaffre}, \& {Timmes}}]{ref:mesa}
{Paxton}, B., {Bildsten}, L., {Dotter}, A., {et~al.} 2011, \apjs, 192, 3

\bibitem[{{Paxton} {et~al.}(2013){Paxton}, {Cantiello}, {Arras}, {Bildsten},
  {Brown}, {Dotter}, {Mankovich}, {Montgomery}, {Stello}, {Timmes}, \&
  {Townsend}}]{ref:mesa_new}
{Paxton}, B., {Cantiello}, M., {Arras}, P., {et~al.} 2013, \apjs, 208, 4

\bibitem[{{Pietrinferni} {et~al.}(2004){Pietrinferni}, {Cassisi}, {Salaris}, \&
  {Castelli}}]{ref:victor_1}
{Pietrinferni}, A., {Cassisi}, S., {Salaris}, M., \& {Castelli}, F. 2004, \apj,
  612, 168

\bibitem[{{Pourbaix} {et~al.}(2002){Pourbaix}, {Nidever}, {McCarthy}, {Butler},
  {Tinney}, {Marcy}, {Jones}, {Penny}, {Carter}, {Bouchy}, {Pepe}, {Hearnshaw},
  {Skuljan}, {Ramm}, \& {Kent}}]{ref:cenAB_masses}
{Pourbaix}, D., {Nidever}, D., {McCarthy}, C., {et~al.} 2002, \aap, 386, 280

\bibitem[{{Quirion} {et~al.}(2010){Quirion}, {Christensen-Dalsgaard}, \&
  {Arentoft}}]{ref:pip_seek}
{Quirion}, P.-O., {Christensen-Dalsgaard}, J., \& {Arentoft}, T. 2010, \apj,
  725, 2176

\bibitem[{{Ram{\'{\i}}rez} {et~al.}(2009){Ram{\'{\i}}rez}, {Mel{\'e}ndez}, \&
  {Asplund}}]{ref:metal_16CygAB}
{Ram{\'{\i}}rez}, I., {Mel{\'e}ndez}, J., \& {Asplund}, M. 2009, \aap, 508, L17

\bibitem[{{Rogers} \& {Nayfonov}(2002)}]{ref:opal_2002}
{Rogers}, F.~J. \& {Nayfonov}, A. 2002, \apj, 576, 1064

\bibitem[{{Rogers} {et~al.}(1996){Rogers}, {Swenson}, \&
  {Iglesias}}]{ref:opal_1996}
{Rogers}, F.~J., {Swenson}, F.~J., \& {Iglesias}, C.~A. 1996, \apj, 456, 902

\bibitem[{{Santos} {et~al.}(2004){Santos}, {Bouchy}, {Mayor}, {Pepe}, {Queloz},
  {Udry}, {Lovis}, {Bazot}, {Benz}, {Bertaux}, {Lo Curto}, {Delfosse},
  {Mordasini}, {Naef}, {Sivan}, \& {Vauclair}}]{ref:muArae_spec}
{Santos}, N.~C., {Bouchy}, F., {Mayor}, M., {et~al.} 2004, \aap, 426, L19

\bibitem[{{Silva Aguirre} {et~al.}(2011){Silva Aguirre}, {Ballot}, {Serenelli},
  \& {Weiss}}]{ref:age_msturnof_diff}
{Silva Aguirre}, V., {Ballot}, J., {Serenelli}, A.~M., \& {Weiss}, A. 2011,
  \aap, 529, A63

\bibitem[{{Silva Aguirre} {et~al.}(2013){Silva Aguirre}, {Basu}, {Brand{\~a}o},
  {Christensen-Dalsgaard}, {Deheuvels}, {Do{\u g}an}, {Metcalfe}, {Serenelli},
  {Ballot}, {Chaplin}, {Cunha}, {Weiss}, {Appourchaux}, {Casagrande},
  {Cassisi}, {Creevey}, {Garc{\'{\i}}a}, {Lebreton}, {Noels}, {Sousa},
  {Stello}, {White}, {Kawaler}, \& {Kjeldsen}}]{ref:vsa_2013_ages}
{Silva Aguirre}, V., {Basu}, S., {Brand{\~a}o}, I.~M., {et~al.} 2013, \apj,
  769, 141

\bibitem[{{Silva Aguirre} {et~al.}(2012){Silva Aguirre}, {Casagrande}, {Basu},
  {Campante}, {Chaplin}, {Huber}, {Miglio}, {Serenelli}, {Ballot}, {Bedding},
  {Christensen-Dalsgaard}, {Creevey}, {Elsworth}, {Garc{\'{\i}}a}, {Gilliland},
  {Hekker}, {Kjeldsen}, {Mathur}, {Metcalfe}, {Monteiro}, {Mosser},
  {Pinsonneault}, {Stello}, {Weiss}, {Tenenbaum}, {Twicken}, \&
  {Uddin}}]{ref:vsa_2012_radii}
{Silva Aguirre}, V., {Casagrande}, L., {Basu}, S., {et~al.} 2012, \apj, 757, 99

\bibitem[{{Soriano} \& {Vauclair}(2010)}]{ref:muArae}
{Soriano}, M. \& {Vauclair}, S. 2010, \aap, 513, A49

\bibitem[{{Steigman}(2010)}]{ref:steigman_2010}
{Steigman}, G. 2010, \jcap, 4, 29

\bibitem[{{Stello} {et~al.}(2009{\natexlab{a}}){Stello}, {Chaplin}, {Basu},
  {Elsworth}, \& {Bedding}}]{ref:stello_2009}
{Stello}, D., {Chaplin}, W.~J., {Basu}, S., {Elsworth}, Y., \& {Bedding}, T.~R.
  2009{\natexlab{a}}, \mnras, 400, L80

\bibitem[{{Stello} {et~al.}(2009{\natexlab{b}}){Stello}, {Chaplin}, {Bruntt},
  {Creevey}, {Garc{\'{\i}}a-Hern{\'a}ndez}, {Monteiro}, {Moya}, {Quirion},
  {Sousa}, {Su{\'a}rez}, {Appourchaux}, {Arentoft}, {Ballot}, {Bedding},
  {Christensen-Dalsgaard}, {Elsworth}, {Fletcher}, {Garc{\'{\i}}a}, {Houdek},
  {Jim{\'e}nez-Reyes}, {Kjeldsen}, {New}, {R{\'e}gulo}, {Salabert}, \&
  {Toutain}}]{ref:pip_radius}
{Stello}, D., {Chaplin}, W.~J., {Bruntt}, H., {et~al.} 2009{\natexlab{b}},
  \apj, 700, 1589

\bibitem[{{Torres} {et~al.}(2012){Torres}, {Fischer}, {Sozzetti}, {Buchhave},
  {Winn}, {Holman}, \& {Carter}}]{ref:torres_2012}
{Torres}, G., {Fischer}, D.~A., {Sozzetti}, A., {et~al.} 2012, \apj, 757, 161

\bibitem[{{Toutain} \& {Fr\"{o}hlich}(1992)}]{ref:largesep_sun}
{Toutain}, T. \& {Fr\"{o}hlich}, C. 1992, \aap, 257, 287

\bibitem[{{Turcotte} {et~al.}(1998){Turcotte}, {Richer}, \&
  {Michaud}}]{ref:turcotte}
{Turcotte}, S., {Richer}, J., \& {Michaud}, G. 1998, \apj, 504, 559

\bibitem[{{Valle} {et~al.}(2014){Valle}, {Dell'Omodarme}, {Prada Moroni}, \&
  {Degl'Innocenti}}]{ref:valle}
{Valle}, G., {Dell'Omodarme}, M., {Prada Moroni}, P.~G., \& {Degl'Innocenti},
  S. 2014, \aap, 561, A125

\bibitem[{{Weiss} \& {Schlattl}(2008)}]{ref:garstec}
{Weiss}, A. \& {Schlattl}, H. 2008, \apss, 316, 99

\bibitem[{{White} {et~al.}(2011{\natexlab{a}}){White}, {Bedding}, {Stello},
  {Appourchaux}, {Ballot}, {Benomar}, {Bonanno}, {Broomhall}, {Campante},
  {Chaplin}, {Christensen-Dalsgaard}, {Corsaro}, {Do{\u g}an}, {Elsworth},
  {Fletcher}, {Garc{\'{\i}}a}, {Gaulme}, {Handberg}, {Hekker}, {Huber},
  {Karoff}, {Kjeldsen}, {Mathur}, {Mosser}, {Monteiro}, {R{\'e}gulo},
  {Salabert}, {Silva Aguirre}, {Thompson}, {Verner}, {Morris}, {Sanderfer}, \&
  {Seader}}]{ref:pip_white}
{White}, T.~R., {Bedding}, T.~R., {Stello}, D., {et~al.} 2011{\natexlab{a}},
  \apjl, 742, L3

\bibitem[{{White} {et~al.}(2011{\natexlab{b}}){White}, {Bedding}, {Stello},
  {Christensen-Dalsgaard}, {Huber}, \& {Kjeldsen}}]{ref:white_epsilon}
{White}, T.~R., {Bedding}, T.~R., {Stello}, D., {et~al.} 2011{\natexlab{b}},
  \apj, 743, 161

\bibitem[{{White} {et~al.}(2013){White}, {Huber}, {Maestro}, {Bedding},
  {Ireland}, {Baron}, {Boyajian}, {Che}, {Monnier}, {Pope}, {Roettenbacher},
  {Stello}, {Tuthill}, {Farrington}, {Goldfinger}, {McAlister}, {Schaefer},
  {Sturmann}, {Sturmann}, {ten Brummelaar}, \&
  {Turner}}]{ref:white_interferometry}
{White}, T.~R., {Huber}, D., {Maestro}, V., {et~al.} 2013, \mnras, 433, 1262

\end{thebibliography}


\begin{appendix}

\section{All AME figures}
\label{om_sec:morefigs}

In this appendix we provide the full set of AME figures. Section~\ref{om_subsec:massplots} contains the three mass plots, the six mean-density plots are given in Sect.~\ref{om_subsec:densityplots}, and Sect.~\ref{om_subsec:ageplots} holds the six age plots.

Explanations of the figures are given in the main text, and the figure captions provide the necessary information to use the figures. An overview of which colour corresponds to which mass is provided in Table~\ref{om_tab:colour_mass} below.

\begin{table*}
\caption{Colour - mass correspondence in the AME figures.}             
\label{om_tab:colour_mass}      
\centering          
\begin{tabular}{l|llllllllll}     
\hline\hline       

Colour		& Black	& Light blue	& Blue	& Green	& Yellow	& Red	& Magenta/Purple
			& Black	& Light blue	& Blue \\
$M/M_\sun$	& $0.7$	& $0.8$			& $0.9$	& $1.0$	& $1.1$		& $1.2$	& $1.3$
			& $1.4$	& $1.5$			& $1.6$ \\

\hline
\end{tabular}
\end{table*}


\subsection{mass plots}
\label{om_subsec:massplots}

\begin{figure}[h!]
	\centering
	\includegraphics[width=\hsize]{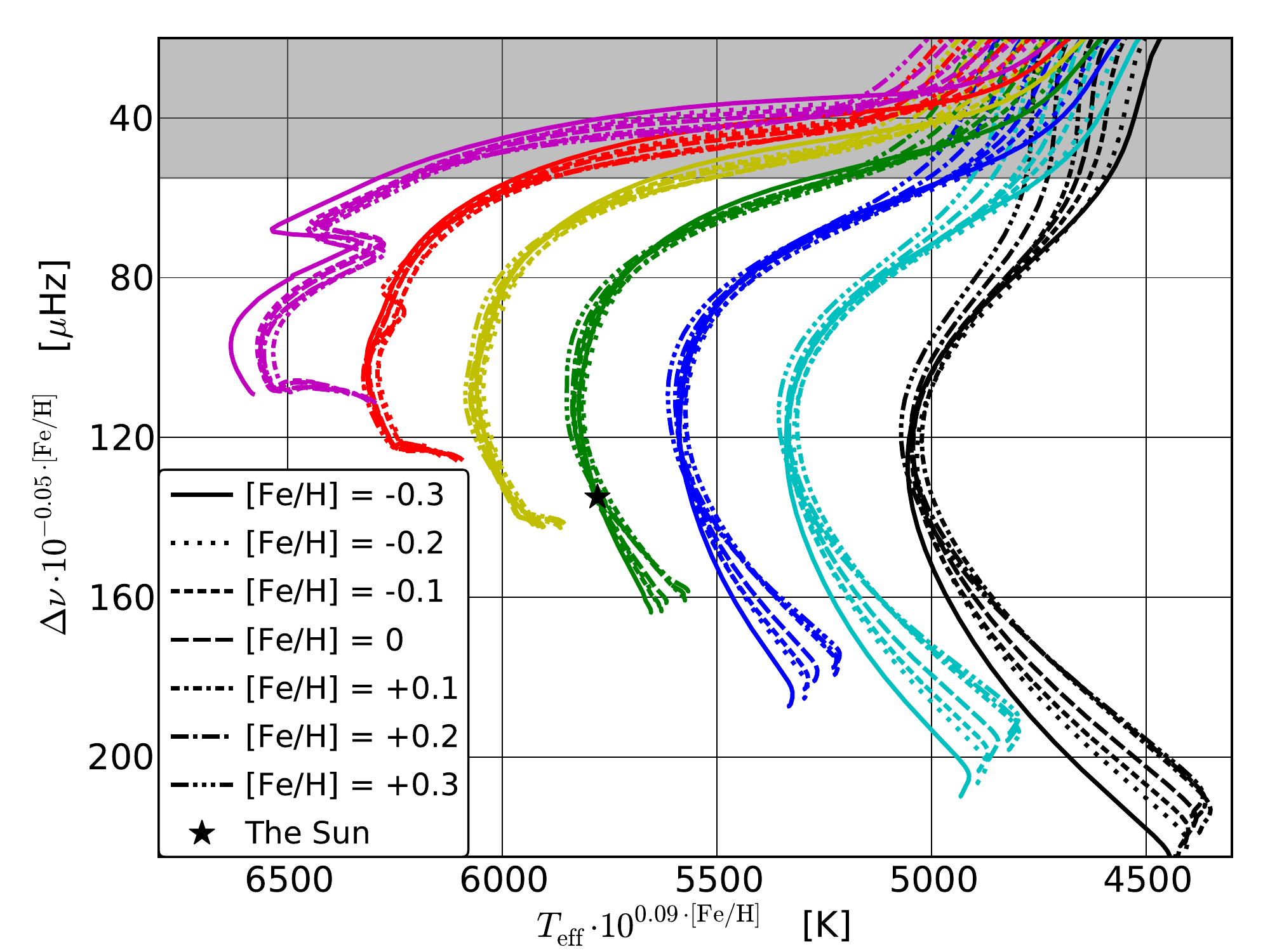}
	\caption{This plot can be used to determine the stellar mass. The colours give the masses of the lines ranging from $0.7 M_\sun$ (black) in steps of $0.1 M_\sun$ to $1.3 M_\sun$ (magenta/purple). The different line types are explained in the legend. The star shows the location of the Sun in the figure. The shaded area in the top of the figure indicates the region where \figref{om_fig:m_lm_ldnu} should be used instead.}
	\label{om_fig:m_lm}
\end{figure}

\begin{figure}[h!]
	\centering
	\includegraphics[width=\hsize]{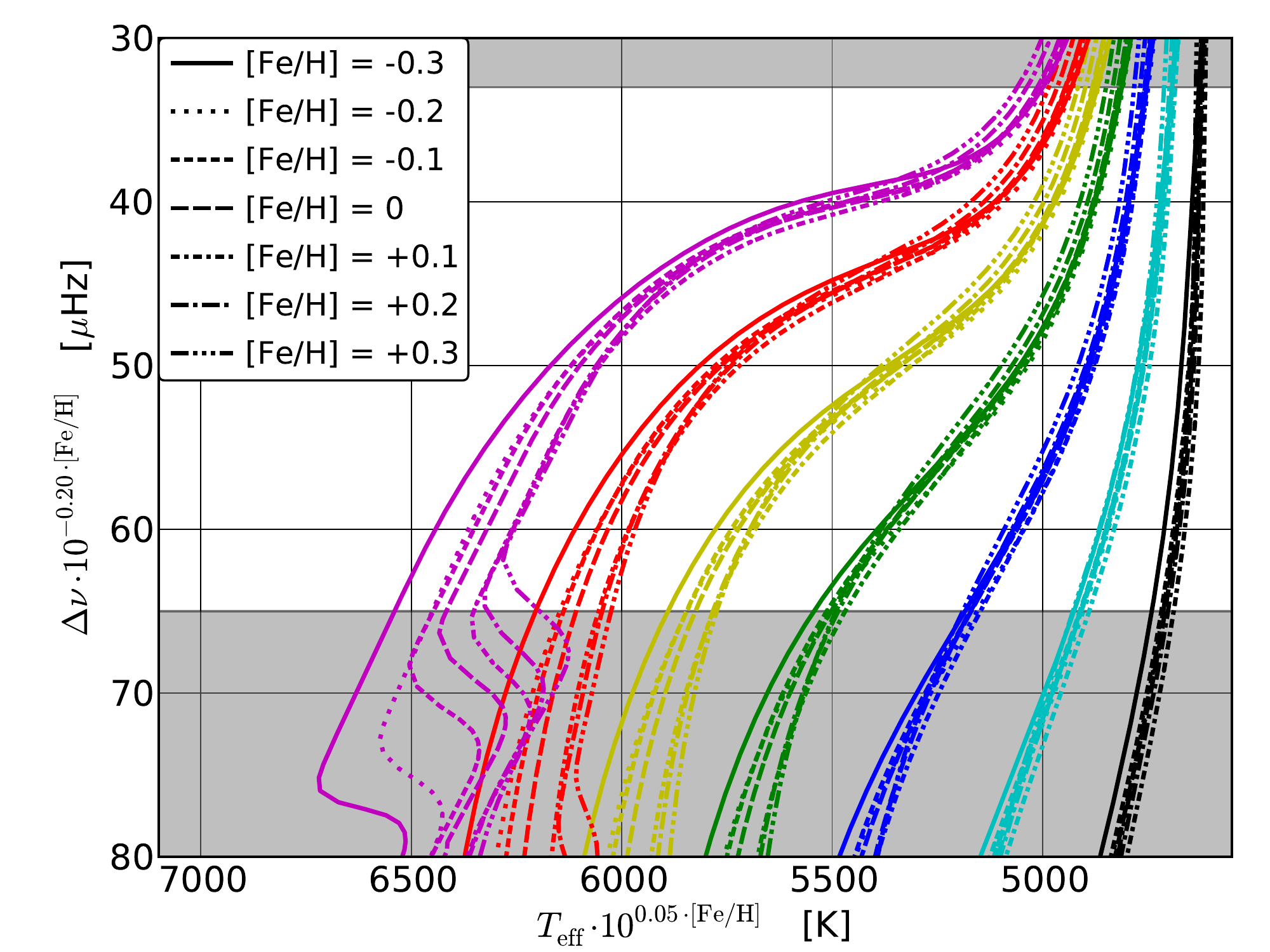}
	\caption{This plot can be used to determine the stellar mass. The colours give the masses of the lines ranging from $0.7 M_\sun$ (black) in steps of $0.1 M_\sun$ to $1.3 M_\sun$ (magenta/purple). The different line types are explained in the legend. The shaded areas in the figure indicate where \figref{om_fig:m_lm} should be used instead (bottom) and where the lines are hard to disentangle (top).}
	\label{om_fig:m_lm_ldnu}
\end{figure}

\begin{figure}[h!]
	\centering
	\includegraphics[width=\hsize]{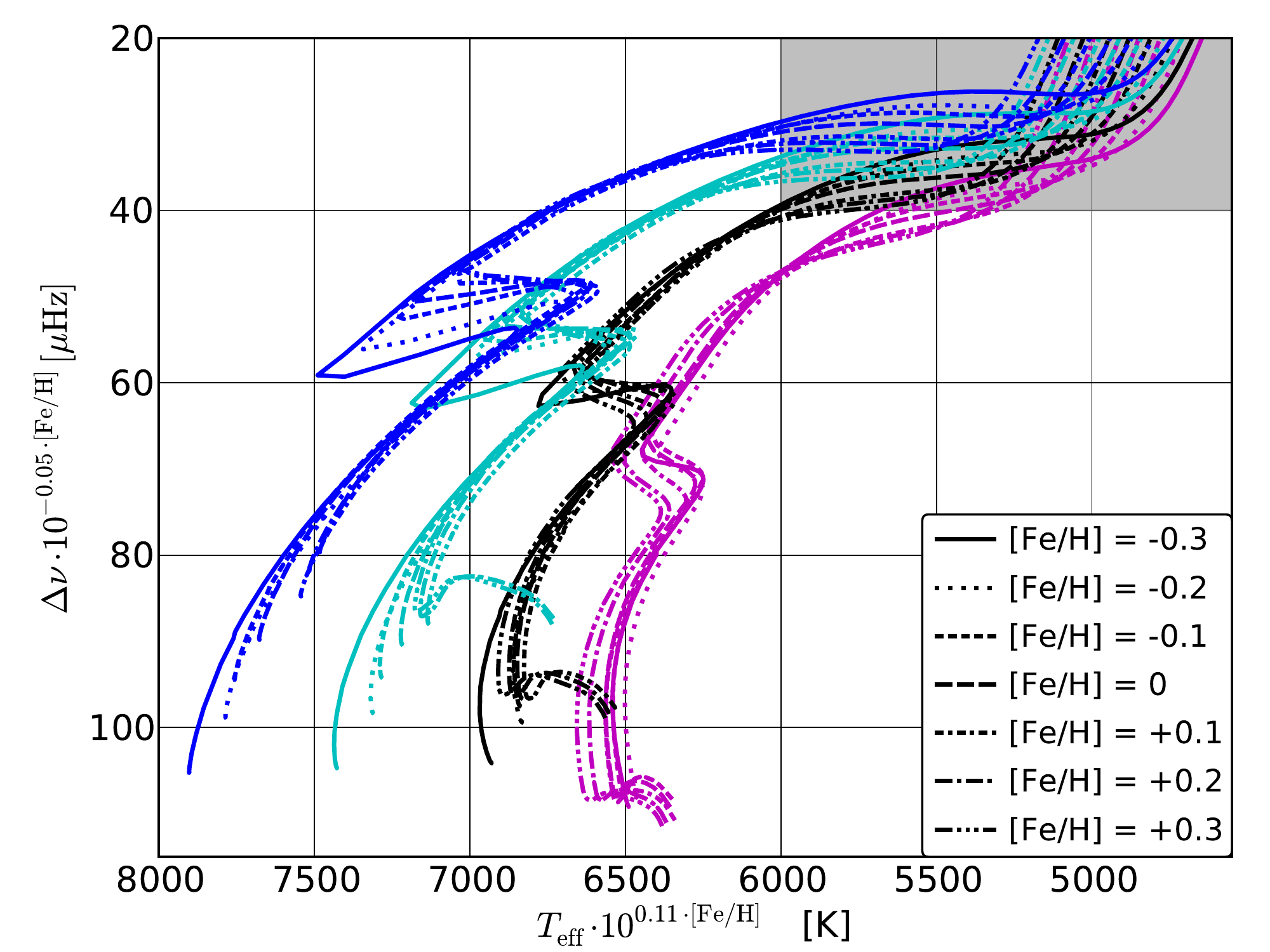}
	\caption{This plot can be used to determine the stellar mass. The colours give the masses of the lines ranging from $1.3 M_\sun$ (magenta/purple) in steps of $0.1 M_\sun$ to $1.6 M_\sun$ (blue). The different line types are explained in the legend. The shaded area in the figure indicates where the lines are hard to disentangle. In the middle of the plot, where the tracks mix (ordinate values of $\sim 60 \ \micro\hertz$), is where a star will go through the MS turn-off.}
	\label{om_fig:m_hm}
\end{figure}

The Figs.~\ref{om_fig:m_lm},~\ref{om_fig:m_lm_ldnu}, and~\ref{om_fig:m_hm} show the plots which can be used to determine the stellar mass.

\subsection{Mean-density plots}
\label{om_subsec:densityplots}

\begin{figure}[h!]
	\centering
	\includegraphics[width=\hsize]{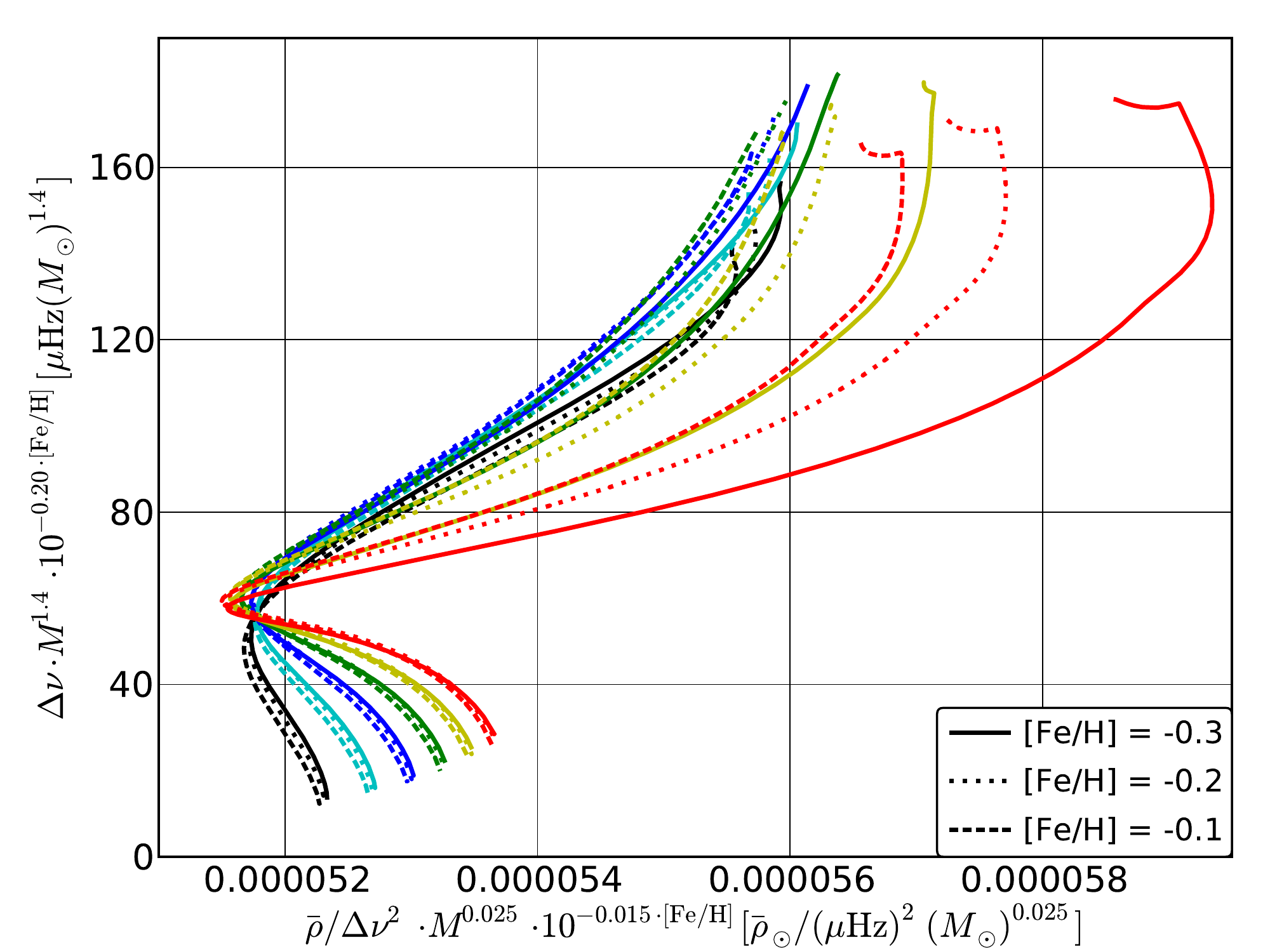}
	\caption{This plot can be used to determine the stellar mean-density for stars with $-0.30 \ \mathrm{dex} \leq [\mathrm{Fe}/\mathrm{H}] \leq -0.10 \ \mathrm{dex}$. The colours give the masses of the lines ranging from $0.7 M_\sun$ (black) in steps of $0.1 M_\sun$ to $1.2 M_\sun$ (red). The different line types are explained in the legend.}
	\label{om_fig:rho_lm_m03m01}
\end{figure}

\begin{figure}[h!]
	\centering
	\includegraphics[width=\hsize]{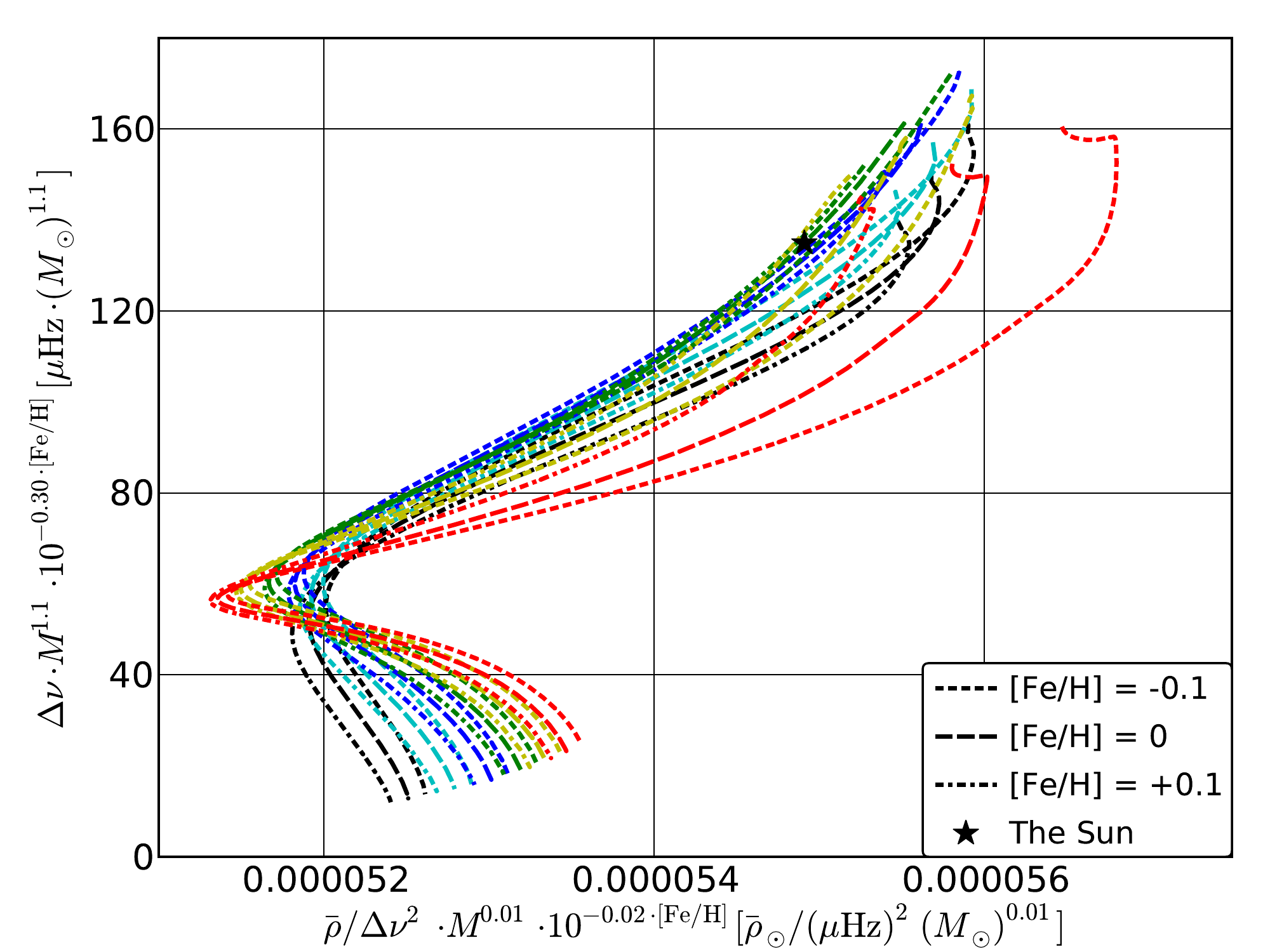}
	\caption{This plot can be used to determine the stellar mean-density for stars with $-0.10 \ \mathrm{dex} \leq [\mathrm{Fe}/\mathrm{H}] \leq +0.10 \ \mathrm{dex}$. The colours give the masses of the lines ranging from $0.7 M_\sun$ (black) in steps of $0.1 M_\sun$ to $1.2 M_\sun$ (red). The different line types are explained in the legend. The star indicates the position of the Sun.}
	\label{om_fig:rho_lm_m01p01}
\end{figure}

\begin{figure}[h!]
	\centering
	\includegraphics[width=\hsize]{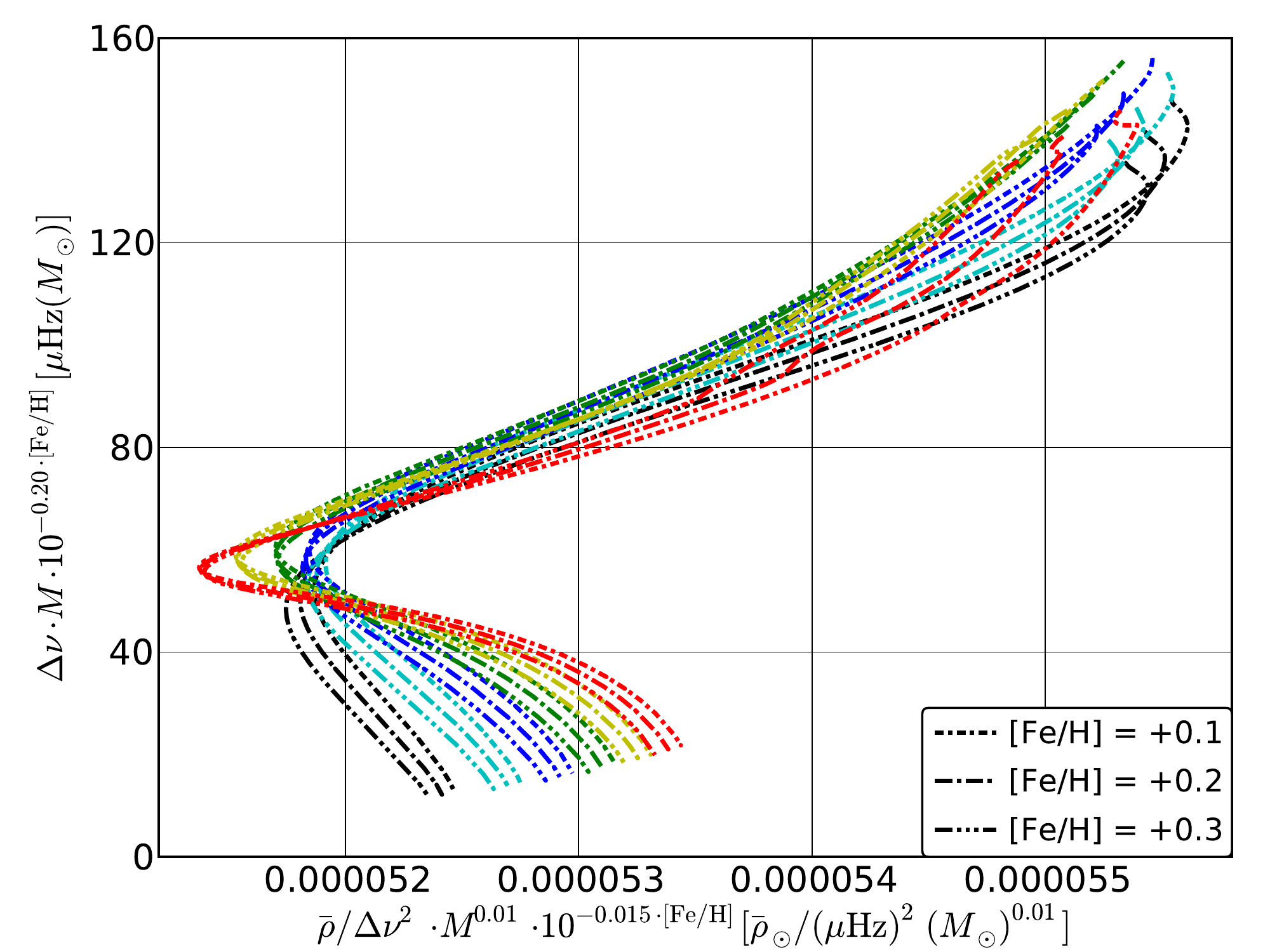}
	\caption{This plot can be used to determine the stellar mean-density for stars with $+0.10 \ \mathrm{dex} \leq [\mathrm{Fe}/\mathrm{H}] \leq +0.30 \ \mathrm{dex}$. The colours give the masses of the lines ranging from $0.7 M_\sun$ (black) in steps of $0.1 M_\sun$ to $1.2 M_\sun$ (red). The different line types are explained in the legend.}
	\label{om_fig:rho_lm_p01p03}
\end{figure}

\begin{figure}[h!]
	\centering
	\includegraphics[width=\hsize]{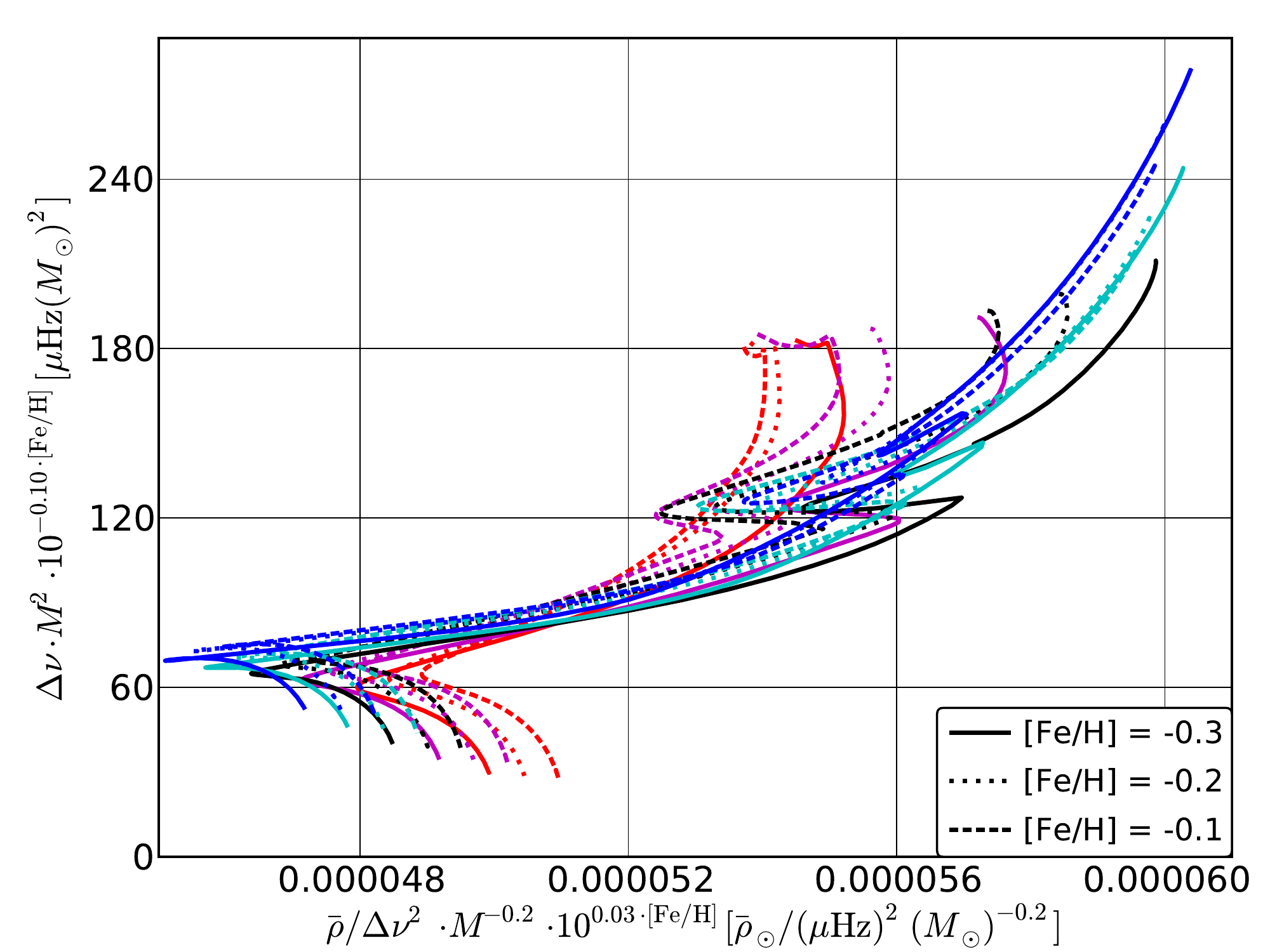}
	\caption{This plot can be used to determine the stellar mean-density for stars with $-0.30 \ \mathrm{dex} \leq [\mathrm{Fe}/\mathrm{H}] \leq -0.10 \ \mathrm{dex}$. The colours give the masses of the lines ranging from $1.2 M_\sun$ (red) in steps of $0.1 M_\sun$ to $1.6 M_\sun$ (blue). The different line types are explained in the legend.}
	\label{om_fig:rho_hm_m03m01}
\end{figure}

\begin{figure}[h!]
	\centering
	\includegraphics[width=\hsize]{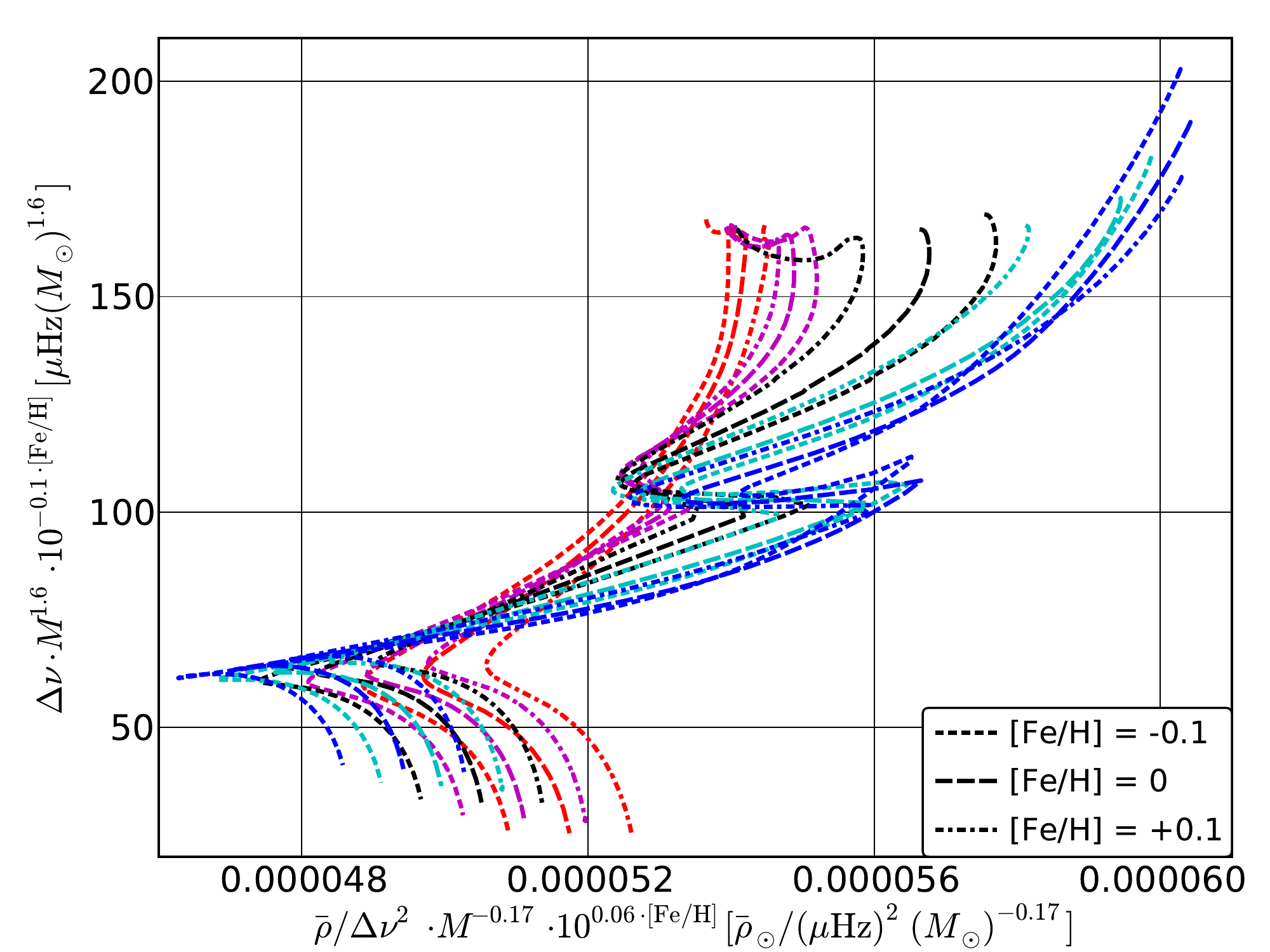}
	\caption{This plot can be used to determine the stellar mean-density for stars with $-0.10 \ \mathrm{dex} \leq [\mathrm{Fe}/\mathrm{H}] \leq +0.10 \ \mathrm{dex}$. The colours give the masses of the lines ranging from $1.2 M_\sun$ (red) in steps of $0.1 M_\sun$ to $1.6 M_\sun$ (blue). The different line types are explained in the legend.}
	\label{om_fig:rho_hm_m01p01}
\end{figure}

\begin{figure}[h!]
	\centering
	\includegraphics[width=\hsize]{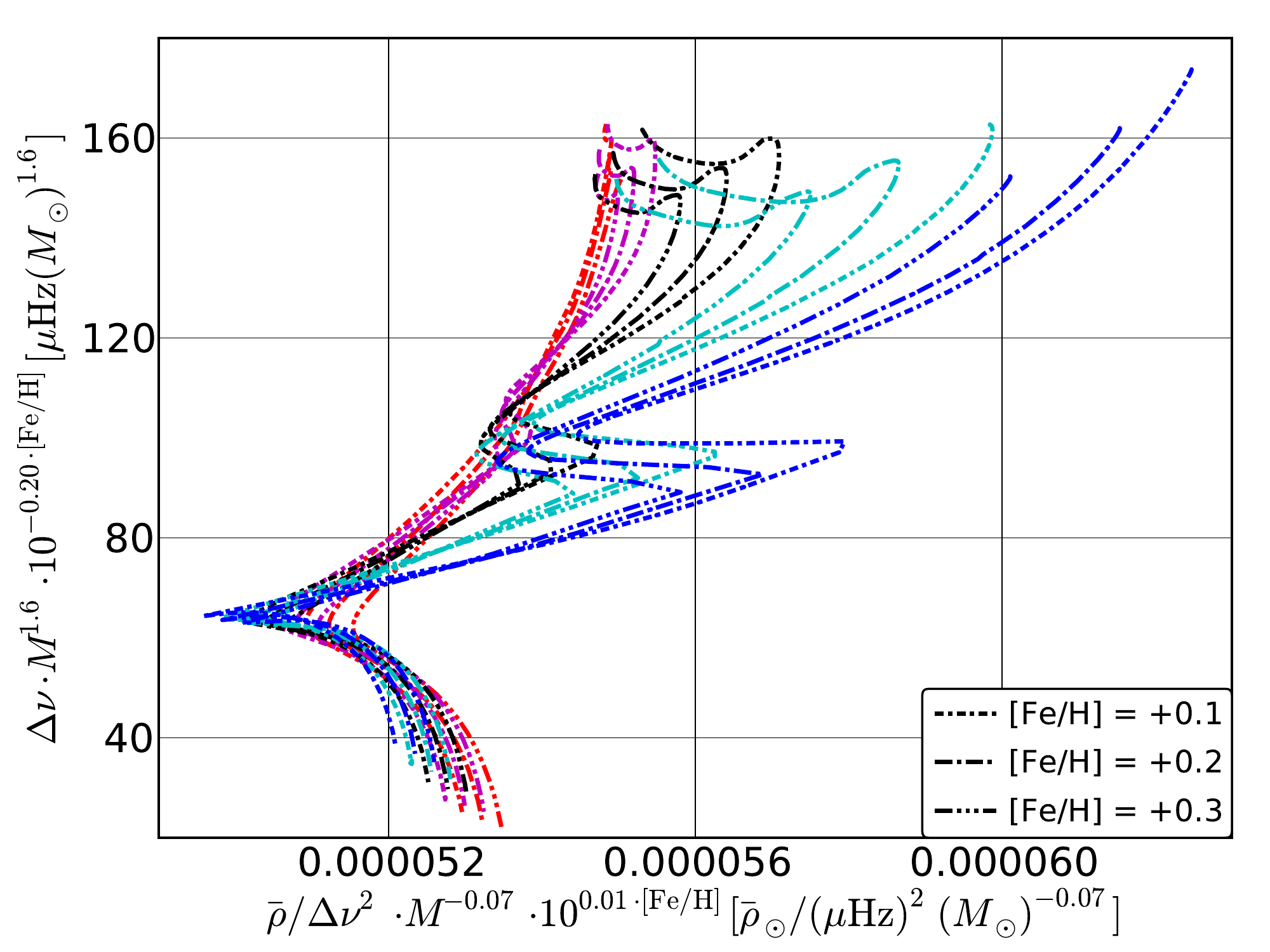}
	\caption{This plot can be used to determine the stellar mean-density for stars with $+0.10 \ \mathrm{dex} \leq [\mathrm{Fe}/\mathrm{H}] \leq +0.30 \ \mathrm{dex}$. The colours give the masses of the lines ranging from $1.2 M_\sun$ (red) in steps of $0.1 M_\sun$ to $1.6 M_\sun$ (blue). The different line types are explained in the legend.}
	\label{om_fig:rho_hm_p01p03}
\end{figure}

Figures~\ref{om_fig:rho_lm_m03m01} to~\ref{om_fig:rho_hm_p01p03} show the plots which can be used to obtain the stellar mean-density.

\subsection{Age plots}
\label{om_subsec:ageplots}

\begin{figure}[h!]
	\centering
	\includegraphics[width=\hsize]{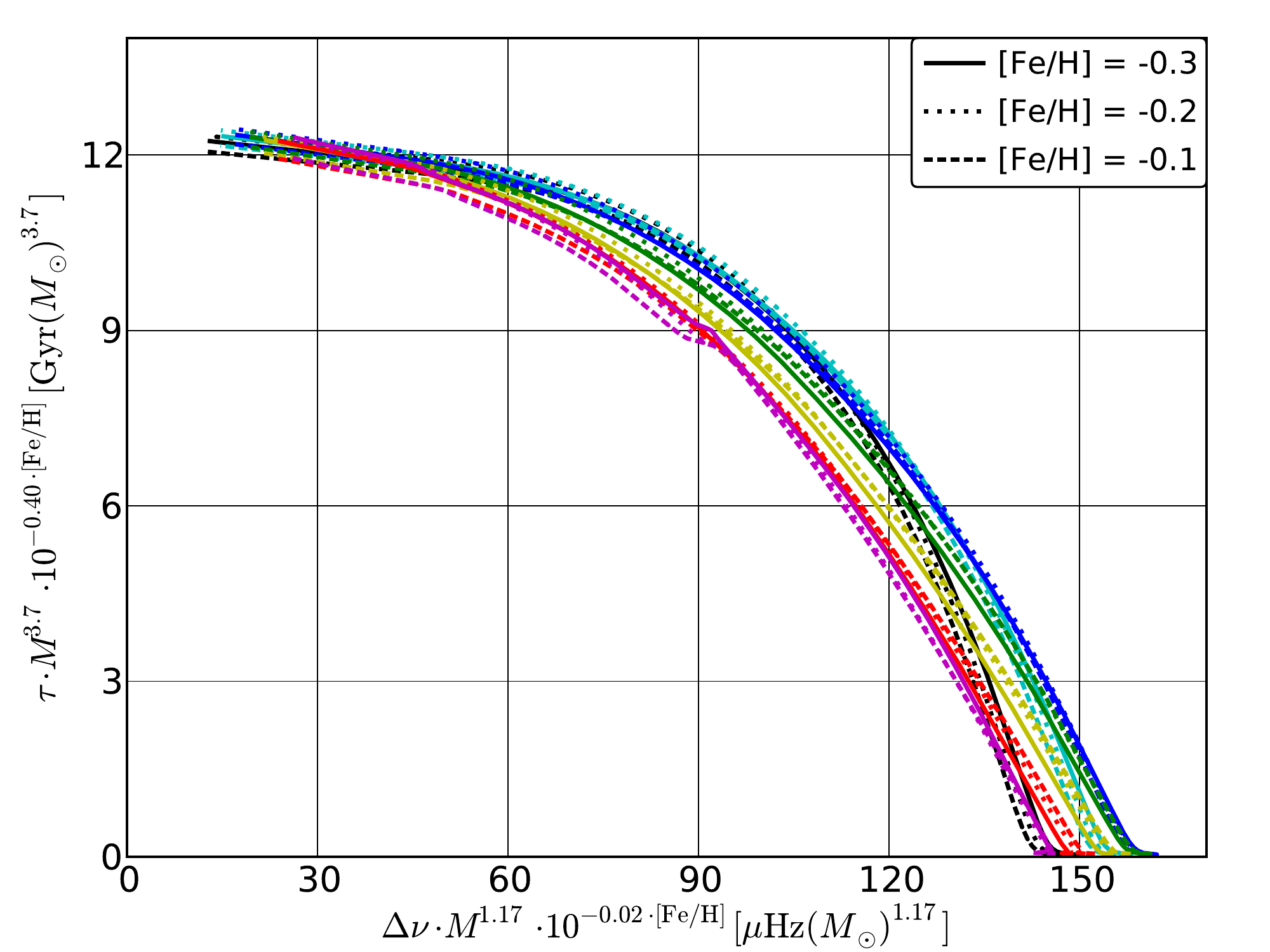}
	\caption{This plot can be used to determine the stellar age for stars with $-0.30 \ \mathrm{dex} \leq [\mathrm{Fe}/\mathrm{H}] \leq -0.10 \ \mathrm{dex}$. The colours give the masses of the lines ranging from $0.7 M_\sun$ (black) in steps of $0.1 M_\sun$ to $1.3 M_\sun$ (magenta/purple). The different line types are explained in the legend.}
	\label{om_fig:tau_lm_m03m01}
\end{figure}

\begin{figure}[h!]
	\centering
	\includegraphics[width=\hsize]{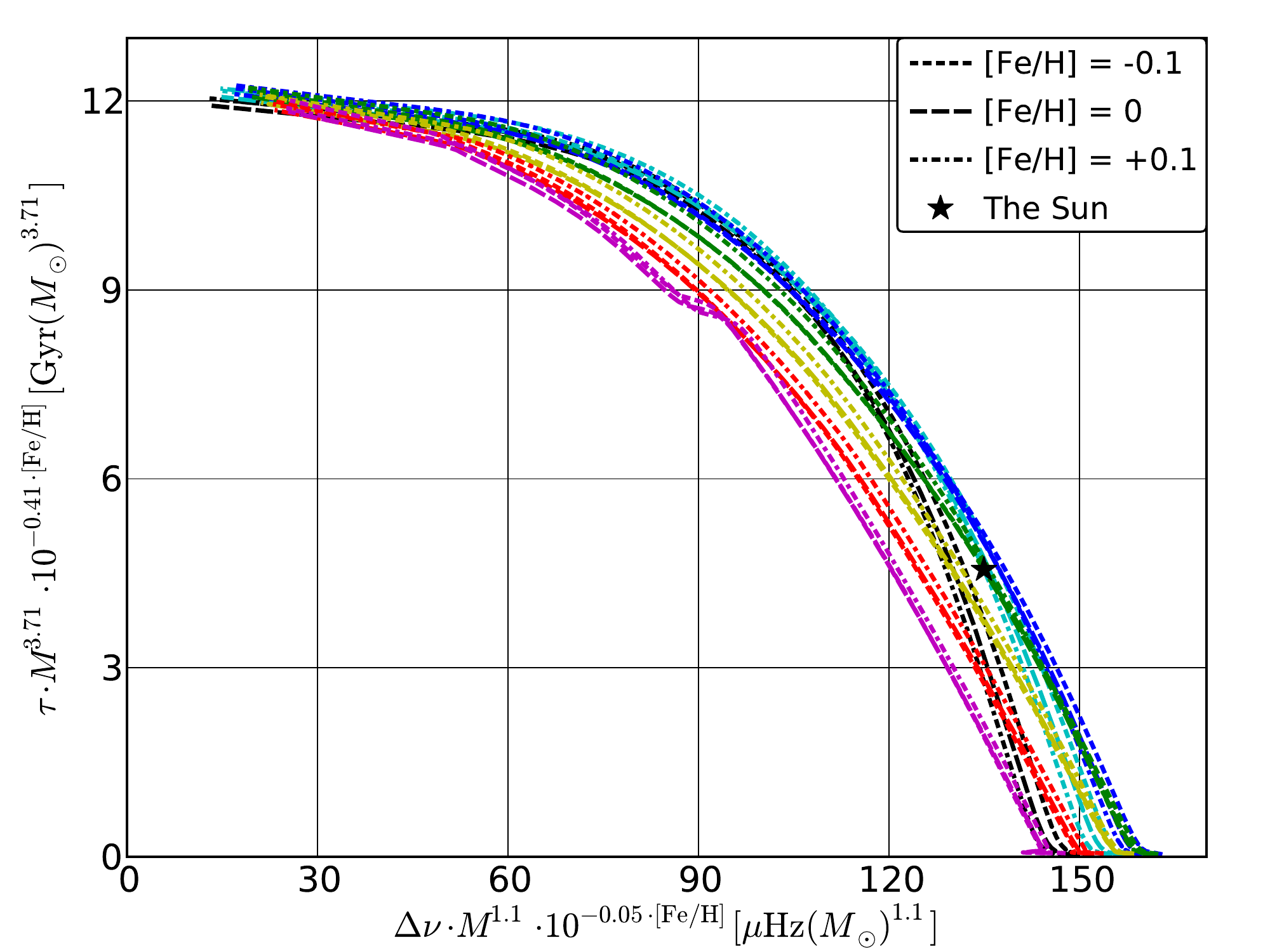}
	\caption{This plot can be used to determine the stellar age for stars with $-0.10 \ \mathrm{dex} \leq [\mathrm{Fe}/\mathrm{H}] \leq +0.10 \ \mathrm{dex}$. The colours give the masses of the lines ranging from $0.7 M_\sun$ (black) in steps of $0.1 M_\sun$ to $1.3 M_\sun$ (magenta/purple). The different line types are explained in the legend. The star shows the location of the Sun.}
	\label{om_fig:tau_lm_m01p01}
\end{figure}

\begin{figure}[h!]
	\centering
	\includegraphics[width=\hsize]{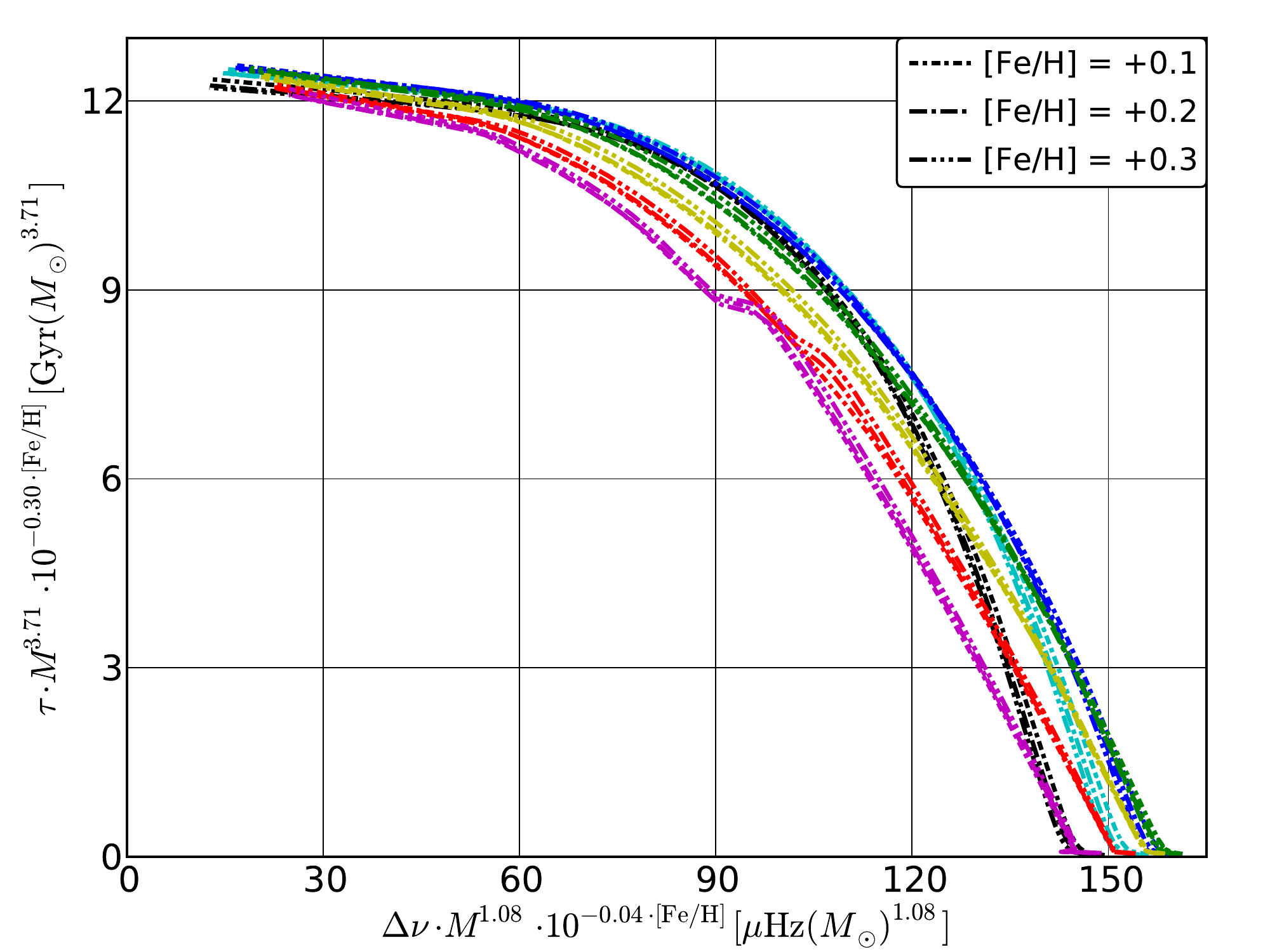}
	\caption{This plot can be used to determine the stellar age for stars with $+0.10 \ \mathrm{dex} \leq [\mathrm{Fe}/\mathrm{H}] \leq +0.30 \ \mathrm{dex}$. The colours give the masses of the lines ranging from $0.7 M_\sun$ (black) in steps of $0.1 M_\sun$ to $1.3 M_\sun$ (magenta/purple). The different line types are explained in the legend.}
	\label{om_fig:tau_lm_p01p03}
\end{figure}

\begin{figure}[h!]
	\centering
	\includegraphics[width=\hsize]{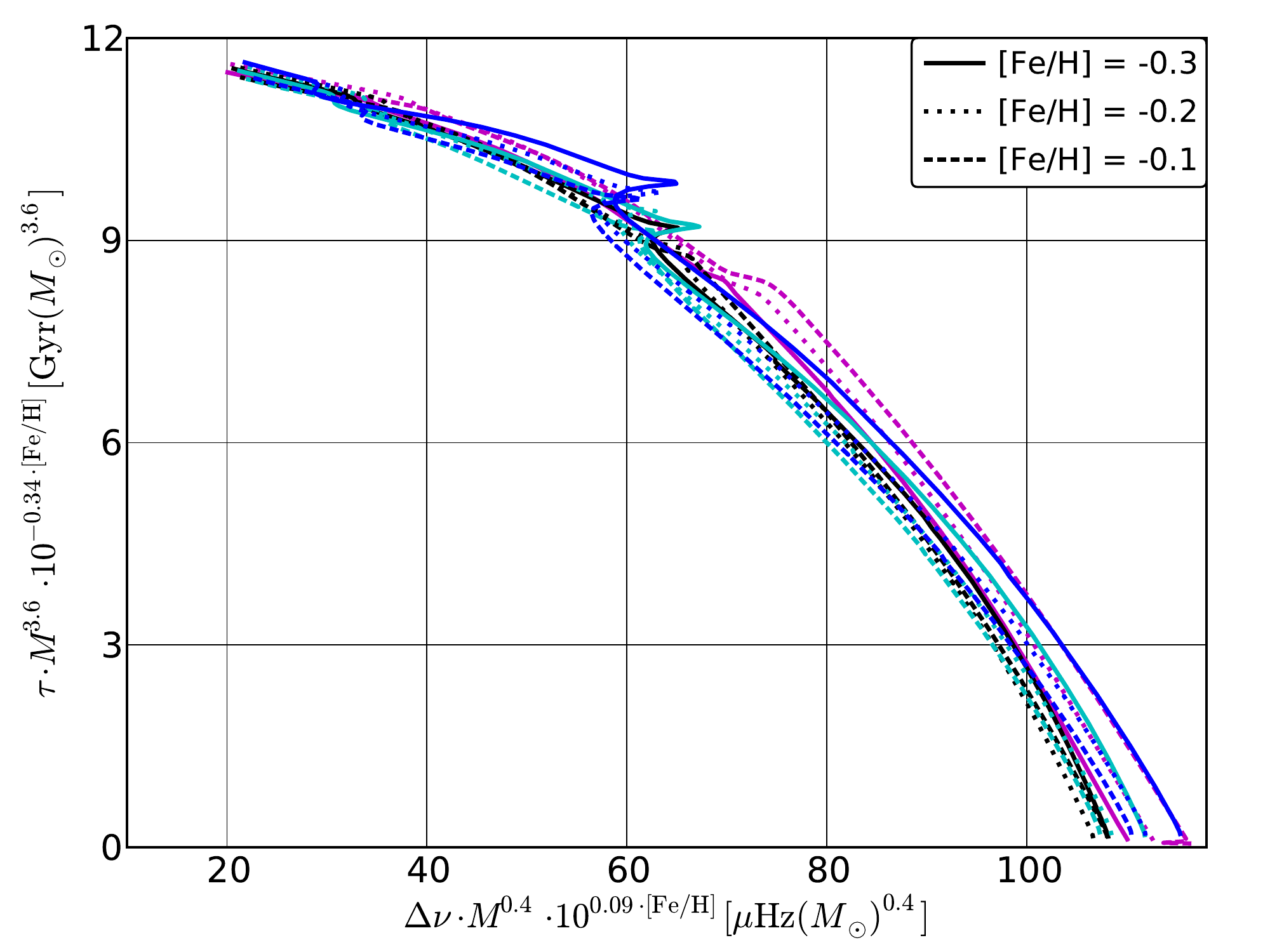}
	\caption{This plot can be used to determine the stellar age for stars with $-0.30 \ \mathrm{dex} \leq [\mathrm{Fe}/\mathrm{H}] \leq -0.10 \ \mathrm{dex}$. The colours give the masses of the lines ranging from $1.3 M_\sun$ (magenta/purple) in steps of $0.1 M_\sun$ to $1.6 M_\sun$ (blue). The different line types are explained in the legend.}
	\label{om_fig:tau_hm_m03m01}
\end{figure}

\begin{figure}[h!]
	\centering
	\includegraphics[width=\hsize]{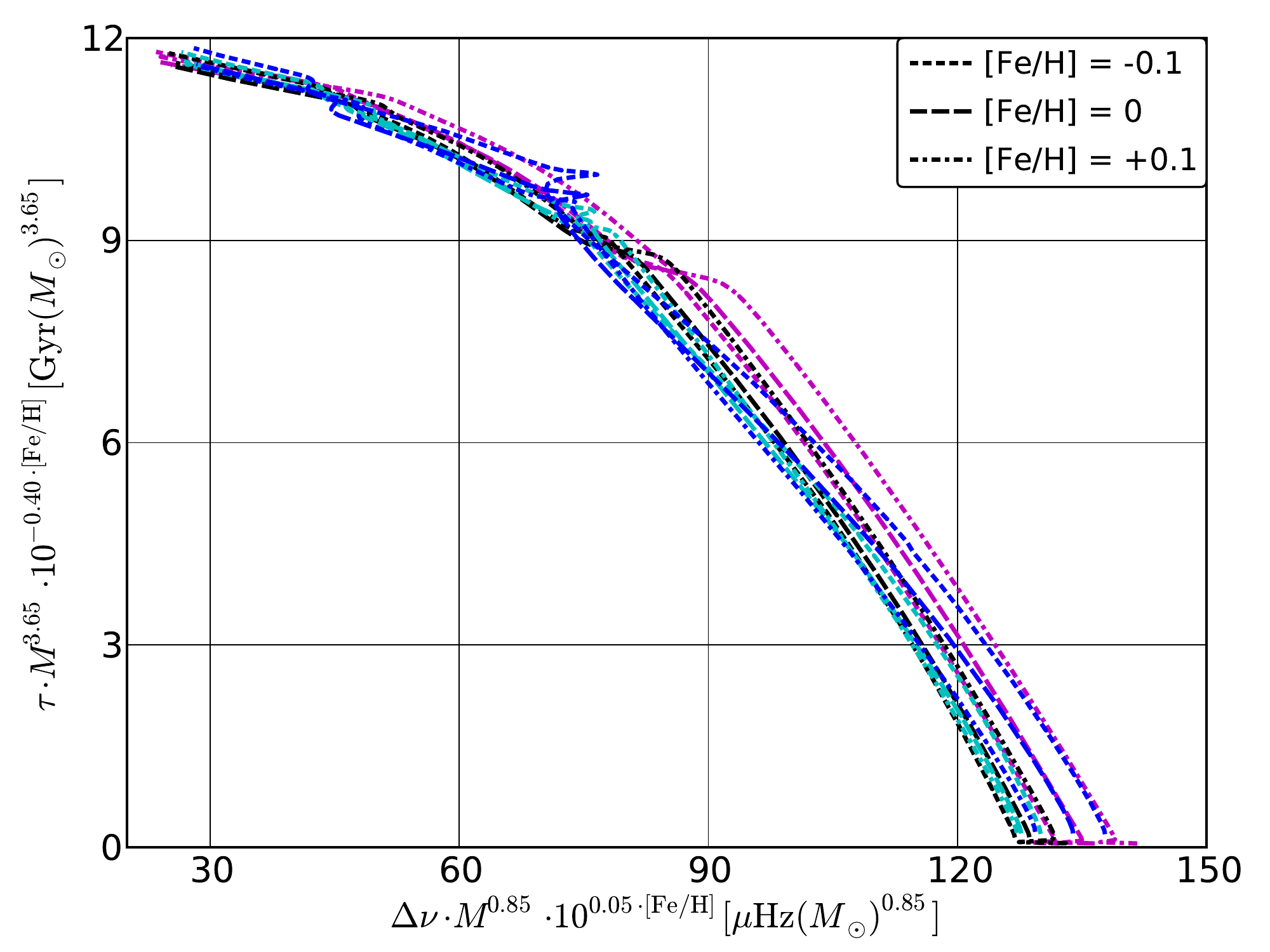}
	\caption{This plot can be used to determine the stellar age for stars with $-0.10 \ \mathrm{dex} \leq [\mathrm{Fe}/\mathrm{H}] \leq +0.10 \ \mathrm{dex}$. The colours give the masses of the lines ranging from $1.3 M_\sun$ (magenta/purple) in steps of $0.1 M_\sun$ to $1.6 M_\sun$ (blue). The different line types are explained in the legend.}
	\label{om_fig:tau_hm_m01p01}
\end{figure}

\begin{figure}[h!]
	\centering
	\includegraphics[width=\hsize]{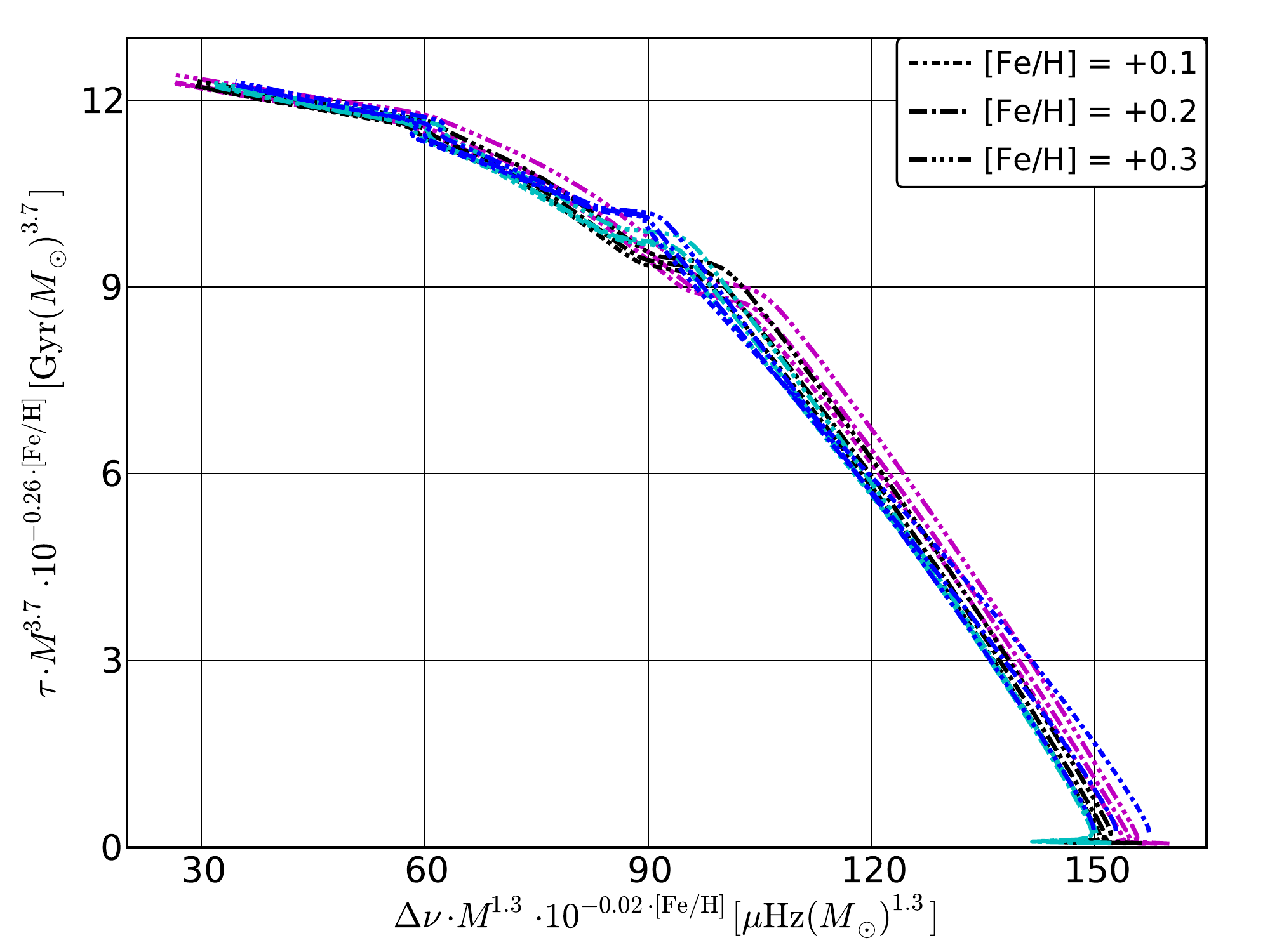}
	\caption{This plot can be used to determine the stellar age for stars with $+0.10 \ \mathrm{dex} \leq [\mathrm{Fe}/\mathrm{H}] \leq +0.30 \ \mathrm{dex}$. The colours give the masses of the lines ranging from $1.3 M_\sun$ (magenta/purple) in steps of $0.1 M_\sun$ to $1.6 M_\sun$ (blue). The different line types are explained in the legend.}
	\label{om_fig:tau_hm_p01p03}
\end{figure}

Figures~\ref{om_fig:tau_lm_m03m01} to~\ref{om_fig:tau_hm_p01p03} give the plots which can be used to estimate the stellar age.

\end{appendix}


\end{document}